\documentclass[12pt,preprint2]{aastex63}
\usepackage{multirow}

\usepackage[utf8]{inputenc}
\usepackage{natbib}
\usepackage{graphicx}
\usepackage{longtable}
\usepackage{enumerate}
\usepackage{amsmath}
\usepackage{hyperref}
\usepackage{CJK}
\usepackage[title]{appendix}
\usepackage{rotating}
\usepackage{longtable}

\def\msun{\ifmmode {\rm M}_{\mathord\odot}\else $M_{\mathord\odot}$\fi}
\def\rsun{\ifmmode {\rm R}_{\mathord\odot}\else $R_{\mathord\odot}$\fi}
\def\lsun{\ifmmode {\rm L}_{\mathord\odot}\else $L_{\mathord\odot}$\fi}

\def\co{$^{12}$CO}
\def\c18o{C$^{18}$O}
\def\h2{H$_{2}$}
\def\13co{$^{13}$CO}
\def\n2hp{$_{2}$H$^{+}$}

\def\radmc{{\sc radmc-3d}}
\def\cm2{cm$^{-2}$}
\def\cmc{cm$^{-3}$}

\newcommand{\kms}{km s$^{-1}$}

\newcommand{\CASI}{{\sc casi}}
\newcommand{\CASItD}{{\sc casi-3d}}

\def\orion{{\sc orion2}}

\shorttitle{}
\shortauthors{}

\begin{document}
\begin{CJK*}{UTF8}{gbsn}

\title{Application of Convolutional Neural Networks to Identify 
Protostellar
Outflows in CO Emission}

\author{Duo Xu}
\affil{Department of Astronomy, The University of Texas at Austin, Austin, TX 78712, USA;}
\email{xuduo117@utexas.edu}

\author{Stella S. R. Offner}
\affil{Department of Astronomy, The University of Texas at Austin, Austin, TX 78712, USA;}
\email{soffner@astro.as.utexas.edu}

\author{Robert Gutermuth}
\affil{Department of Astronomy, University of Massachusetts, Amherst, MA 01003, USA;}

\author{Colin Van Oort}
\affil{University of Vermont, Burlington, VT 05405, USA;}

\begin{abstract}
We adopt the deep learning method \CASItD\ (Convolutional Approach to Structure Identification-3D) to identify protostellar outflows in molecular line spectra. We conduct magneto-hydrodynamics simulations that model forming stars that launch protostellar outflows and use these to generate synthetic observations. We apply the 3D radiation transfer code \radmc\ to model \co\ ($J$=1-0) line emission from the simulated clouds. We train two \CASItD\ models: ME1 is trained to predict only the position of outflows, while MF is trained to predict the fraction of the mass coming from outflows in each voxel. The two models successfully identify all 60 previously visually identified outflows in Perseus. Additionally, \CASItD\ finds 20 new high-confidence outflows. All of these have coherent high-velocity structure, and 17 of them have nearby young stellar objects, while the remaining three are outside the Spitzer survey coverage. The mass, momentum and energy of individual outflows in Perseus predicted by model MF is comparable to the previous estimations. This similarity is due to a cancelation in errors: previous calculations missed outflow material with velocities comparable to the cloud velocity, however, they compensate for this by over-estimating the amount of mass at higher velocities that has contamination from non-outflow gas. We show outflows likely driven by older sources have more high-velocity gas compared to those driven by younger sources.

\end{abstract}


\section{Introduction} 

Protostars launch collimated bipolar outflows along magnetic field lines when accreting mass from their disks. Bipolar outflows eject high-velocity gas into their natal molecular clouds, injecting a substantial amount of energy into their surroundings \citep{2014prpl.conf..451F, 2016ARA&A..54..491B}. Jets and outflows also significantly reduce protostellar masses and accretion rates, which primarily resolves the low ``core-to-star” efficiency problem \citep{2008ApJ...687..340M,2012ApJ...747...22H,2013MNRAS.431.1719M,2014ApJ...784...61O}. In addition, outflows disperse some of the surrounding gas, reducing the global star formation rate \citep{2007ApJ...659.1394M, 2010ApJ...715.1170A, 2014ApJ...790..128F}. The extra momentum and energy from the outflows compressing and heating the gas can considerably change cloud properties \citep{2015ApJ...815...68M}.

Protostellar outflows also shape molecular cloud chemistry. For example, high-velocity outflows generate molecular bow shocks, which trigger chemical reactions that do not happen in quiescent environments, yielding complex physical and chemical conditions \citep{1996ARA&A..34..111B}. One of the most extreme examples is SiO, which is usually considered a shock tracer, whose abundance is enhanced by several orders of magnitude along the axes of molecular outflows \citep{1991A&A...243L..21B,1992A&A...254..315M}. 

In order to understand the impact of outflows on molecular clouds, especially the effect on their energy budget, a complete census of outflows is needed. Unfortunately, most outflows are asymmetric, and deeply embedded in dense clouds, making them difficult to identify \citep{2010ApJ...715.1170A, 2014ApJ...783...29D}.

Historically, astronomers have identified outflows and separated them from the surrounding gas ``by eye'' \citep{1996ARA&A..34..111B,2005ApJ...625..864Z,2008AJ....136.2391C,2015ApJS..219...20L}. For example, \citet{2010ApJ...715.1170A} identified 60 outflows in the Perseus molecular cloud by determining high-velocity features from a three-dimensional visualization. They concluded the total outflow energy is sufficient to replenish the dissipation of turbulence. However, \citet{2010ApJ...715.1170A} found no correlation between outflow strength (in terms of mass, momentum or energy) and star formation efficiency (SFE) in different regions of Perseus, which is contrary to the predictions of studies using simulations \citep[e.g.][]{2012ApJ...747...22H,2014ApJ...790..128F}. Incompleteness of the outflow sample or interlopers, i.e., false outflows, that originate from turbulence rather than feedback might be the reason. Alternatively, some additional feedback mechanism, such as stellar winds, rather than outflows may explain the lack of correlation.

Due to the difficulty and subjectivity of identifying outflows visually, different studies have drawn different conclusions about the importance of feedback. For example, \citet{2012MNRAS.425.2641N} identify 20 outflows in the Taurus molecular cloud and conclude that the impact of feedback is negligible compared to the dissipation of turbulence. However, \citet{2015ApJS..219...20L} identify twice as many outflows, whose energy injection rate is 16 times larger than before, yielding the opposite conclusion. Consequently, a complete and high-confidence outflow sample is required to understand the true impact of outflows.

Machine learning makes it possible to systematically and quickly identify outflow features. Several machine learning algorithms have been utilized to identify stellar feedback features \citep{2011ApJ...741...14B, 2014ApJS..214....3B, 2019ApJ...880...83V, 2020ApJ...890...64X,2020ApJS..248...15Z}. Support Vector Machines (SVM) were employed to distinguish a supernova remnant from the ambient gas \citep{2011ApJ...741...14B} and to identify molecular outflows in a dark cloud complex \citep{2020ApJS..248...15Z}. \citet{2020ApJS..248...15Z} adopted SVM to identify outflow features in \co\ and \13co\ emission in Cygnus. SVM is proficient in classification tasks, but it requires manual feature extraction to create a training set, which is arbitrary and likely to omit information compared to adopting complete images or sophisticated 3D models as the training set. Random Forest algorithms, which classify feature vectors by learning a series of decision rules, perform robustly in identifying stellar feedback bubbles in dust emission \citep{2014ApJS..214....3B, 2017ApJ...851..149X}. Similar to SVM, Random Forests necessitate manually extracted features as inputs. Moreover, the accuracy of their classifications are sensitive to the location of stellar feedback bubbles in the input images \citep{2014ApJS..214....3B, 2017ApJ...851..149X, 2019MNRAS.488.1141J}. Convolutional neural networks (CNNs) are a powerful new approach being applied to identify structures or objects, such as exoplanets \citep{2018AJ....155...94S} and stellar feedback bubbles \citep{2019ApJ...880...83V, 2020ApJ...890...64X}. CNNs are not sensitive to the position of the objects in the data, so it is straightforward to apply them to large surveys. Most importantly, CNNs do not require manually extracted features as inputs. Instead, CNNs extract features automatically from the data by applying different filters on different layers during training. A sophisticated algorithm, Convolutional Approach to Shell/Structure Identification \citep[\CASI,][]{2019ApJ...880...83V, 2020ApJ...890...64X} based on CNNs, was recently developed and successfully applied to identify bubbles produced by stellar feedback in both 2D and 3D simulation and observational data. \CASItD\ can identify bubble-like structures in position-position-velocity (PPV) molecular line spectra cubes, achieving a 96\% voxel-level accuracy \citep{2020ApJ...890...64X}. Furthermore,  \CASItD\ successfully infers hidden information in the data cube, e.g., the fraction of mass coming from feedback, which provides a more accurate feedback mass estimation. \citet{2020ApJ...890...64X} show that the newly calculated mass associated with shell features in the Taurus molecular cloud is an order of magnitude smaller than the previous estimates. This result underscores the power of CNNs to identify structures and constrain non-linear and complex physical processes like stellar feedback signatures.

In this paper, we adopt the deep learning method \CASItD\ to systemically identify protostellar outflows in CO data 
and conduct statistical studies. We describe \CASItD\ and how we generate the training set from synthetic observations in Section~\ref{Method}. In Section~\ref{Results}, we present the performance of the CNN model in identifying protostellar outflows in both synthetic and observational data. We discuss the application of the CNN model in different physical environments in Section~\ref{Discussion} and summarize our results and conclusions in Section~\ref{Conclusions}.

\section{Method}
\label{Method}

\subsection{Magneto-Hydrodynamics (MHD) Simulations} 
\label{Magneto-Hydrodynamics (MHD) Simulations}

We conduct magneto-hydrodynamics simulations with \orion\ \citep{2012ApJ...745..139L} to model forming stars that launch protostellar outflows. The simulation box is $2\times 2\times 2$ pc$^3$, with a total mass of $M=301.5~\msun$, mean particle density of $654$ cm$ ^{-3} $ and three-dimensional Mach number of 6.6, which places the simulated cloud on the line-width size relation, $\sigma_{\rm 1D}=0.72R^{0.5}_{\rm pc}$ \kms \citep{2007ARA&A..45..565M}. These conditions represent a piece of a typical low-mass star-forming region, such as the Perseus molecular cloud, in which we intend to identify protostellar outflows. We treat the gas as an ideal gas that perfectly thermally couples with the dust. The calculations use a basegrid of 256$^{3}$ with four adaptive mesh refinement (AMR) levels, where the minimum cell size is 100 AU. We initialize the density and velocity fields by driving the simulation gas for two mach crossing times without gravity but adding random large scale perturbations with Fourier modes $1\le k\le 2$ \citep{1999ApJ...524..169M}. Sink particles are formed when the gas density exceeds the Jeans criterion at the finest AMR level \citep{2004ApJ...611..399K}. The sink particles represent individual stars and adopt a sub-grid stellar evolution model that launches protostellar bipolar outflows \citep{2009ApJ...703..131O,2011ApJ...740..107C,2014MNRAS.439.3420M}. The outflow launching velocity is determined by the Keplerian velocity at the stellar surface, $v_{K}=\sqrt{GM_{*}/R_{*}}$. We use a ``tracer'' field to track the material launched by the outflows \citep{2017ApJ...847..104O}. A detailed description of the outflow method can be found in \citet{2011ApJ...740..107C}. 

\begin{figure*}[hbt!]
\centering
\includegraphics[width=0.98\linewidth]{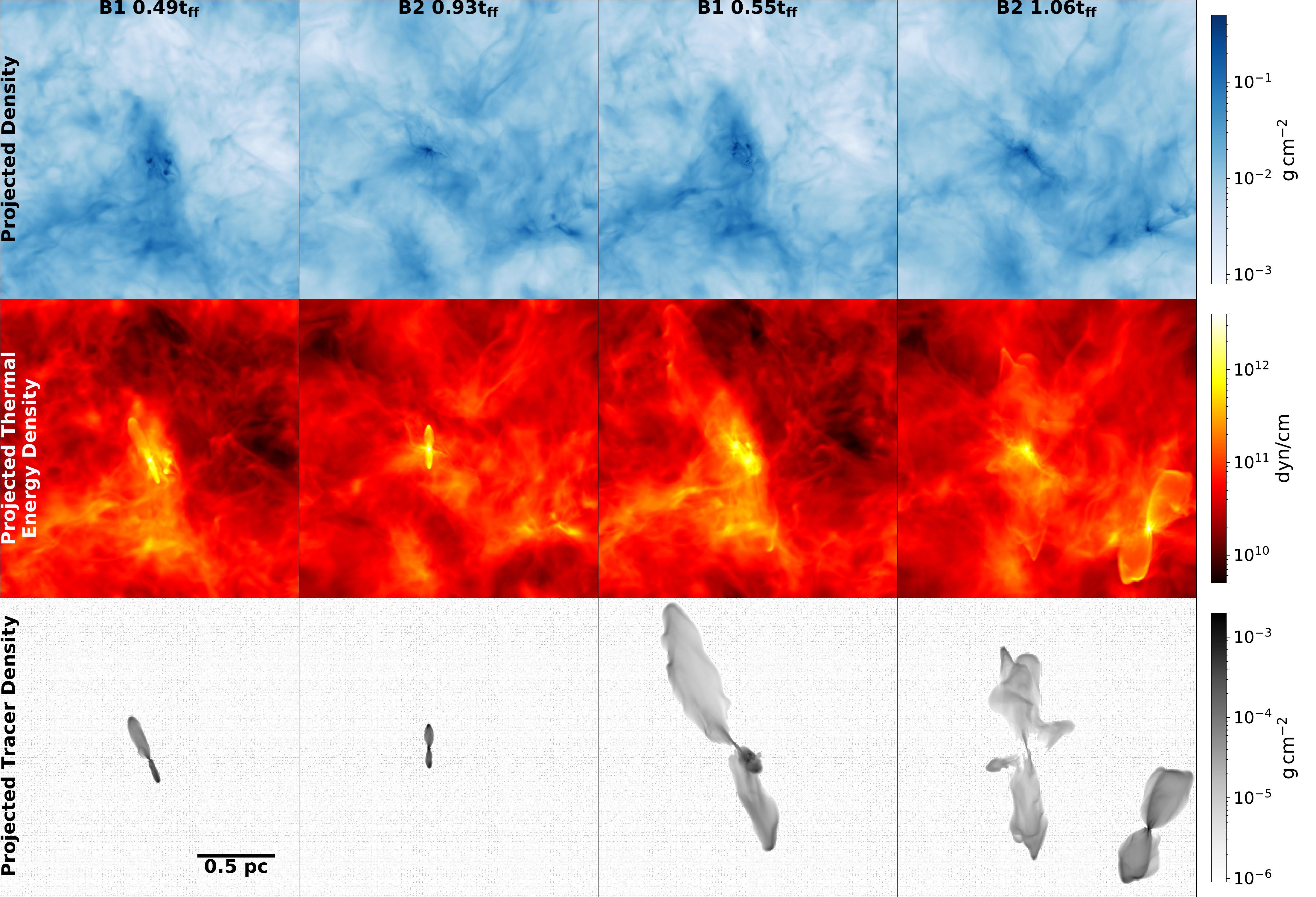}
\caption{Projected density (upper row), projected thermal density (middle row) and project tracer density (lower row) of different simulations at different evolutionary stages.}
\label{fig.simulation-density-energy-tracer}
\end{figure*}

We conduct three different simulations with different magnetic fields and turbulent driving patterns. We carry out the analysis at several simulation times to capture different evolutionary stages of the protostars and study the impact of these outflows on the surrounding molecular cloud. Table~\ref{Model Properties} shows the physical properties of the simulations. Figure~\ref{fig.simulation-density-energy-tracer} illustrates the projected density, projected thermal density and projected tracer density of different simulations. 

\begin{table*}[]
\begin{center}
\caption{Model Properties$^{a}$ \label{Model Properties}}
\begin{tabular}{ccccc}
\hline
Model & $B$ $(\mu G)$  &$\mu_{\Phi}$  & $N_{\rm seed}$ & $t_{\rm run}$ ($t_{\rm ff}$)    \\
 \hline
  {B1}   & {3.2} & {8} & {2}   & (0.49, 0.52, 0.55) $^{b}$  \\
  {B2}   & {0.8} & {32} & {1}   & (0.93, 1.06, 1.09) $^{c}$  \\
\hline
\multicolumn{5}{p{0.65\linewidth}}{Notes:}\\
\multicolumn{5}{p{0.65\linewidth}}{
$^{a}$ Model name, initial mean magnetic field, global mass-to-flux ratio ($\mu_{\Phi} = M_{gas} / M_{\Phi} = 2\pi G^{1/2} M_{\rm gas}/(B L^2)$), the number of different turbulent driving patterns used for each model and the evolutionary time in free-fall times. All models have $L$ = 2 pc, $M$ = 301.5 \msun, $T_{i}$ = 10 K and a free-time time of 1.31 Myr.}\\
\multicolumn{5}{p{0.65\linewidth}}{
$^{b}$ The first star particle is formed right before 0.49$t_{\rm ff}$.}\\
\multicolumn{5}{p{0.65\linewidth}}{
$^{c}$ The first star particle is formed right before 0.93$t_{\rm ff}$.}\\
\end{tabular}
\end{center}
\end{table*}

\subsection{CNN Architecture}
\label{CNN Architecture}

We adopt the same CNN architecture, \CASItD, from \citet{2020ApJ...890...64X}, which combines both residual networks \citep{He2016} and a ``U-net'' \citep{Ronneberger2015}. \CASItD\ has two parts. The ``encoder'' part extracts the features from the input data, then the ``decoder'' part translates these features into another data cube that represents the quantity being modeled by the network. We adopt the same hyper-parameters as \citet{2020ApJ...890...64X} to train our models to identify protostellar outflows within molecular clouds. The choice of hyper-parameters is discussed in detail in \citet{2020ApJ...890...64X}.

\subsection{Training Sets}
\label{Training Sets}

\subsubsection{Synthetic Observations} 
\label{Synthetic Observations}
We apply the publicly available radiation transfer code \radmc\ \citep{2012ascl.soft02015D} to model the \co\ ($J$=1-0) and 
\13co\ ($J$=1-0) line emission of the simulation gas. Since most outflows are not detectable in \13co\ in high-velocity channels due to the limited sensitivity of most single dish telescopes \citep{2010ApJ...715.1170A}, we use \co\ emission to identify the outflows. To better calculate the mass of outflows where \co\ is optically thick, we combine both \co\ and \13co\ if there is distinct \13co\ emission in the corresponding position. Otherwise, we use \co\ only to calculate the mass. 

To construct the synthetic observations, we adopt the simulation density, temperature and velocity distributions for the \radmc\ inputs. In the radiative transfer, we assume \h2 is the only collisional partner of \co. We take 62 as the fiducial abundance ratio between \co\ and \13co\ and 10$^{-4}$ as the abundance ratio between \co\ and \h2 \citep{2010ApJ...715.1170A}. However, when $T>1000$~K or $n(\rm H_{2})<$~50~\cmc\, where \co\ and \13co\ are likely to be dissociated, we set the abundance to zero. In addition, we also set the \co\ and \13co\ abundance to zero in conditions where it would freeze out onto dust grains ($n(\rm H_{2})>10^{5}$~\cmc) or where it would be dissociated by strong shocks ($|v| > 20$~km/s) \citep{2011piim.book.....D}. 

To produce a variety of physical and chemical outflow conditions for the training set, we produce synthetic observation data cubes where both \co\ and \13co\ abundances are reduced by a factor of 2 and a factor of 10. These data represent regions where some of the CO may be dissociated due to a stronger radiation field. We adopt 10\,K and 14\,K for the kinetic temperatures based on observations that suggests these values are representative of Perseus \citep[e.g.][]{1978ApJ...222..881G,2006ApJ...643..932R,2009ApJ...696..298F}. We also follow the same method as \citet{2020ApJ...890...64X} to increase the training set by considering thin clouds with thicknesses of 0.7\,pc and 1.4\,pc. This has the effect of producing data cubes in which outflow lobes and cavities are more distinct. Table~\ref{Synthetic Observation Properties} lists the physical and chemical properties of the synthetic observations. 

\begin{table*}[]
\begin{center}
\caption{Synthetic Observation Properties$^{a}$ \label{Synthetic Observation Properties}}
\begin{tabular}{cccccc}
\hline
Synthetic Observation Name & \co/\h2 & $T_{\rm k} (K)$ & FoV (pc $\times$ pc) & Thickness (pc)   & Label Color \\
 \hline
  {T10\_a1-4}   & {$10^{-4}$} & {10} & {(2$\times$2, 1$\times$1, 0.5$\times$0.5)}   & (2, 1.4, 0.7)  & green\\
  {T14\_a1-4}   & {$10^{-4}$} & {14} & {(2$\times$2, 1$\times$1, 0.5$\times$0.5)}   & (2, 1.4, 0.7)  & red \\
  {T10\_a5-5}   & {$5\times 10^{-5}$} & {10} & {(2$\times$2, 1$\times$1, 0.5$\times$0.5)}   & (2, 1.4, 0.7) & yellow \\
  {T14\_a5-5}   & {$5\times 10^{-5}$} & {14} & {(2$\times$2, 1$\times$1, 0.5$\times$0.5)}   & (2, 1.4, 0.7)  & black\\
  {T10\_a1-5}   & {$10^{-5}$} & {10} & {(2$\times$2, 1$\times$1, 0.5$\times$0.5)}   & (2, 1.4, 0.7)  & blue \\
  {T14\_a1-5}   & {$10^{-5}$} & {14} & {(2$\times$2, 1$\times$1, 0.5$\times$0.5)}   & (2, 1.4, 0.7)   & cyan\\
\hline
\multicolumn{6}{p{0.95\linewidth}}{Notes:}\\
\multicolumn{6}{p{0.95\linewidth}}{
$^{a}$ Synthetic observation name, \co\ to \h2\ abundance, kinetic temperature, field of view of the synthetic observation, thickness of the cloud in the synthetic observations and the color of the symbols in the plots in Section~\ref{Assessing Model Accuracy Using Synthetic Observations}.  }\\
\end{tabular}
\end{center}
\end{table*}

\subsubsection{Training Data} 
\label{Training Data}

We build two training tasks following the procedure of \citet{2020ApJ...890...64X}. The training target in model ME1 is \co\ emission, which is associated with outflows, while the training target in model MF is the fraction of the mass that comes from stellar feedback. 

To construct the training target in model ME1, we define the position of the protostellar outflows using the tracer field that indicates the amount of gas launched by the outflow subgrid model at each position. We impose two criteria. First, if more than 10\% of the mass comes from protostellar outflows in a voxel, we treat the voxel as a part of an outflow structure. Second, in order to better capture the morphology of the gas associated with the outflows, we define a voxel as belonging to an outflow if the gas temperature is over 12 K and adjacent to gas where over 10\% of the mass is launched material \citep{2020ApJ...890...64X}. We compare different models trained on data with different thresholds for the outflow mass in Appendix~\ref{Exploring Different Outflow Definitions}.

The training data for model ME1 is the \co\ emission that is only coming from the outflow gas. We mask the positions of pristine (non-outflow) gas and set the \co\ abundance to be 0 in the masked region. We then compute the radiative transfer to obtain the \co\ emission that is only coming from the protostellar outflows, which we refer to as the \co\ feedback cube.

To build the training data for model MF, we calculate the fraction of feedback mass by converting the raw density from position-position-position (PPP) space to position-position-velocity (PPV) space. The fraction ranges from 0 to 1, and it is not necessarily proportional to the actual \co\ or \13co\ intensity. If the \co\ or \13co\ emission is optically thin, the column density is proportional to the emission intensity. 

A more detailed description about how we generate training data for these two models can be found in \citet{2020ApJ...890...64X}.

\subsubsection{Data Augmentation}
\label{Data Augmentation}

We adopt outputs with different magnetic fields, different turbulent patterns and different evolutionary stages to create synthetic observations. Multiple stars form and launch outflows in each simulated cloud. We construct a ``zoomed in" synthetic observation centered on each protostar with an image size of 0.5\,pc$\times$0.5\,pc and 1\,pc$\times$1\,pc. We also conduct synthetic observations with an image size of 2\,pc$\times$2\,pc, which contains multiple outflows. These synthetic observations span early and late evolutionary stages. By constructing a training set with outflows of different sizes, we reinforce the ability of the model to detect outflows on different scales. In addition to the different image sizes, we resample the synthetic observations with two different velocity resolutions: at low resolution with an interval of 0.25\,km/s and at high-resolution with an interval of 0.125\,km/s. To enhance the diversity of the training set, we conduct radiative transfer from six different angular views and rotate the images randomly from 0$^{\circ}$ to 360$^{\circ}$. 

\begin{figure*}[hbt!]
\centering
\includegraphics[width=0.98\linewidth]{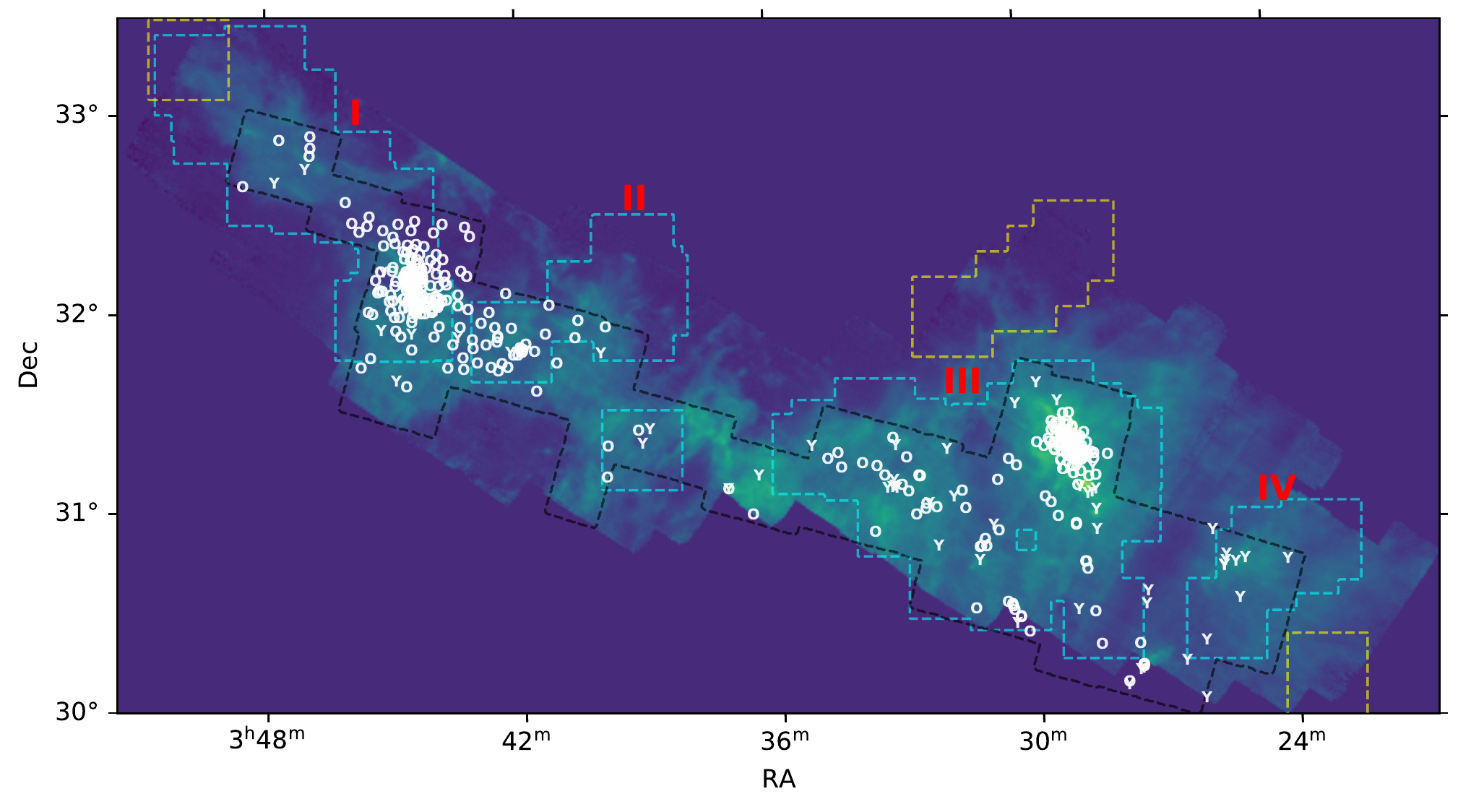}
\caption{Intensity of \co\ integrated over all velocity channels. ``Y'' and ``O'' indicates the location of YSOs: ``Y'' for early evolutionary stage YSOs, i.e., protostars and ``O'' for late evolutionary stage YSOs, i.e., pre-main sequence stars with disks \citep{Gutermuth.in.prep,2020arXiv200505466P}. The cyan dashed lines enclose sub-regions searched by our models. The yellow dashed lines indicate regions we include in our negative training set. The black dashed lines illustrate the coverage of YSOs in \citet{Gutermuth.in.prep}.  }
\label{fig.pred-perseus-lm-region-mask}
\end{figure*}

To help \CASItD\ distinguish outflows from high velocity blobs produced from supersonic turbulence in the molecular cloud and prevent false detections, we construct a negative training set. We conduct synthetic observations on turbulent simulations including noise, which do not contain feedback sources. In addition, we pick several regions in Perseus where there are no young stellar objects. Figure~\ref{fig.pred-perseus-lm-region-mask} shows these non-feedback regions enclosed by yellow dashed lines. 

\begin{figure*}[hbt!]
\centering
\includegraphics[width=0.98\linewidth]{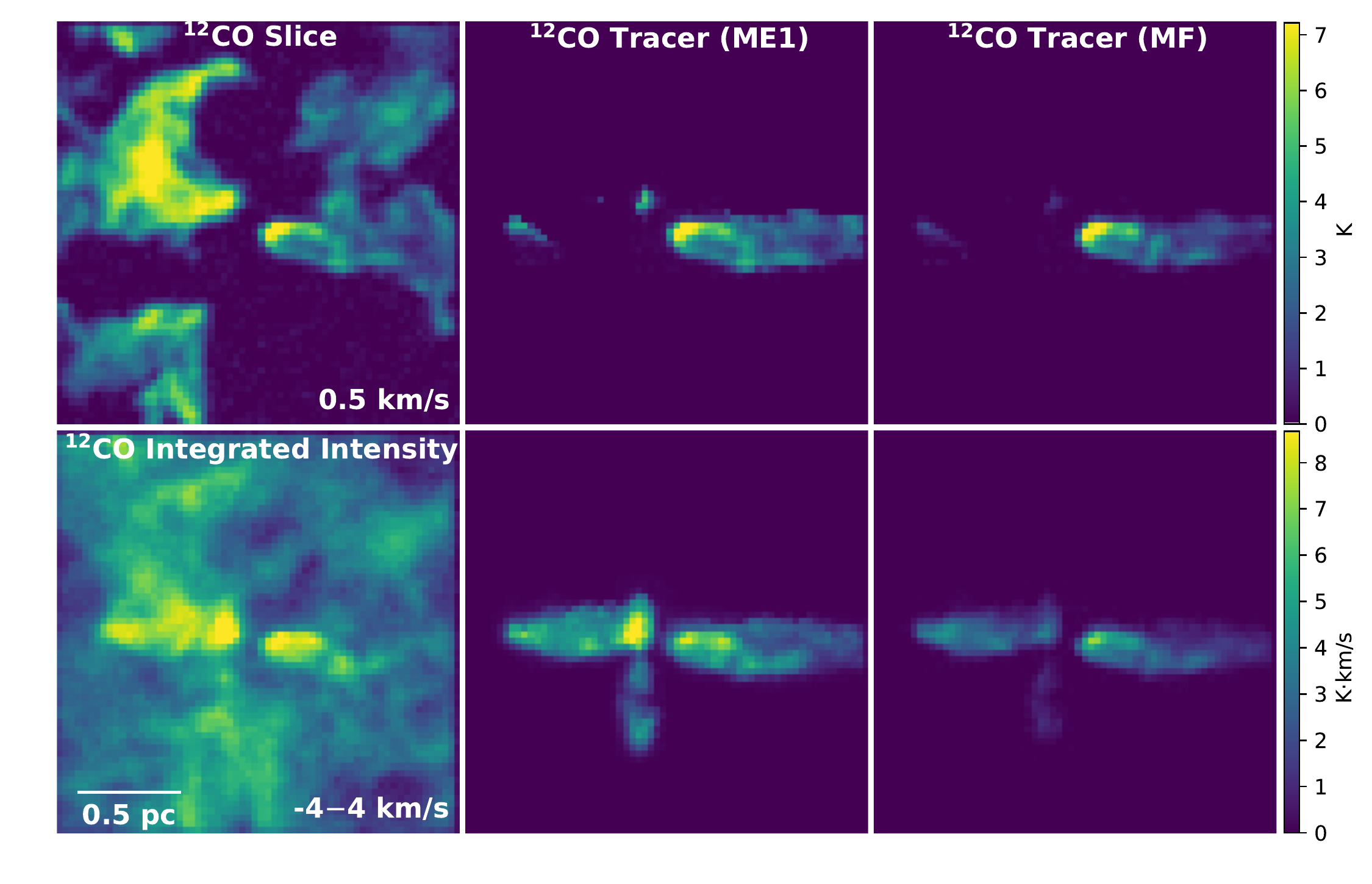}
\caption{\co\ synthetic observation of a simulated outflow and the outflow tracer field. Upper row, the first panel: \co\ emission at V=0.5 km/s. Second and third panels: ground truth \co\ emission from the outflow for model ME1 and MF. The lower row: same as the upper row but indicates the \co\ integrated intensity between -4 to 4 km/s, and integrated ground truth \co\ emission for models ME1 and MF. Note that the actual target for model MF is the fraction of mass associated with outflows, whose value ranges from 0 to 1, but here we make the plot with the fraction times the \co\ emission for fairly comparison.}
\label{fig.synt-co-slice-iim}
\end{figure*}

To make the synthetic cubes closer to the real observational data described in Section~\ref{Perseus Data}, we assume the synthetic images are at a distance of 250 pc and are observed with the Five College Radio Astronomy Observatory (FCRAO) \citep{2006AJ....131.2921R}.  We convolve them with a telescope beam of 50$^{\prime\prime}$ and add noise with a mean-square-error of 0.17 K, which is consistent with the noise level in the Perseus \co\ data in Section~\ref{Perseus Data}. Moreover, since many observed sources have velocities that are offset from the cloud velocity and/or not at 0 km/s, we randomly shift the central velocity of the cubes between -3 to 3 \kms. Figure~\ref{fig.synt-co-slice-iim} shows an example of a synthetic observation of a simulated outflow and corresponding target images like those included in the training set.

In total, we generate 23,715 synthetic data cubes, among which 3,483 contain no feedback sources. We adopt 14,229 of the data cubes as a training set, 4,743 data cubes as a test set and 4,743 data cubes as a validation set. The validation set allows us to evaluate how well the model has been trained. The test set assesses the accuracy of the final model.

\subsection{Perseus Data}
\label{Perseus Data}

We use the publicly available \co\ ($J$=1-0) and \13co\ ($J$=1-0) Perseus data from the COordinated Molecular Probe Line Extinction Thermal Emission (COMPLETE) Survey of Star Forming Regions \citep{2006AJ....131.2921R}. The Perseus \co\ and \13co\ maps were observed between 2003 and 2005 using the 13.7 m FCRAO Telescope. The maps are about $6^{\circ}.25\times 3^{\circ}$ with a beam size of 46$^{\prime\prime}$. The \co\ and \13co\ data have a mean root-mean-square (RMS) antenna temperature of 0.25 K and 0.2 K, respectively. We resampled the spectra with a lower velocity resolution of 0.125 km/s, which matches the velocity resolution in the training set. The noise levels for the new \co\ and \13co\ spectra is reduced by a factor of square root of 2, which are 0.17 K and 0.14 K, respectively.

We adopt the outflow catalog from \citet{2010ApJ...715.1170A}, in which 60 outflow candidates are identified in the Perseus molecular cloud by determining high-velocity features from a three-dimensional visualization using the COMPLETE \co\ maps. \citet{2010ApJ...715.1170A} call these high-velocity features “COMPLETE Perseus Outflow Candidates” (CPOC). Some high-velocity features at the same position are split into two CPOCs, a red-shifted lobe and a blue-shifted lobe. \citet{2010ApJ...715.1170A} also note that these 60 outflows are an underestimate of the true number of outflows. Outflows that are smaller than the beam size of the \co\ map or that have weak high-velocity wings cannot be detected by their technique. Consequently, outflows with smaller size and lower velocity along the line-of-sight are missed.

To search for the potential outflow driving sources, we use the draft YSO catalog for Perseus from SESNA \citep[Spitzer Extended Solar Neighborhood Archive,][]{Gutermuth.in.prep} used by \citet{2020arXiv200505466P}. SESNA is a uniform retreatment of most of the Spitzer surveys of nearby molecular clouds that uses an updated implementation of the data treatment, source catalog construction, and YSO identification and classification processes of \citet{2009ApJS..184...18G}. As in that work, the resulting YSOs are classified into one of four groups (``deeply embedded protostars'', ``Class I YSOs'', ``Class II YSOs'', and ``transition disks'') using a series of color-color and color-magnitude diagram selections that largely mitigate extra-galactic contamination and are relatively unbiased in a wide range of reddening conditions \citep{2007ApJ...663.1069F}. For further analysis, the former two groups are merged to encompass the dusty-envelope-bearing protostars that we call ``Younger YSOs'' throughout this work. Similarly, the latter two groups are merged to list the protoplanetary-disk-bearing pre-main sequence stars that we call ``Older YSOs'' herein. As is the case with all mid-IR YSO surveys, diskless ``Class III'' YSOs cannot be distinguished from field stars and are thus ignored here. Residual contamination comes from two sources. For the draft SESNA catalog we adopt, the residual extra-galactic contamination is $9 \pm 1$ sources per square degree, with even split between younger and older YSO types, based on applying the full SESNA treatment to 16 square degrees of so-called ``blank field" galaxy survey data (the ``Elias-N1'' and ``Bo\"otes'' fields). These sources are generally assumed to be statistically uniform in their spatial distribution on the sky.  In contrast, the other contaminant is class contamination, whereby YSOs are misclassified as younger versus older. \citet{2009ApJS..184...18G} estimated that up to 3.5\% of older YSOs may be reported as younger YSOs as a result of edge-on viewing angle. These will thus be spatially correlated with elevated densities of older YSOs. We do not attempt to treat either of these residual contaminant types further in this work.

\section{Results}
\label{Results}

\subsection{Assessing Model Accuracy Using Synthetic Observations}
\label{Assessing Model Accuracy Using Synthetic Observations}

\begin{figure*}[hbt!]
\centering
\includegraphics[width=0.98\linewidth]{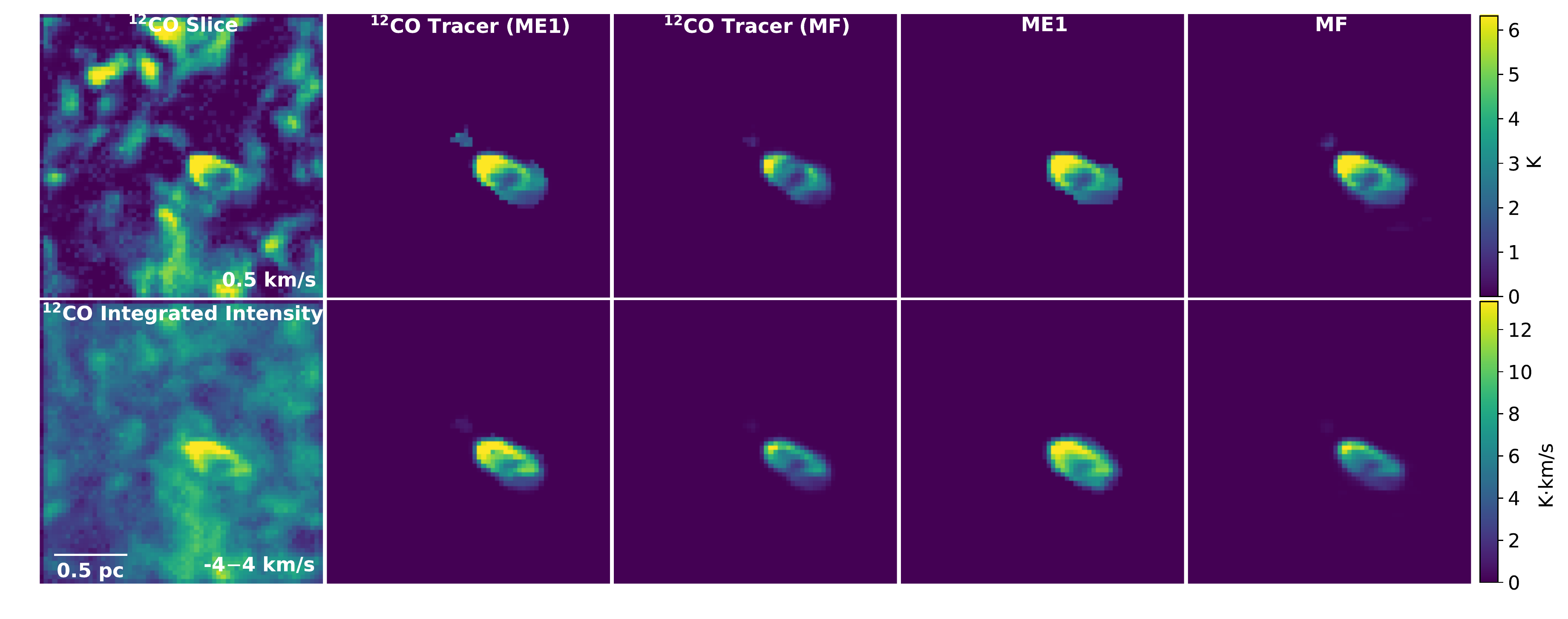}
\caption{Results of the models ME1 and MF applied to a synthetic outflow. Upper row, the first panel: \co\ emission at V=0.5 km/s. Second and third panels: ground truth \co\ tracer for models ME1 and MF, respectively. Fourth and fifth panels: prediction from models ME1 and MF, respectively. The lower row: the same as the upper row but indicates the \co\ integrated intensity between -4 to 4 km/s, integrated ground truth \co\ tracer for models ME1 and MF, and integrated prediction from models ME1 and MF, respectively. }
\label{fig.synth-cnn-tracer-model}
\end{figure*}

In this section we use the synthetic data to assess how accurately physical properties can be determined from the identified outflows. We apply both models to the outflow synthetic observations in the test set. Figure~\ref{fig.synth-cnn-tracer-model} shows an example of the model performance on a test synthetic outflow. Both models ME1 and MF capture the outflow features clearly. 

We follow the same strategy in \citet{2010ApJ...715.1170A} to calculate the outflow mass by combining both \co\ and \13co\ data. If there is distinct \13co\ emission in the corresponding position, we assume the \13co\ emission line is optically thin and has an excitation temperature of 25 K to calculate the mass \citep{2010ApJ...715.1170A,2012MNRAS.425.2641N,2015ApJS..219...20L}. Under the assumption of LTE, the mass estimation goes linearly with the excitation temperature. From previous feedback mass estimates, the choice of excitation temperature ranges from 10 K to 50 K. This could potentially introduce a factor of two uncertainty in the mass estimation. Here we take the widely used and moderate value of 25 K as the excitation temperature for all outflow calculations \citep{2010ApJ...715.1170A,2012MNRAS.425.2641N,2015ApJS..219...20L}. If there is no reliable \13co\ emission in the corresponding position, we use \co\ to derive the mass by assuming the \co\ line is optically thin.

We take 62 as the abundance ratio between \co\ and \13co\ and 10$^{-4}$ as the abundance ratio between \co\ and \h2 \citep{2010ApJ...715.1170A}. As mentioned in Section~\ref{Synthetic Observations}, we also conduct \co\ synthetic observations with three different abundance ratios. We adopt the corresponding abundance ratios to calculate the mass of the outflows. 

\begin{figure}[hbt!]
\centering
\includegraphics[width=0.98\linewidth]{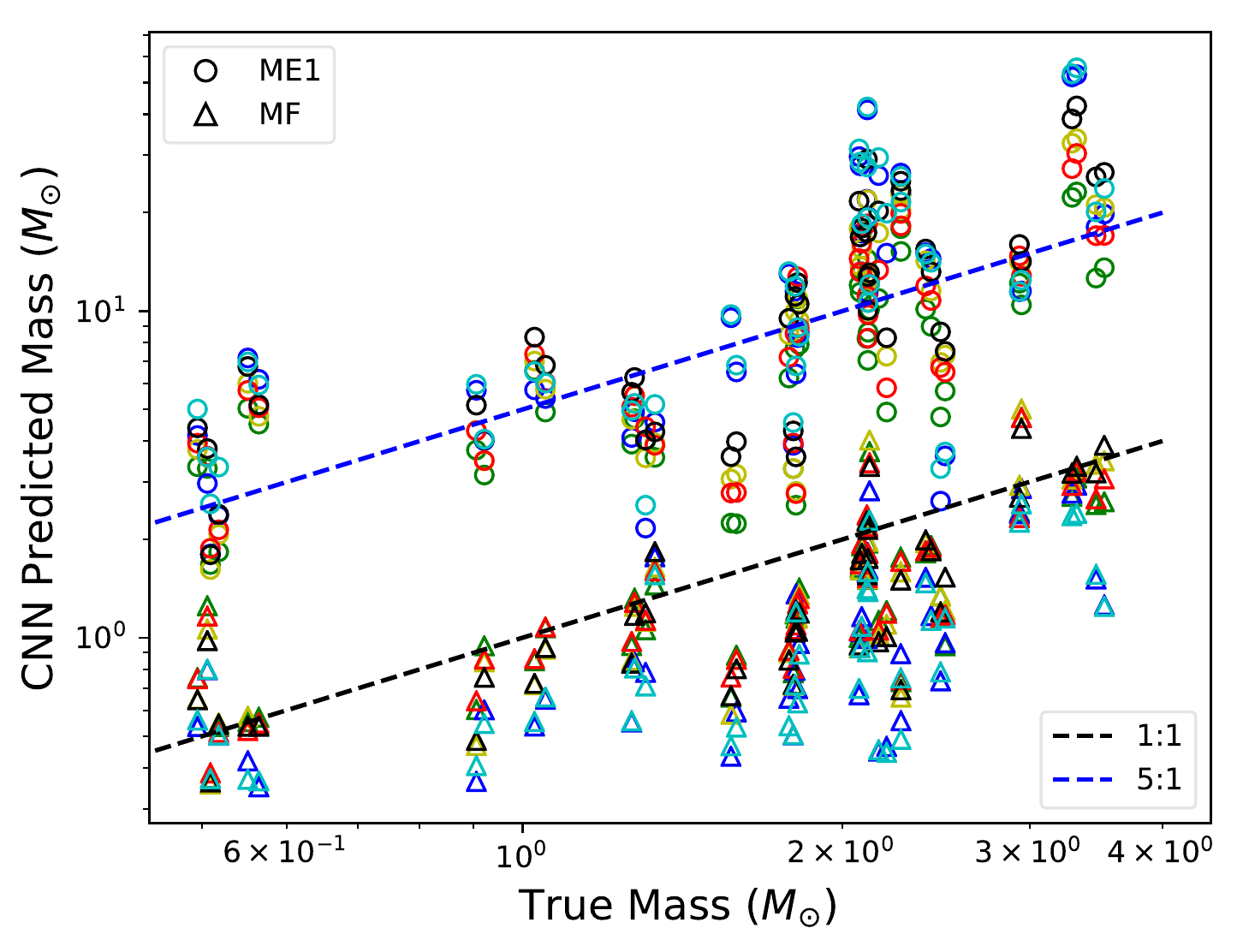}
\caption{Relation between the \CASItD\ predicted outflow mass and the true mass for different outflows. Circle symbols indicate the mass calculated by model ME1. Triangle symbols represent the mass calculated by model MF. Different symbol colors indicate synthetic observations with different physical and chemical conditions as listed in Table~\ref{Synthetic Observation Properties}. The black dashed line indicates where \CASItD\ correctly predicts the true mass. The blue dashed line has a slope of 5. {We investigate the uncertainty of the mass estimates predicted by both models in Appendix~\ref{Performance on Test Set Data}.}}
\label{fig.tacer-comp-cnn2}
\end{figure}

Figure~\ref{fig.tacer-comp-cnn2} shows the mass estimated from the two models, ME1 and MF. We also plot the true feedback mass, which we estimate directly by adding the mass contained in all cells where there is feedback as defined in Section~\ref{Training Data}. We find ME1 overestimates the outflow mass by a factor of 5 or more, while MF correctly predicts the outflow mass within a scatter of a factor of two.

Different symbol colors in Figure~\ref{fig.tacer-comp-cnn2} indicate different outflow physical conditions, e.g. different kinetic temperatures and different \co\ and \13co\ abundance ratios. The scatter at a given true mass value suggests the uncertainty in converting from \co\ and \13co\ emission to mass; multiple synthetic observations with different physical and chemical conditions of a simulated outflow have the same mass associated with feedback but have different \co\ and \13co\ emission. Model MF correctly predicts the outflow mass within a reasonable uncertainty under different physical and chemical conditions. {We quantitatively evaluate the two model performance in Appendix~\ref{Quantitatively Evaluating Model Performance}.}

\begin{figure}[hbt!]
\centering
\includegraphics[width=0.98\linewidth]{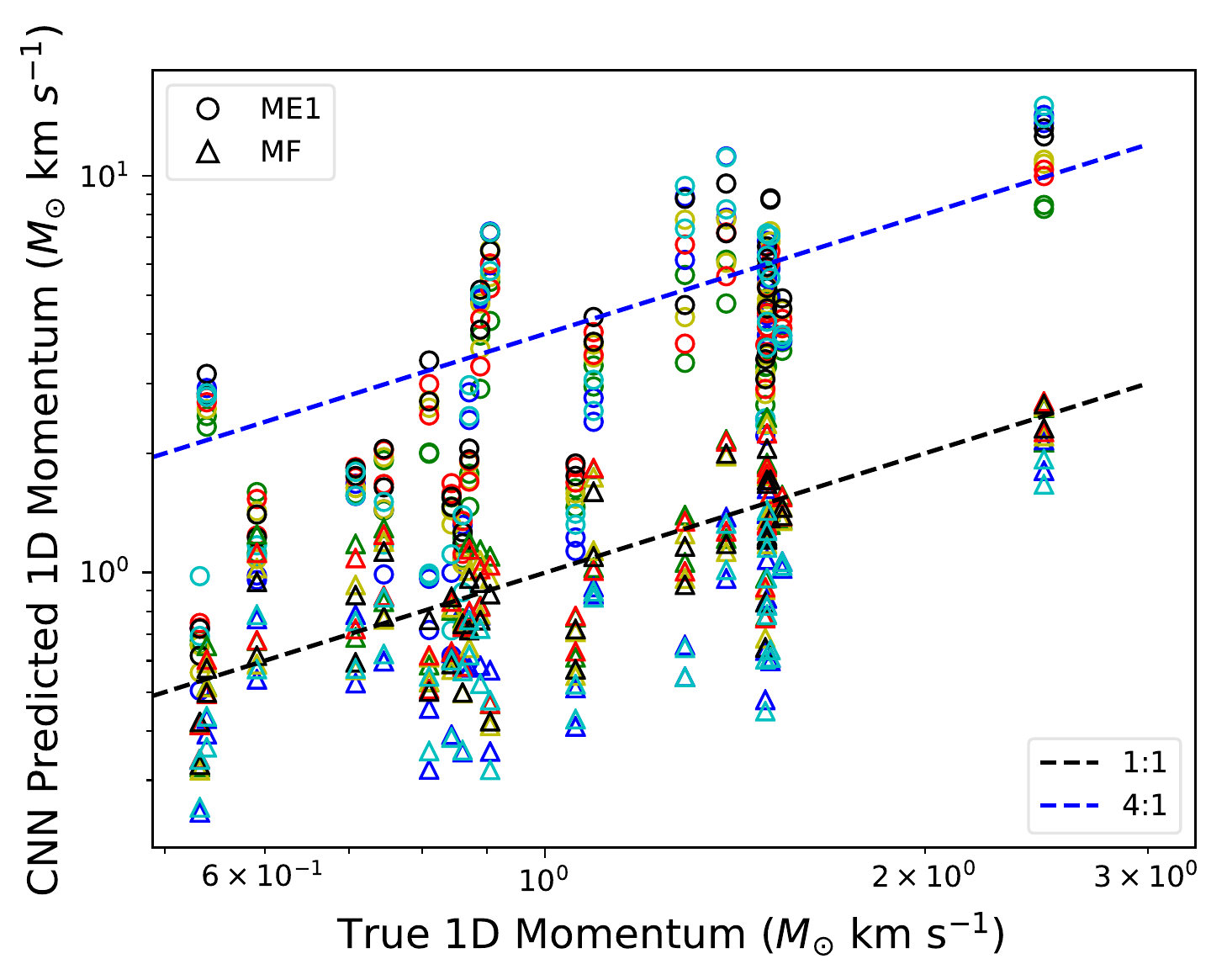}
\caption{Relation between the \CASItD\ predicted outflow momentum for particular lines of sight and the true 1D momentum for different outflows. Circle symbols indicate the momentum calculated by model ME1. Triangle symbols represent the momentum calculated by model MF. The black dashed line indicates where \CASItD\ correctly predicts the true momentum. The blue dashed line has a slope of 4.}
\label{fig.tacer-comp-cnn2-momentum}
\end{figure}

We compare the 1D line of sight momentum between the model predictions and the true simulation feedback in Figure~\ref{fig.tacer-comp-cnn2-momentum}.  We define the 1D momentum as the sum of the gas mass in each channel multiplied by the channel velocity, where we have shifted the mean cloud velocity to zero. Model ME1 overestimates the 1D momentum by a factor of 4. In contrast, model MF correctly predicts the 1D momentum within 30\% relative error. This uncertainty is the same order of magnitude as the uncertainty in the outflow mass predicted by model MF, which reflects the challenge of converting from \co\ and \13co\ emission to mass under different physical and chemical conditions. 

\begin{figure}[hbt!]
\centering
\includegraphics[width=0.98\linewidth]{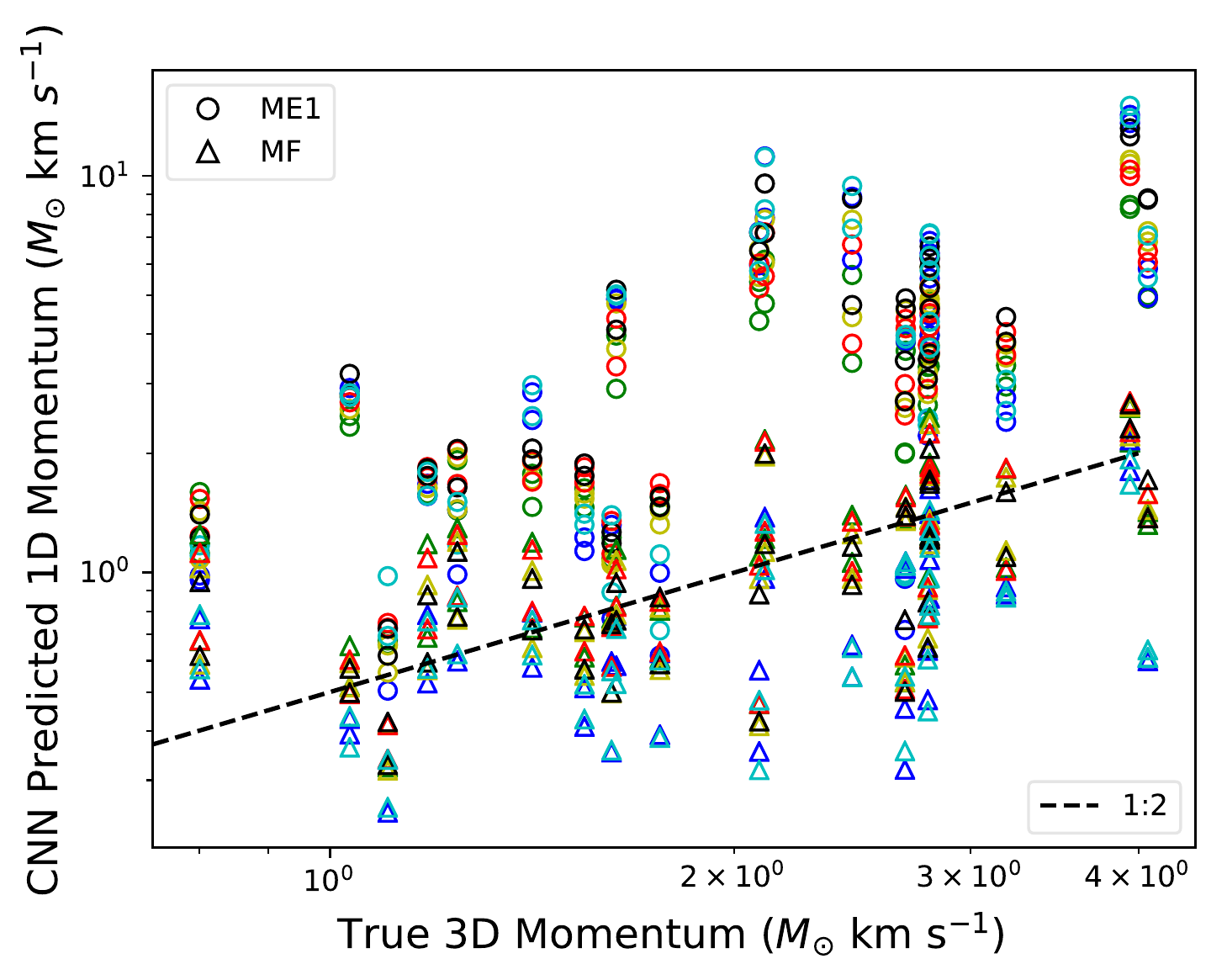}
\caption{Relation between the \CASItD\ predicted outflow momentum and the true 3D momentum for different outflows. Circle symbols indicate the momentum calculated by model ME1. Triangle symbols represent the momentum calculated by model MF. The black dashed line has a slope of 1/2. }
\label{fig.tacer-comp-cnn2-momentum-3d}
\end{figure}

\begin{figure}[hbt!]
\centering
\includegraphics[width=0.98\linewidth]{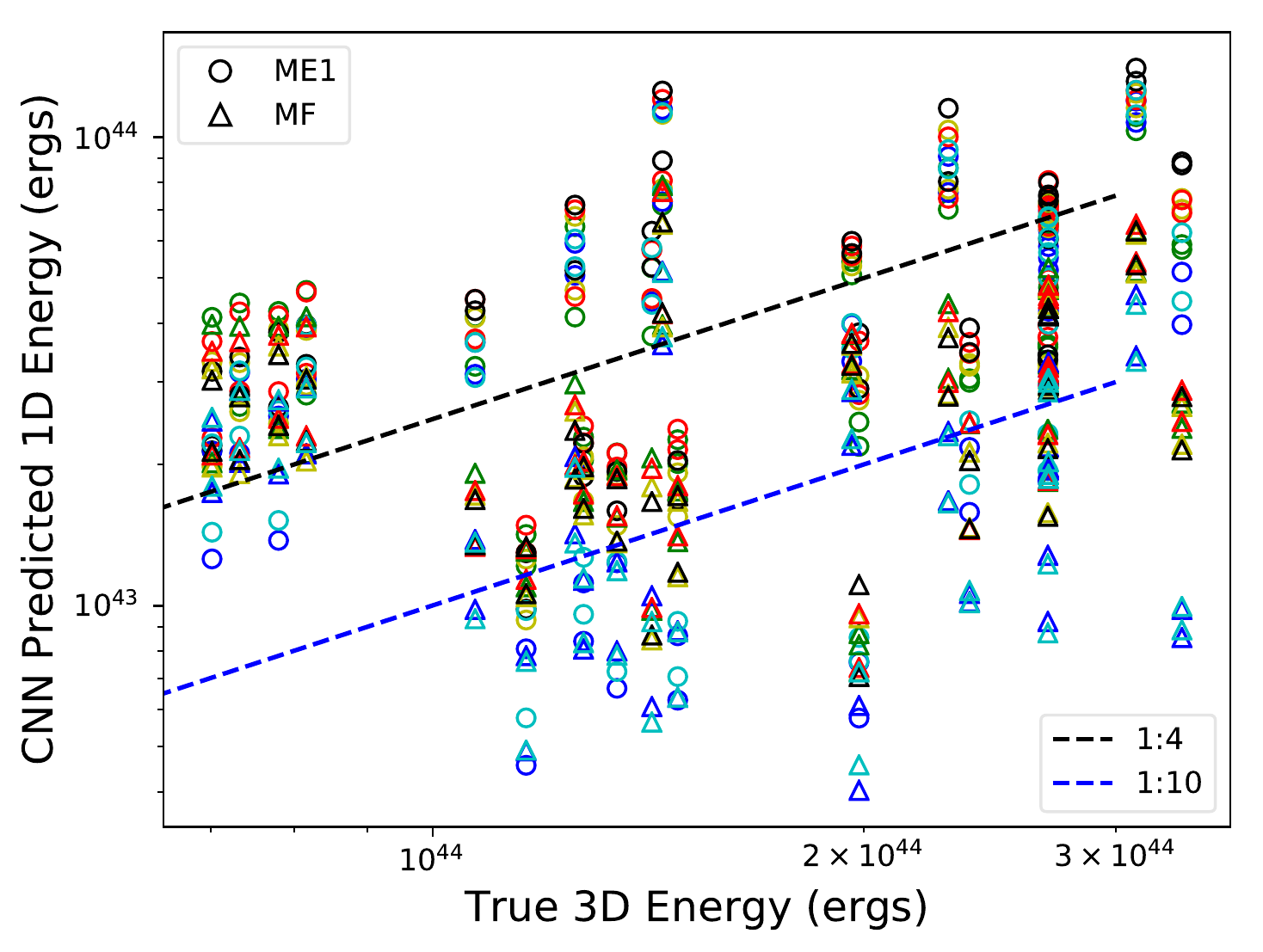}
\caption{Relation between the \CASItD\ predicted outflow energy and the true feedback energy for different outflows. Circle symbols indicate the energy calculated by ME1. Triangle symbols represent the mass calculated by model MF. The black dashed line has a slope of 1/4 and the blue dashed line has a slope of 1/10. }
\label{fig.tacer-comp-cnn2-energy}
\end{figure}

If we assume the inclination angle between the outflow axis and the line-of-sight is $\theta$, the 3D momentum would be expected to be a factor of $\frac{1}{ {\rm cos\, } \theta}$ larger than the 1D momentum, while the 3D kinetic energy would be a factor of $\frac{1}{ {\rm cos^{2}\, } \theta}$ larger than the 1D kinetic energy. Figures~\ref{fig.tacer-comp-cnn2-momentum-3d} and \ref{fig.tacer-comp-cnn2-energy} show the 1D momentum and energy, respectively, predicted by the two models compared to the respective 3D quantities calculated from the simulation. We see a significant scatter in these two figures. The 1D momentum and 1D energy deviate from the linear trend of their 3D true value, indicating diverse range of inclination angles, which is usually the case. 

Figures~\ref{fig.tacer-comp-cnn2-momentum-3d} illustrates the 3D momentum is a factor of 2 higher than the 1D momentum predicted by model MF. This corresponds to an average inclination angle of $60 ^{\circ}$ in our test sample. The scatter along the one to one line suggests the variation of inclination angles for different outflows. 

Figure~\ref{fig.tacer-comp-cnn2-energy} demonstrates that outflow inclination uncertainty together with uncertainties in the physical conditions produce a factor of 10 uncertainty in the 3D kinetic energy of individual outflows. For example, the correction factor for an outflow with an inclination angle of $35^{\circ}$ is an order of magnitude larger than that with an inclination angle of $75^{\circ}$. Consequently, it is not possible to accurately calculate the 3D momentum and 3D energy for an individual outflow only from the line of sight velocity information. Statistically however, an average inclination angle between $30 ^{\circ}$ and $60 ^{\circ}$ approximately reproduces the total true 3D momentum and 3D energy within a factor of two and a factor of four uncertainty, respectively. Thus, our results verify the statement on the uncertainty of outflow energy in Perseus by \citet{2010ApJ...715.1170A}, who adopt an average outflow inclination of $45^{\circ}$.

\subsection{Identifying Individual Outflows in Perseus Using \CASItD}
\label{Identifying Individual Outflows in Perseus Using CASItD}

\begin{figure*}[hbt!]
\centering
\includegraphics[width=0.48\linewidth]{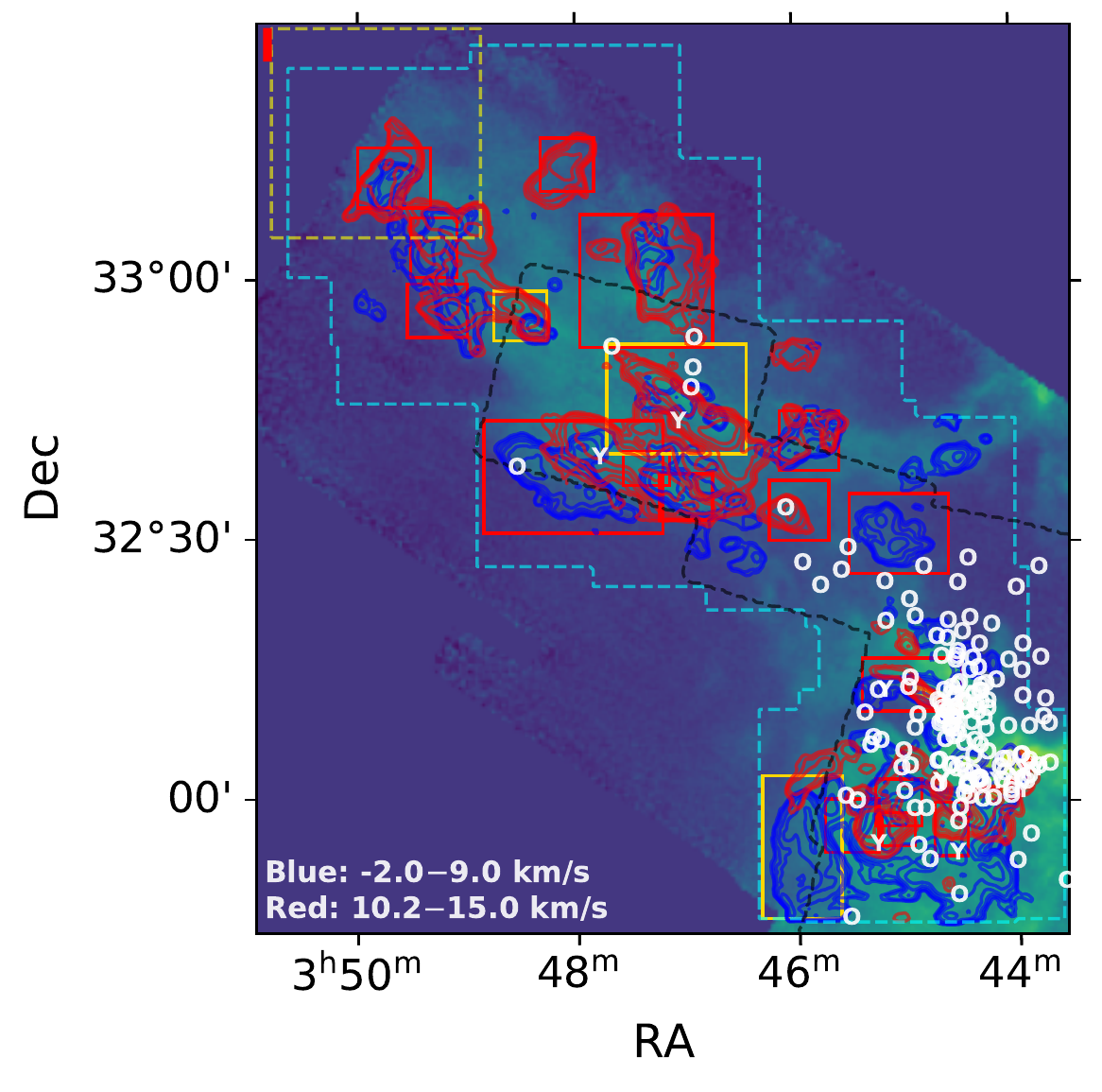}
\includegraphics[width=0.48\linewidth]{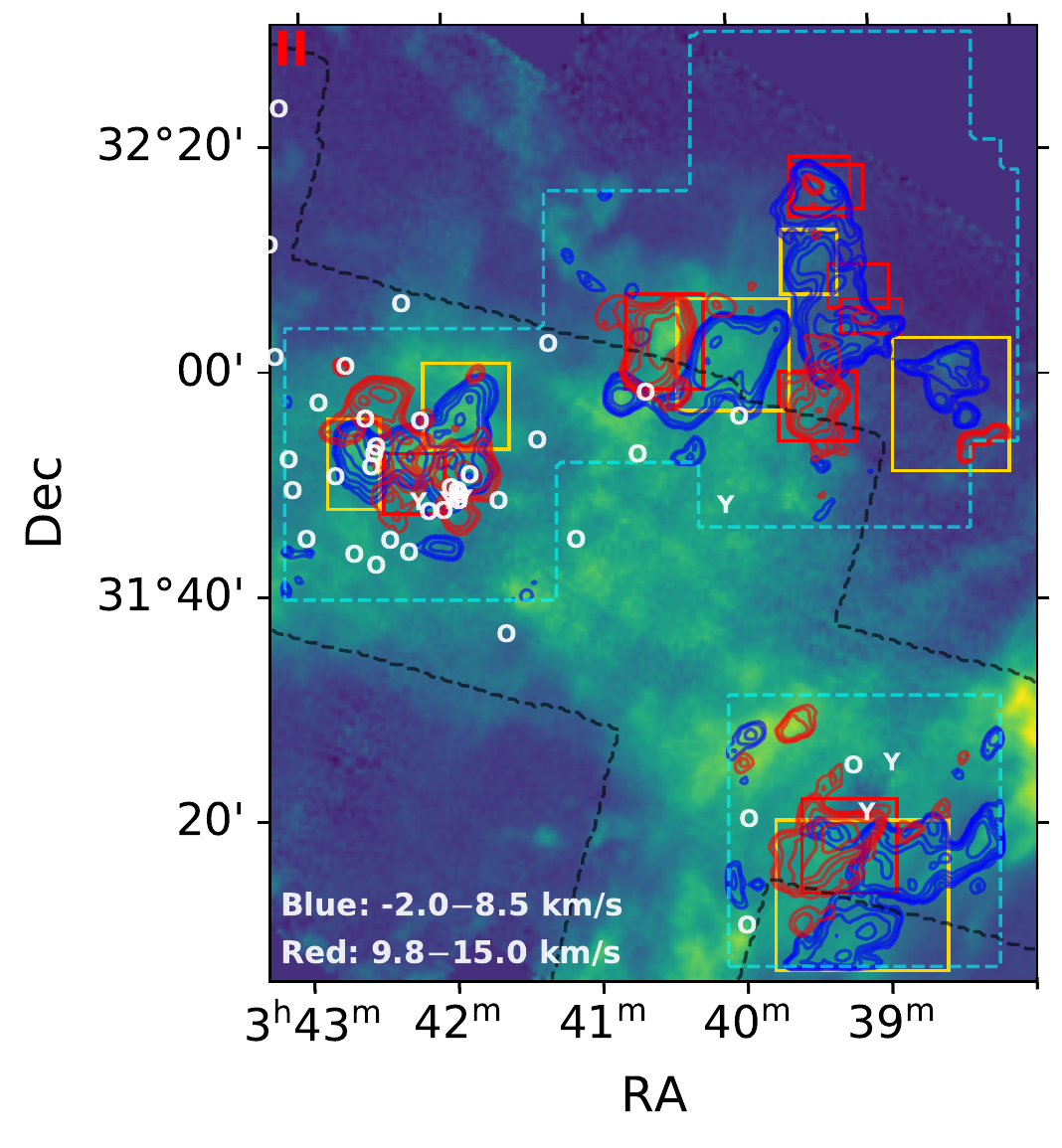}
\includegraphics[width=0.48\linewidth]{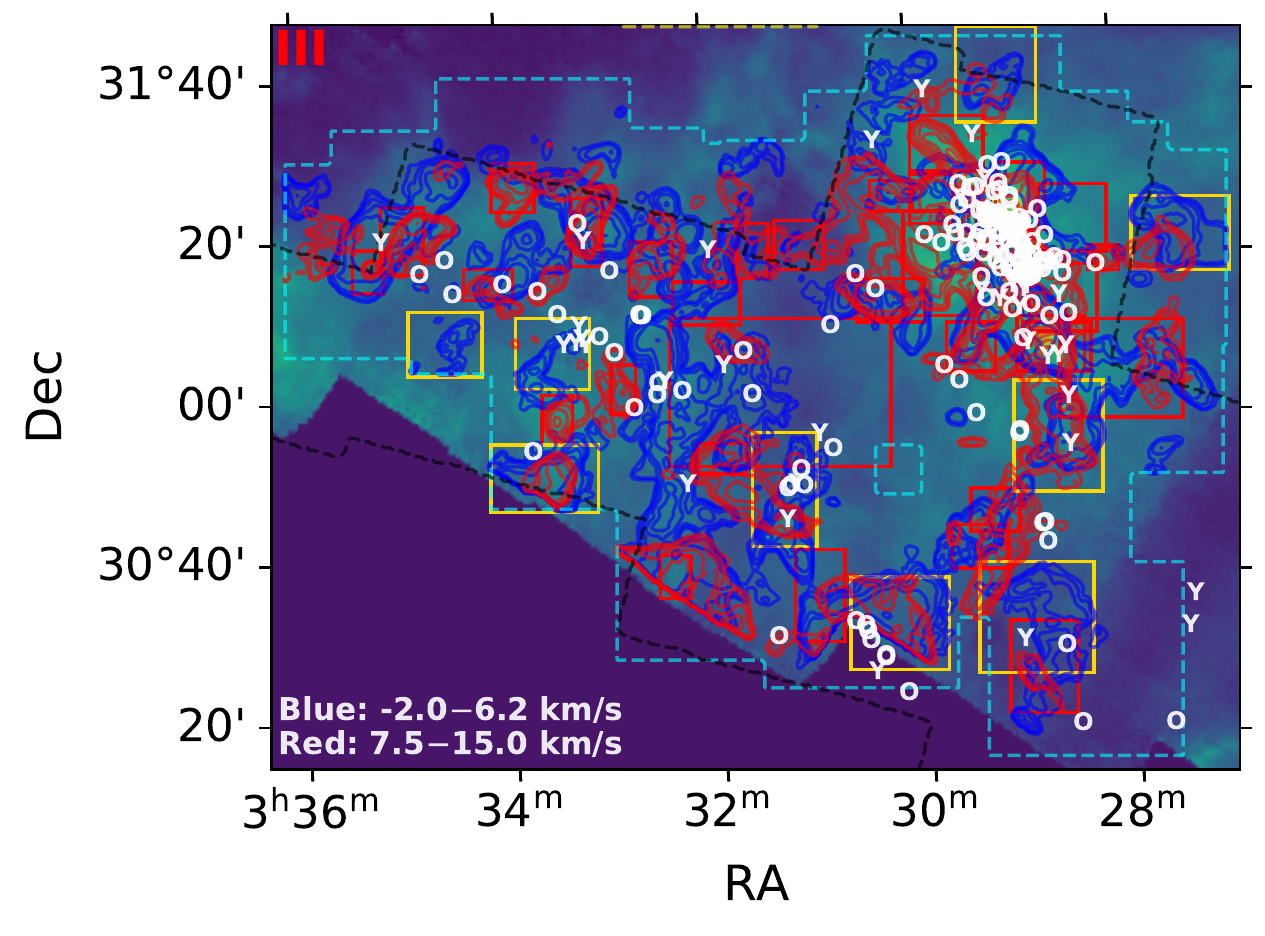}
\includegraphics[width=0.48\linewidth]{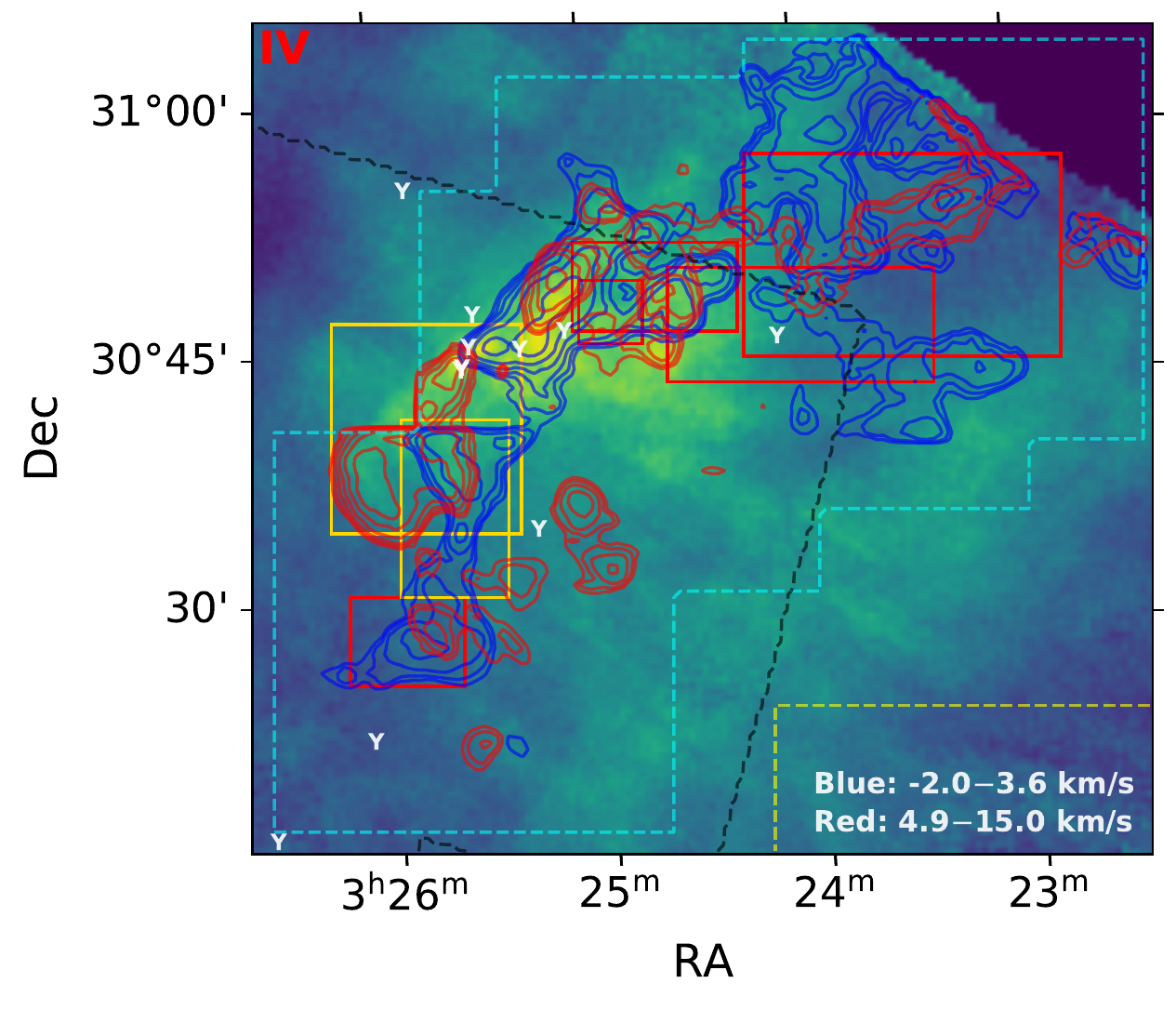}
\caption{Intensity of \co\ integrated over all velocity channels for the four sub-regions, overlaid with the model ME1 prediction (red and blue contours). Red boxes indicate the position of outflows previously identified by \citet{2010ApJ...715.1170A}. Letters ``Y'' and ``O'' mark YSO positions, as described in Figure~\ref{fig.pred-perseus-lm-region-mask}. Yellow boxes indicate the position of 20 newly identified outflows. The cyan dashed lines enclose sub-regions searched by our models. }
\label{fig.pred-perseus-lm-region-ME1}
\end{figure*}

\begin{figure*}[hbt!]
\centering
\includegraphics[width=0.48\linewidth]{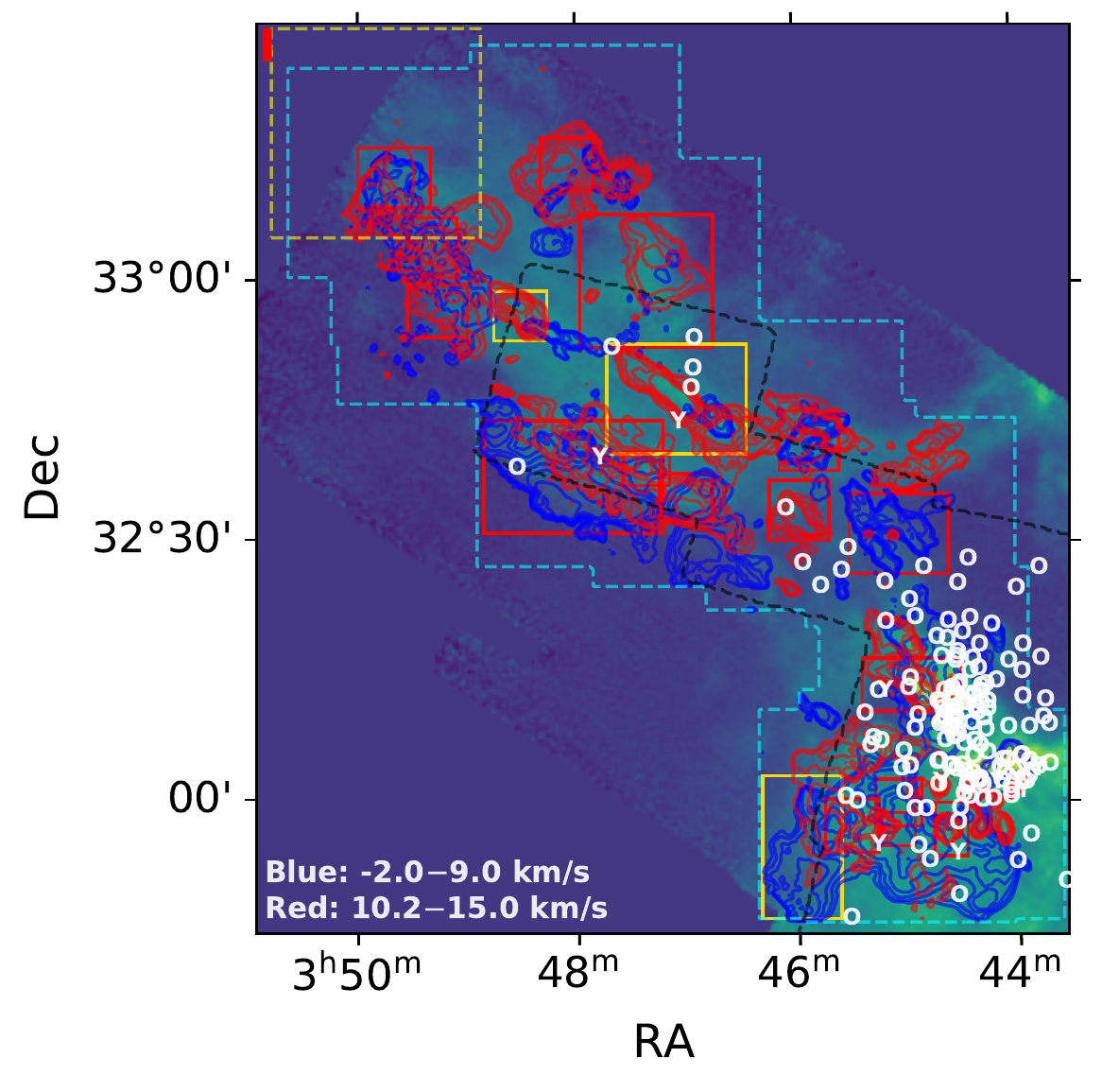}
\includegraphics[width=0.48\linewidth]{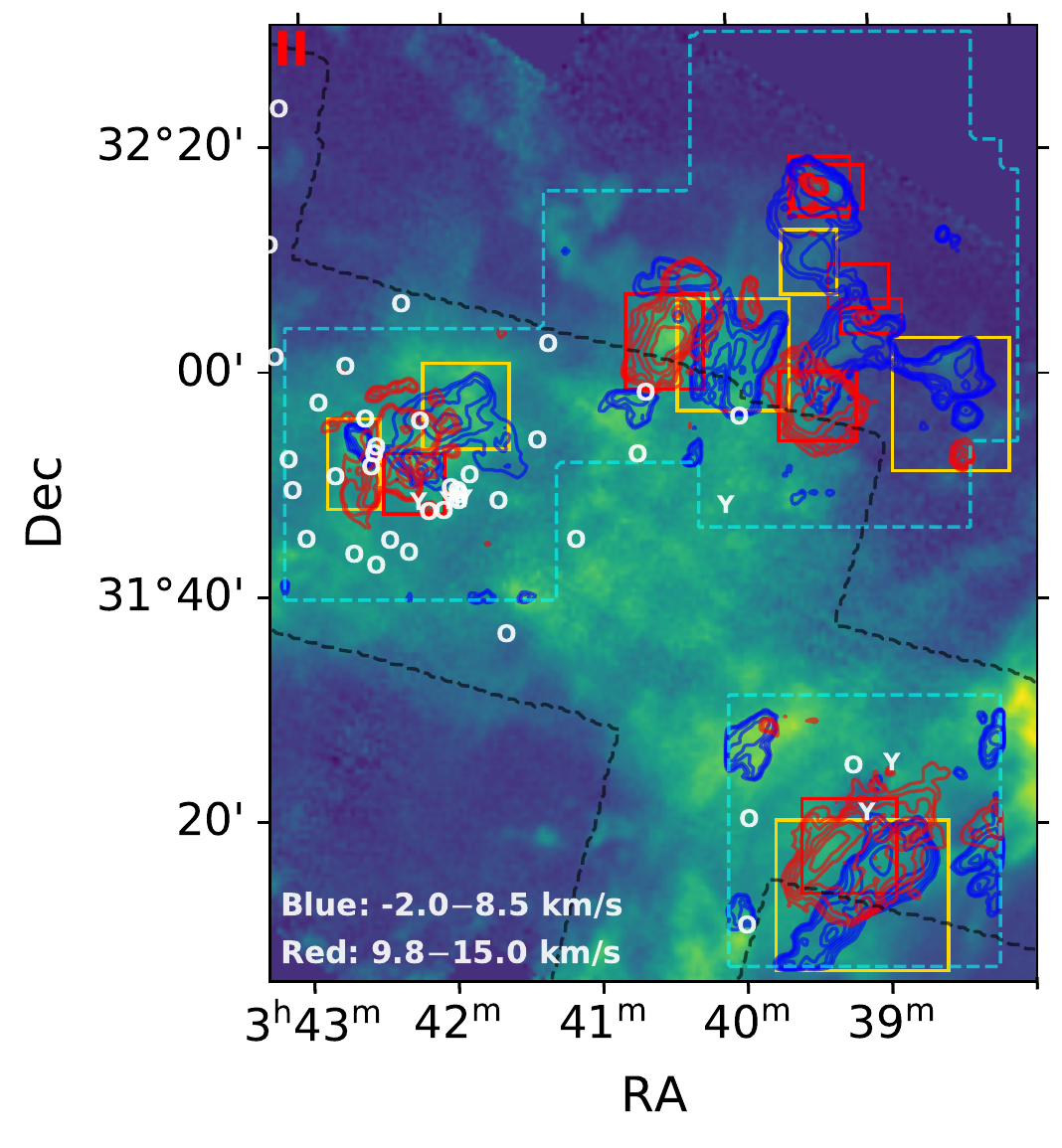}
\includegraphics[width=0.48\linewidth]{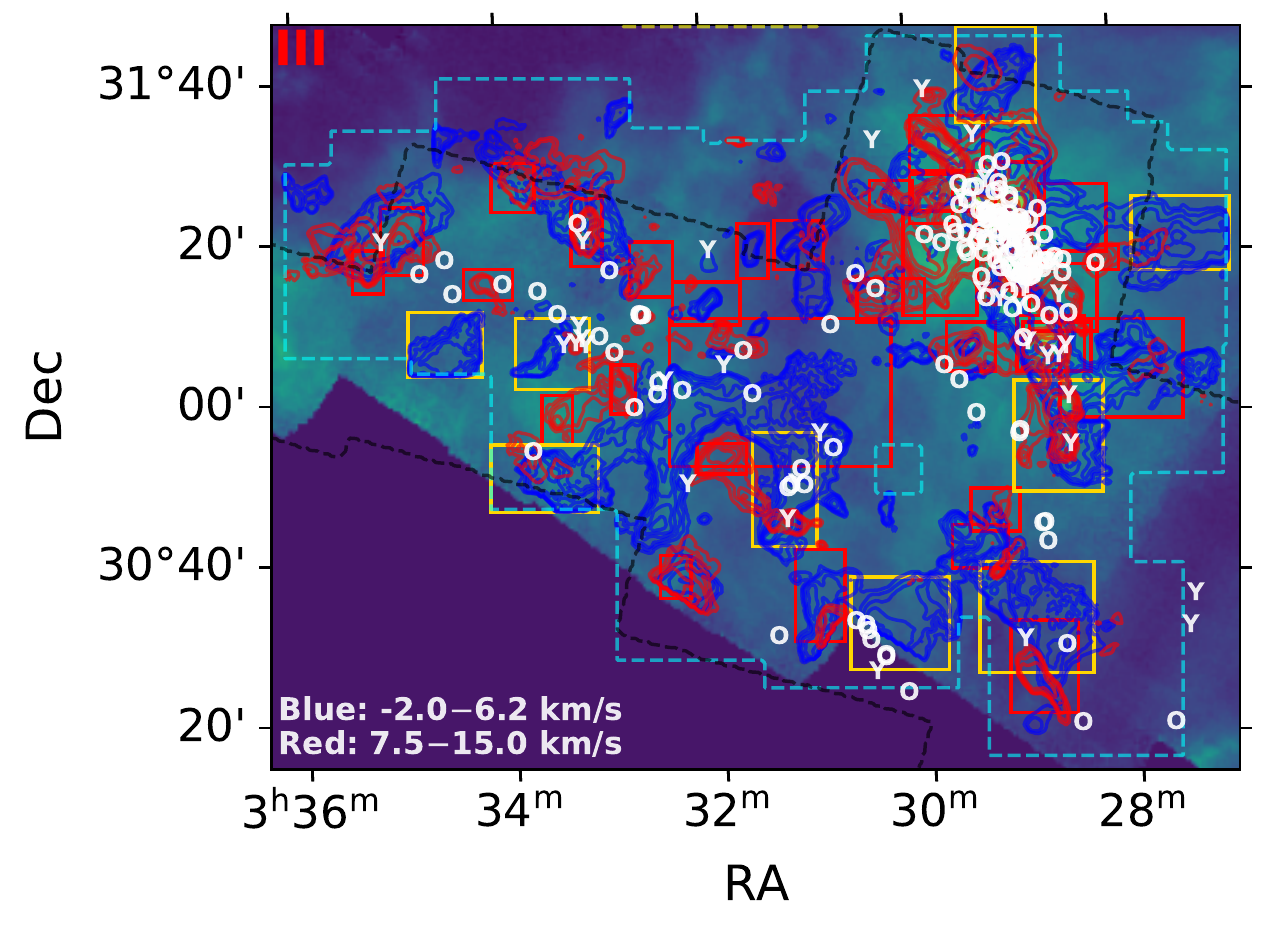}
\includegraphics[width=0.48\linewidth]{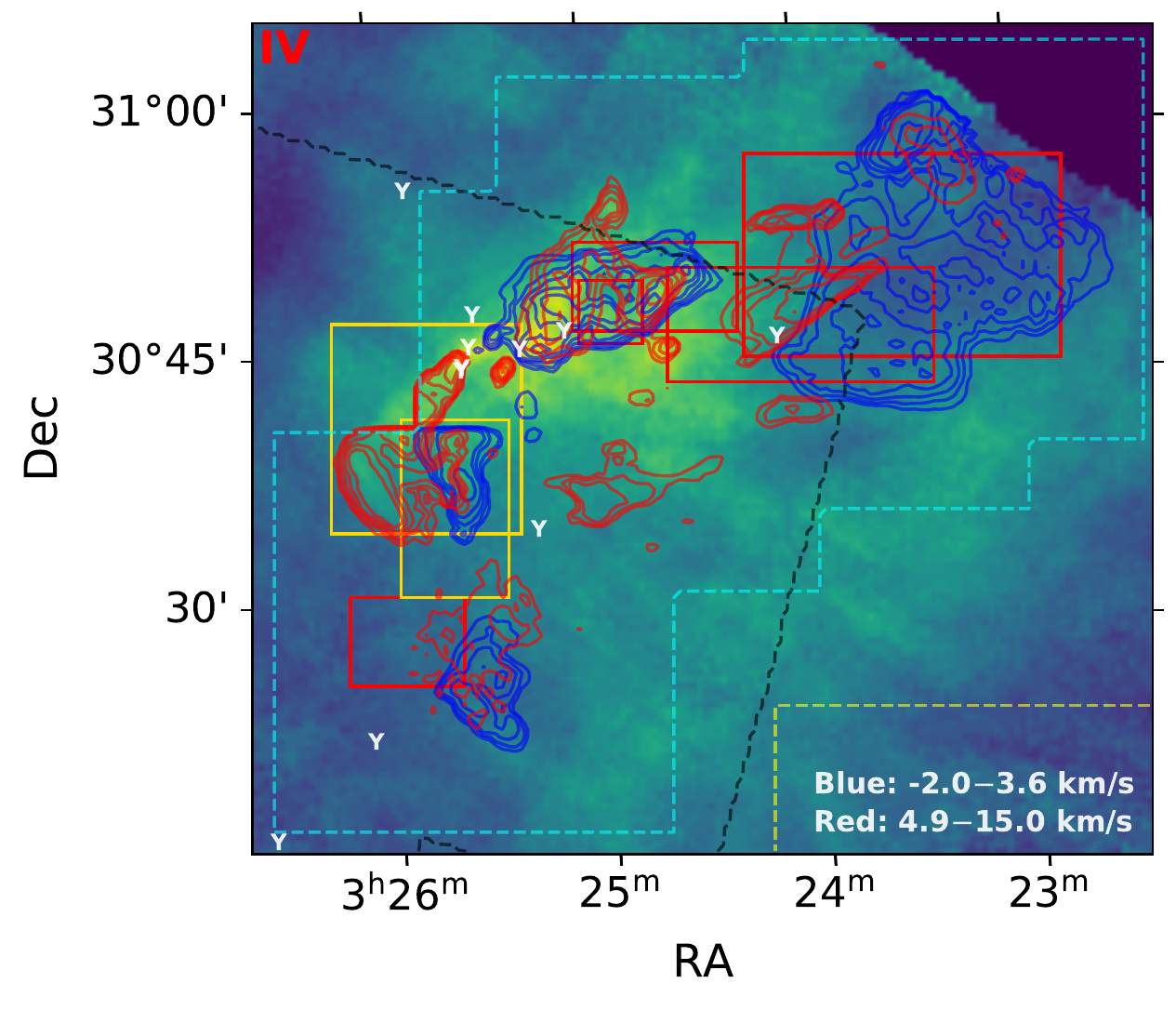}
\caption{The same as Figure~\ref{fig.pred-perseus-lm-region-ME1} but displaying the prediction of model MF (red and blue contours).}
\label{fig.pred-perseus-lm-region-MF}
\end{figure*}

\begin{figure*}[hbt!]
\centering
\includegraphics[width=0.98\linewidth]{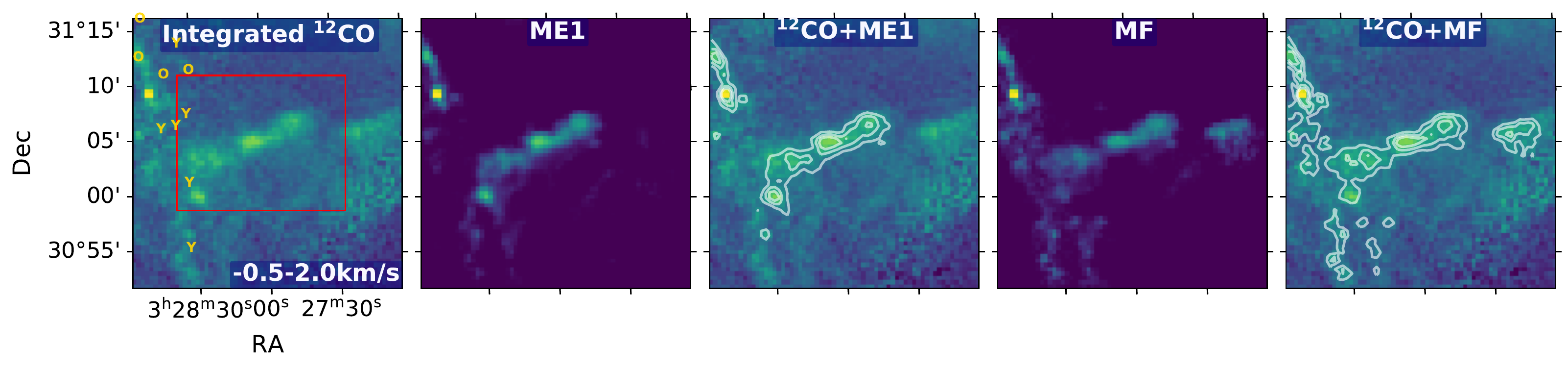}
\caption{Results of models ME1 and MF applied to previously identified Perseus outflow CPOC 7. First panel: integrated intensity of \co\ over the outflow velocity channels indicated by \citet{2010ApJ...715.1170A}. The red box shows the position of the outflow. Letters ``Y'' and ``O'' mark YSO positions, as described in Figure~\ref{fig.pred-perseus-lm-region-mask}. Second panel: predicted intensity integrated along the velocity axis from model ME1. Third panel: integrated intensity of \co\ over the outflow velocity channels overlaid with the model ME1 prediction in white contours. Fourth and fifth panels: same as the second and the third panels but displaying the prediction by model MF.}
\label{fig.pred-perseus-outflow-7}
\end{figure*}

To visually assess the performance of our CNN models on observational data, we apply models ME1 and MF to a catalog of outflows previously identified in the Perseus \co\ data. \citet{2010ApJ...715.1170A} identified 60 outflows by identifying high-velocity features in a three-dimensional visualization. We apply models ME1 and MF to sub-regions of the Perseus data containing these outflows. Figures~\ref{fig.pred-perseus-lm-region-ME1} and \ref{fig.pred-perseus-lm-region-MF} show the integrated intensity of \co\ over the entire velocity range (-2 km/s -- 15 km/s) of four sub-regions, overlaid with the model ME1 prediction and that overlaid with the model MF prediction, respectively. Figure~\ref{fig.pred-perseus-outflow-7} demonstrates the encouraging performance of both models on previously identified outflow CPOC 7. Both models ME1 and MF capture the morphology of the outflow very well. 

\begin{figure*}[hbt!]
\centering
\includegraphics[width=0.48\linewidth]{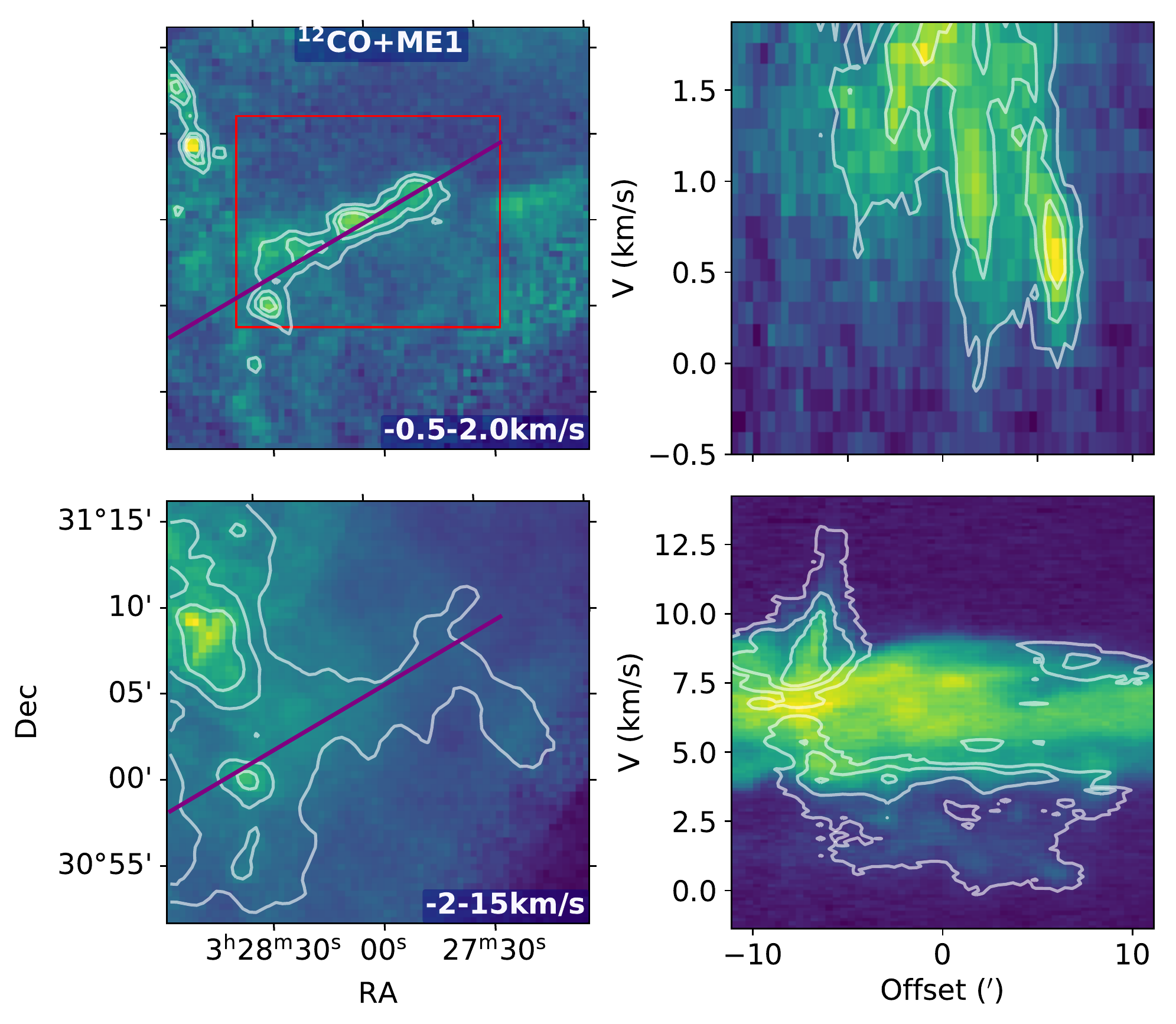}
\includegraphics[width=0.48\linewidth]{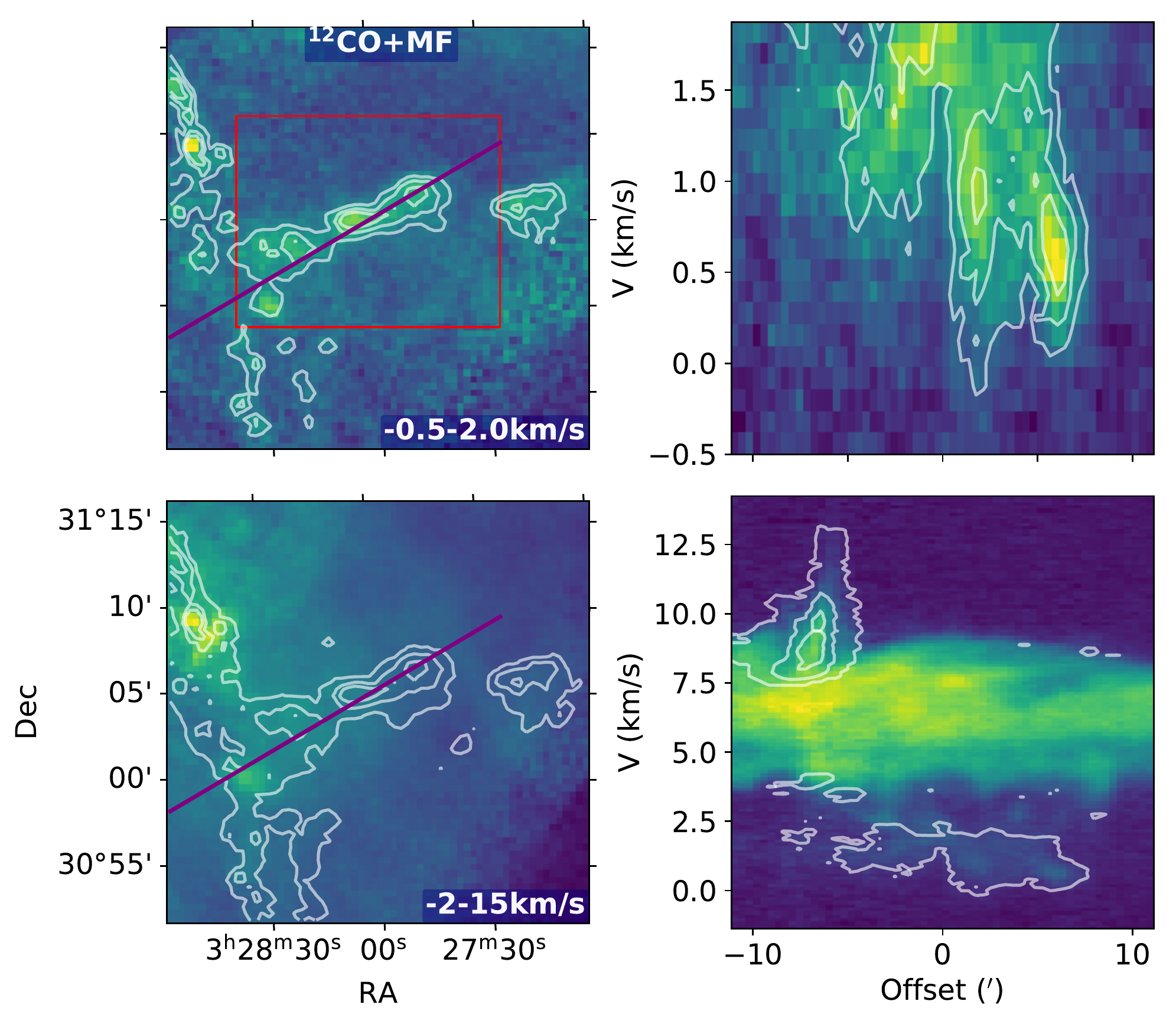}
\caption{Position-velocity diagram of \co\ emission toward the outflow in Figure~\ref{fig.pred-perseus-outflow-7}. In the left sub-figure, upper left panel: integrated intensity of \co\ over the outflow velocity channels (from -0.5 km/s to 2 km/s) overlaid with the model ME1 prediction in white contour. The red box shows the position of the outflow. The purple line illustrates the cutting direction of the position-velocity diagram. Upper right panel: position-velocity diagram of \co\ emission in the outflow velocity channels overlaid with the model ME1 prediction in white contour. Lower left panel: \co\ intensity integrated over the full velocity range (from -2 km/s to 15km/s) overlaid with the integrated model ME1 prediction (white contours). Lower right panel: \co\ position-velocity diagram overlaid with the model ME1 prediction (white contours). The right four panels are the same as the left but show the prediction by model MF. }
\label{fig.pred-perseus-outflow-7-pv}
\end{figure*}

Figure~\ref{fig.pred-perseus-outflow-7-pv} shows the position-velocity diagram of \co\ emission for the outflow in Figure~\ref{fig.pred-perseus-outflow-7}. Coherent high-velocity structures are identified by both the ME1 and MF models. Most outflows are only distinct in high-velocity channels. The lower panels of Figure~\ref{fig.pred-perseus-outflow-7-pv} indicate that if we integrate the \co\ emission over the full velocity range, -2 km/s to 15 km/s, the outflow morphology vanishes in the bright cloud emission. However, both models ME1 and MF are still able to find the outflow when the entire velocity range is searched rather than just the range of channels containing the visually identified outflow.

As shown in Figure~\ref{fig.pred-perseus-outflow-7-pv}, model ME1 predicts a wider velocity coverage compared to that predicted by model MF. This is because at high-velocity channels, the fraction of mass associated with feedback is higher compared to that of the rest frame gas. Model MF is more likely to highlight a higher fraction of mass associated with feedback voxels than model ME1. On the other hand, model ME1 predicts the \co\ emission associated with feedback. Generally there is more \co\ emission at the rest frame of the cloud, which is produced by the molecular cloud itself, than at high velocities. Model ME1 is likely to identify the stronger emission velocities associated with feedback that span near the rest frame velocity. This in turn demonstrates that model ME1 likely overestimates the mass associated with feedback. 

\begin{figure*}[hbt!]
\centering
\includegraphics[width=0.98\linewidth]{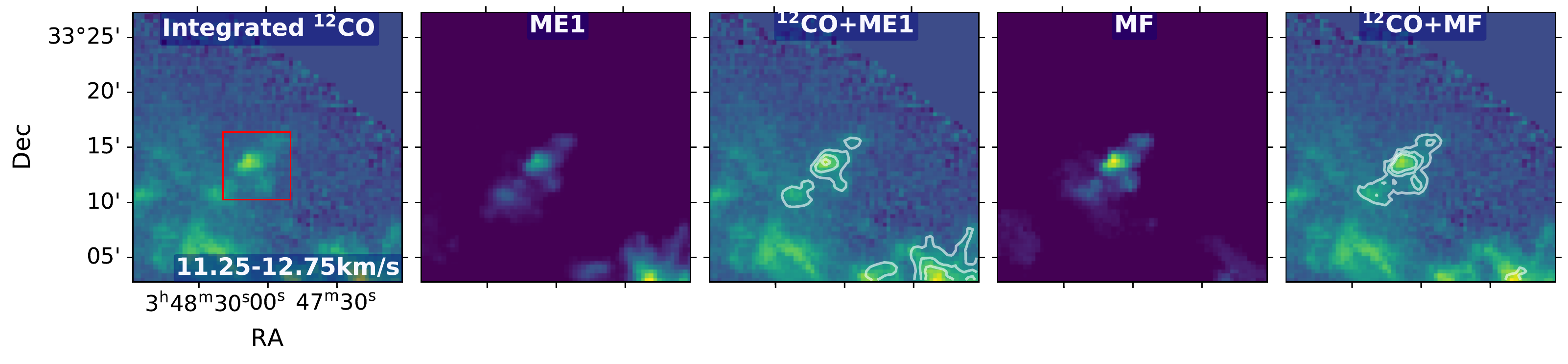}
\caption{Results of models ME1 and MF applied to previously identified Perseus outflow CPOC 57. First panel: integrated intensity of \co\ over the outflow velocity channels indicated by \citet{2010ApJ...715.1170A}. The red box shows the position of the outflow. Second panel: predicted intensity integrated along the velocity axis from model ME1. Third panel: integrated intensity of \co\ over the outflow velocity channels overlaid with the model ME1 prediction in white contour. Fourth and fifth panels: same as the second and the third panels but show the prediction by model MF.}
\label{fig.pred-perseus-outflow-57}
\end{figure*}

Among the 60 previously identified outflows, about one third lack a clearly associated driving source. Consequently, some of these candidates may not in fact be true outflows. Figure~\ref{fig.pred-perseus-outflow-57} shows an example of an outflow without nearby YSOs but which is identified as an outflow in \citet{2010ApJ...715.1170A}. Similar to human identification, models ME1 and MF successfully capture the morphology of this outflow, which in turn demonstrates that \CASItD\ identifies outflows via similar characteristics that humans have used. This lends confidence that these candidates are in fact real outflows. So, either these are true outflows, whose driving source has not yet been located, or these emission features are excellent facsimiles of actual outflows. Without additional data, such as a complete YSO census, it is impossible to distinguish coherent high velocity features that may be caused by turbulence from stellar feedback. We discuss the presence of YSOs in detail in Section~\ref{Relation Between Outflow Properties and Candidate Driving Sources}.

To summarize, both \CASItD\ models successfully identify all 60 perviously identified outflows\footnote{The velocity range of CPOC14 is 9.0-10.0 km/s rather than 9.8-10.8 km/s indicated in \citet{2010ApJ...715.1170A}.}. We note Figures~\ref{fig.pred-perseus-outflow-7} and \ref{fig.pred-perseus-outflow-57} represent typical identifications among the 60 outflows.

\subsection{New Outflow Candidates Identified in Perseus}
\label{New Outflow Candidates Identified in Perseus}

Apart from the previously identified outflows, we find 20 new candidate outflows. Figure~\ref{fig.pred-perseus-lm-region-ME1} shows that model ME1 successfully predicts all previously identified outflows (red boxes). However, Figure~\ref{fig.pred-perseus-lm-region-ME1} shows there are outflows identified by model ME1 that were not previously identified by \citet{2010ApJ...715.1170A} (yellow boxes). These outflows are also identified by model MF (Figure~\ref{fig.pred-perseus-lm-region-MF}), which gives us confidence that they are indeed real outflows. Visual inspection of the new identifications indicates that they have similar high-velocity gas distributions to the previously identified outflows. Hereafter each of these newly identified high-velocity features is named according to the CPOC numbering established by \citet{2010ApJ...715.1170A}. We list their positions and other properties in Table~\ref{Table Physical Properties of New Outflow Candidates in Perseus}.

\begin{figure*}[hbt!]
\centering
\includegraphics[width=0.98\linewidth]{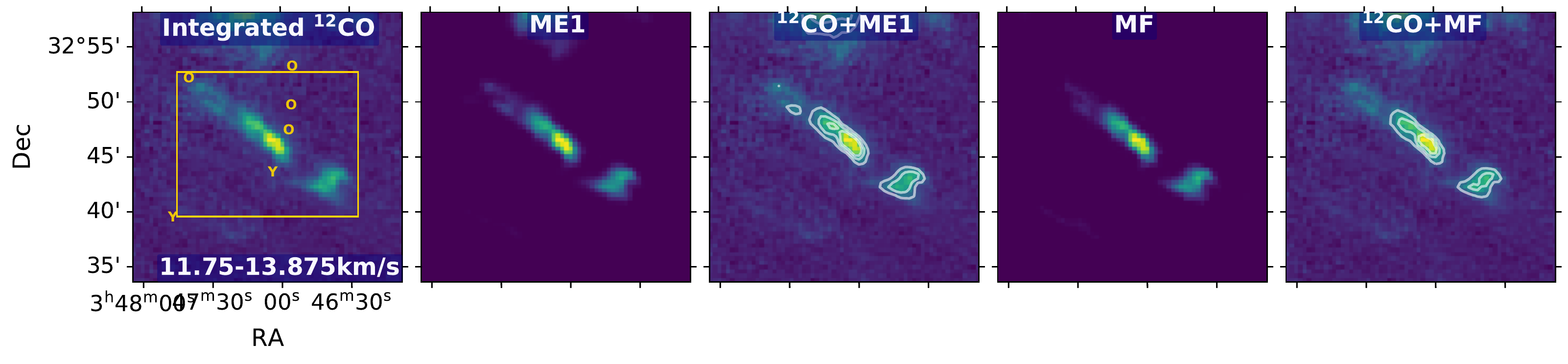}
\caption{New outflow candidate, CPOC 79, identified by models ME1 and MF. First panel: intensity of \co\ integrated over the outflow velocity channels, from 11.5 km/s to 14.0 km/s. The red box shows the position of the outflow. Letters ``Y'' and ``O'' mark YSO positions, as described in Figure~\ref{fig.pred-perseus-lm-region-mask}. Second panel: predicted intensity integrated along the velocity axis from model ME1. Third panel: integrated intensity of \co\ over the outflow velocity channels overlaid with the model ME1 prediction in white contours. Fourth and fifth panels: same as the second and the third panels but displaying the prediction by model MF.}
\label{fig.pred-perseus-outflow-new1}
\end{figure*}

\begin{figure*}[hbt!]
\centering
\includegraphics[width=0.48\linewidth]{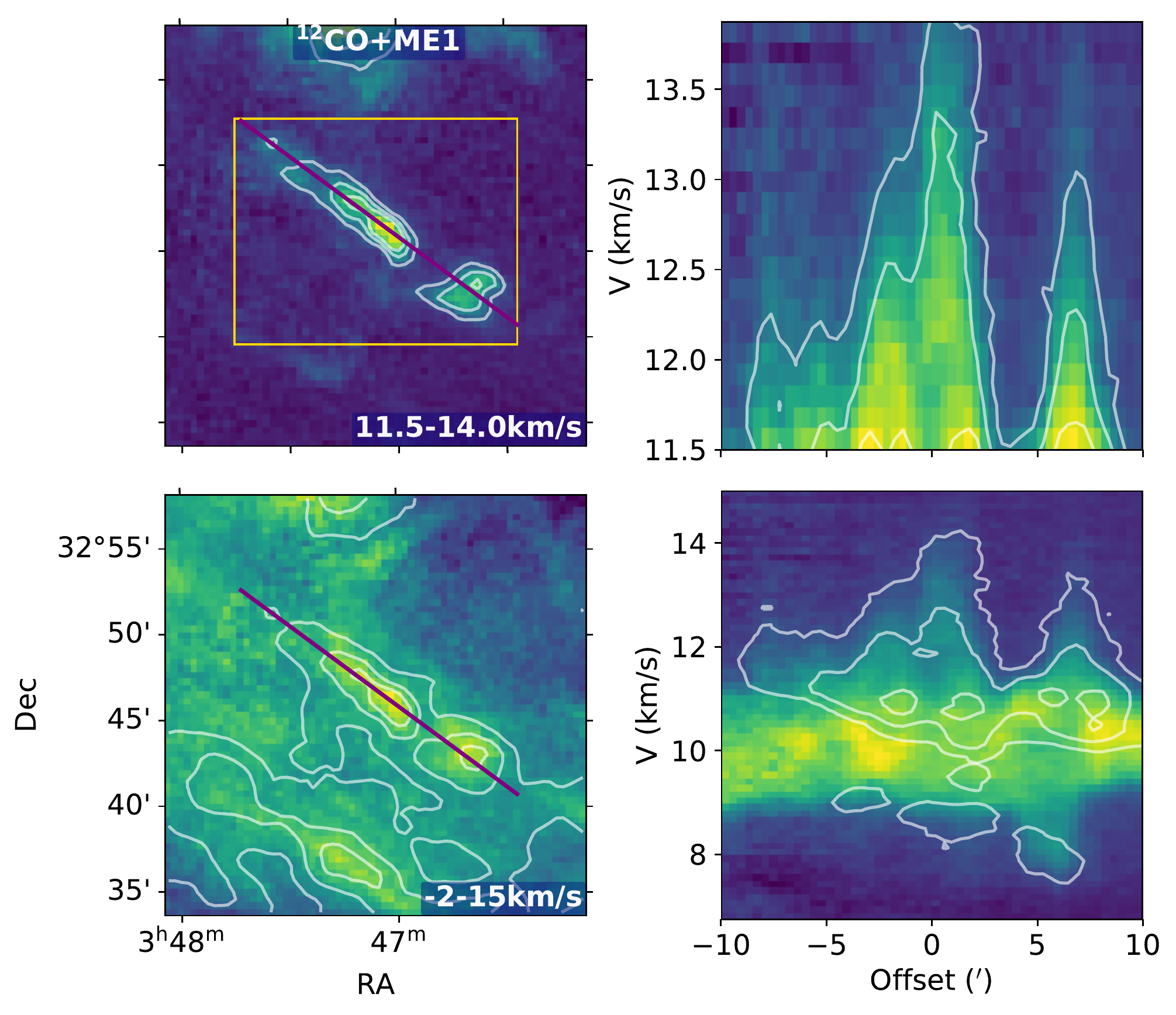}
\includegraphics[width=0.48\linewidth]{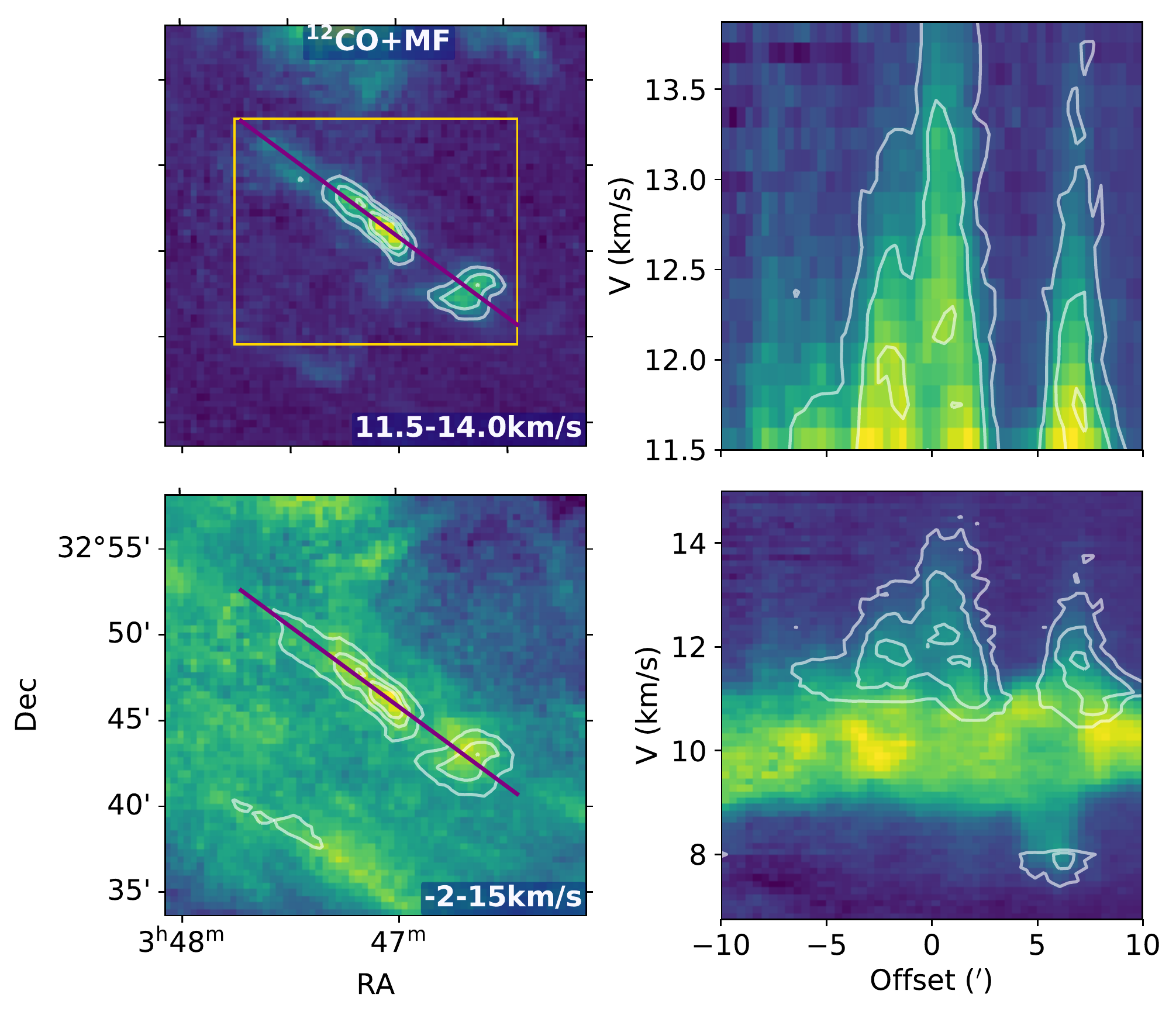}
\caption{The same as Figure~\ref{fig.pred-perseus-outflow-7-pv} but toward the new candidate outflow shown in Figure~\ref{fig.pred-perseus-outflow-new1}. }
\label{fig.pred-perseus-outflow-new1-pv}
\end{figure*}

Figure~\ref{fig.pred-perseus-outflow-new1} shows an example of newly identified outflow CPOC 79. Both models ME1 and MF capture the morphology of this new candidate. In addition, several YSOs are found nearby, one of which is likely the driving source. Figure~\ref{fig.pred-perseus-outflow-new1-pv} shows the position-velocity diagram of the \co\ emission for this outflow. Figure~\ref{fig.pred-perseus-outflow-new1-pv} shows distinct coherent high-velocity structures, which are similar to those for previously identified outflows (e.g., Figure \ref{fig.pred-perseus-outflow-7-pv}). Both Models ME1 and MF identify all 20 new outflow candidates.

Our \CASItD\ models indicate some of the previously identified outflows are more extended than previously found. For example, CPOC 4 and 5 (region IV in Figure~\ref{fig.pred-perseus-lm-region-ME1} and \ref{fig.pred-perseus-lm-region-MF}) appear to be part of much more extended outflow lobes. Both models show the red and blue counterparts. CPOC 4 and 5 and newly identified CPOC 61 and 62 are likely driven by a cluster of young sources, which creates a 2 pc scale combined outflow.

\begin{table*}[]
\begin{center}
\caption{Physical Properties of New Outflow Candidates in Perseus  \label{Table Physical Properties of New Outflow Candidates in Perseus}}
\begin{tabular}{cccccccc}
\hline 
Source  & RA  \, \, \, \,  Dec & Area &Velocity  & Velocity &   Mass  &   Momentum   & Energy \\
Name  &  (J2000) &  (arcmin) &(ME1, km/s)  & (MF, km/s)&   ($\msun$)   &   ($\msun$ km/s)   &  ($10^{43}$ ergs)\\
\hline
CPOC 61   &    03$^{h}$25$^{m}$43.9$^{s}$ +30$^{\circ}$35$^{\prime}$45.0$^{\prime\prime}$   &   6$\times$10   &   -2-5.1   &   0.1-1.4   &   0.09   &   0.28   &   1.9\\
CPOC 62   &    03$^{h}$25$^{m}$49.8$^{s}$ +30$^{\circ}$40$^{\prime}$21.0$^{\prime\prime}$   &   11$\times$13   &   3.8-8.9   &   5.3-9.9   &   0.13   &   0.40   &   2.8\\
CPOC 63   &    03$^{h}$27$^{m}$25.0$^{s}$ +31$^{\circ}$21$^{\prime}$22.0$^{\prime\prime}$   &   12$\times$9   &   -1.1-4.4   &   -0.4-4.3   &   0.35   &   1.64   &   16.5\\
CPOC 64   &   03$^{h}$28$^{m}$42.7$^{s}$+30$^{\circ}$56$^{\prime}$04.0$^{\prime\prime}$   &   11$\times$13   &   0.5-11.8   &   0.1-13.0   &   0.75   &   1.83   &   10.2\\
CPOC 65   &   03$^{h}$29$^{m}$00.3$^{s}$+30$^{\circ}$33$^{\prime}$27.0$^{\prime\prime}$   &   14$\times$13   &   1.8-6.1   &   2.4-7.4   &   0.82   &   1.19   &   3.8\\
CPOC 66   &    03$^{h}$29$^{m}$08.3$^{s}$ +31$^{\circ}$40$^{\prime}$55.0$^{\prime\prime}$   &   9$\times$12   &   4.1-9.0   &   3.8-7.1   &   0.16   &   0.28   &   1.2\\
CPOC 67   &   03$^{h}$30$^{m}$18.8$^{s}$+30$^{\circ}$32$^{\prime}$41.0$^{\prime\prime}$   &   12$\times$11   &   0.5-8.6   &   0.5-6.1   &   1.60   &   1.63   &   4.7\\
CPOC 68   &    03$^{h}$31$^{m}$23.2$^{s}$ +30$^{\circ}$49$^{\prime}$10.0$^{\prime\prime}$   &   8$\times$14   &   -0.5-6.3   &   -0.5-4.3   &   0.24   &   0.95   &   8.1\\
CPOC 69   &    03$^{h}$33$^{m}$34.1$^{s}$ +31$^{\circ}$06$^{\prime}$02.0$^{\prime\prime}$   &   9$\times$9   &   0.1-4.1   &   -0.3-2.9   &   0.12   &   0.57   &   5.6\\
CPOC 70   &   03$^{h}$33$^{m}$42.2$^{s}$+30$^{\circ}$50$^{\prime}$42.0$^{\prime\prime}$   &   13$\times$8   &   -1.3-14.3   &   -2-6.6   &   0.13   &   0.44   &   4.1\\
CPOC 71   &   03$^{h}$34$^{m}$36.6$^{s}$+31$^{\circ}$07$^{\prime}$11.0$^{\prime\prime}$   &   9$\times$8   &   2.3-4.9   &   -0.1-4.3   &   0.37   &   1.75   &   16.9\\
CPOC 72   &   03$^{h}$38$^{m}$32.0$^{s}$+31$^{\circ}$56$^{\prime}$38.0$^{\prime\prime}$   &   10$\times$12   &   6.8-10.0   &   5.6-10.4   &   0.43   &   0.33   &   0.8\\
CPOC 73   &   03$^{h}$39$^{m}$13.5$^{s}$+31$^{\circ}$12$^{\prime}$56.0$^{\prime\prime}$   &   15$\times$13   &   2.4-7.9   &   2.6-7.5   &   0.59   &   1.18   &   7.0\\
CPOC 74   &   03$^{h}$39$^{m}$30.1$^{s}$+32$^{\circ}$09$^{\prime}$17.0$^{\prime\prime}$   &   5$\times$6   &   6.8-9.9   &   7.1-8.3   &   0.19   &   0.13   &   0.2\\
CPOC 75   &   03$^{h}$40$^{m}$01.8$^{s}$+32$^{\circ}$01$^{\prime}$14.0$^{\prime\prime}$   &   9$\times$9   &   5.6-8.9   &   5.1-7.9   &   1.21   &   2.07   &   7.4\\
CPOC 76   &   03$^{h}$41$^{m}$54.4$^{s}$+31$^{\circ}$56$^{\prime}$38.0$^{\prime\prime}$   &   7$\times$7   &   5.6-10.5   &   6.5-8.1   &   0.46   &   0.71   &   2.2\\
CPOC 77   &   03$^{h}$42$^{m}$41.8$^{s}$+31$^{\circ}$51$^{\prime}$16.0$^{\prime\prime}$   &   4$\times$8   &   5.5-9.1   &   4.9-6.9   &   0.02   &   0.06   &   0.4\\
CPOC 78   &   03$^{h}$46$^{m}$00.3$^{s}$+31$^{\circ}$53$^{\prime}$57.0$^{\prime\prime}$   &   9$\times$16   &   5.6-9.8   &   5.8-7.5   &   0.06   &   0.17   &   1.0\\
CPOC 79   &   03$^{h}$47$^{m}$06.9$^{s}$ +32$^{\circ}$45$^{\prime}$42.0$^{\prime\prime}$   &   16$\times$13   &   9.5-14.1   &   10.3-14.1   &   1.79   &   2.53   &   8.1\\
CPOC 80   &    03$^{h}$48$^{m}$32.4$^{s}$ +32$^{\circ}$55$^{\prime}$17.0$^{\prime\prime}$   &   6$\times$6   &   9.6-12.3   &   10.1-13.1   &   0.11   &   0.17   &   0.6\\
    \hline
    Total &  - & - & - & - & 9.63  & 11.3  & 104 \\
    
\hline
 \multicolumn{8}{p{1\linewidth}}{Notes:}\\
\multicolumn{8}{p{1\linewidth}}{
$^{a}$ Outflow name, position, area, velocity range indicated by model ME1, velocity range indicated by model MF, mass predicted by model MF, 1D momentum predicted by model MF, and 1D energy predicted by model MF. }\\
\end{tabular}
\end{center}
\end{table*}

\subsection{Physical Properties of Outflows in Perseus}
\label{Physical Properties of Outflows in Perseus}

Next, we estimate the masses of the outflows identified by models ME1 and MF and compare them with the previous observational estimates. We calculate the outflow mass, 1D momentum and 1D energy for each model as described in Section~\ref{Assessing Model Accuracy Using Synthetic Observations}. We further consider the outflow properties as defined by the subset of velocity channels previously identified by \citet{2010ApJ...715.1170A} as belonging to the outflow as well as the outflow defined by using all velocity channels.

\begin{figure}[hbt!]
\centering
\includegraphics[width=0.98\linewidth]{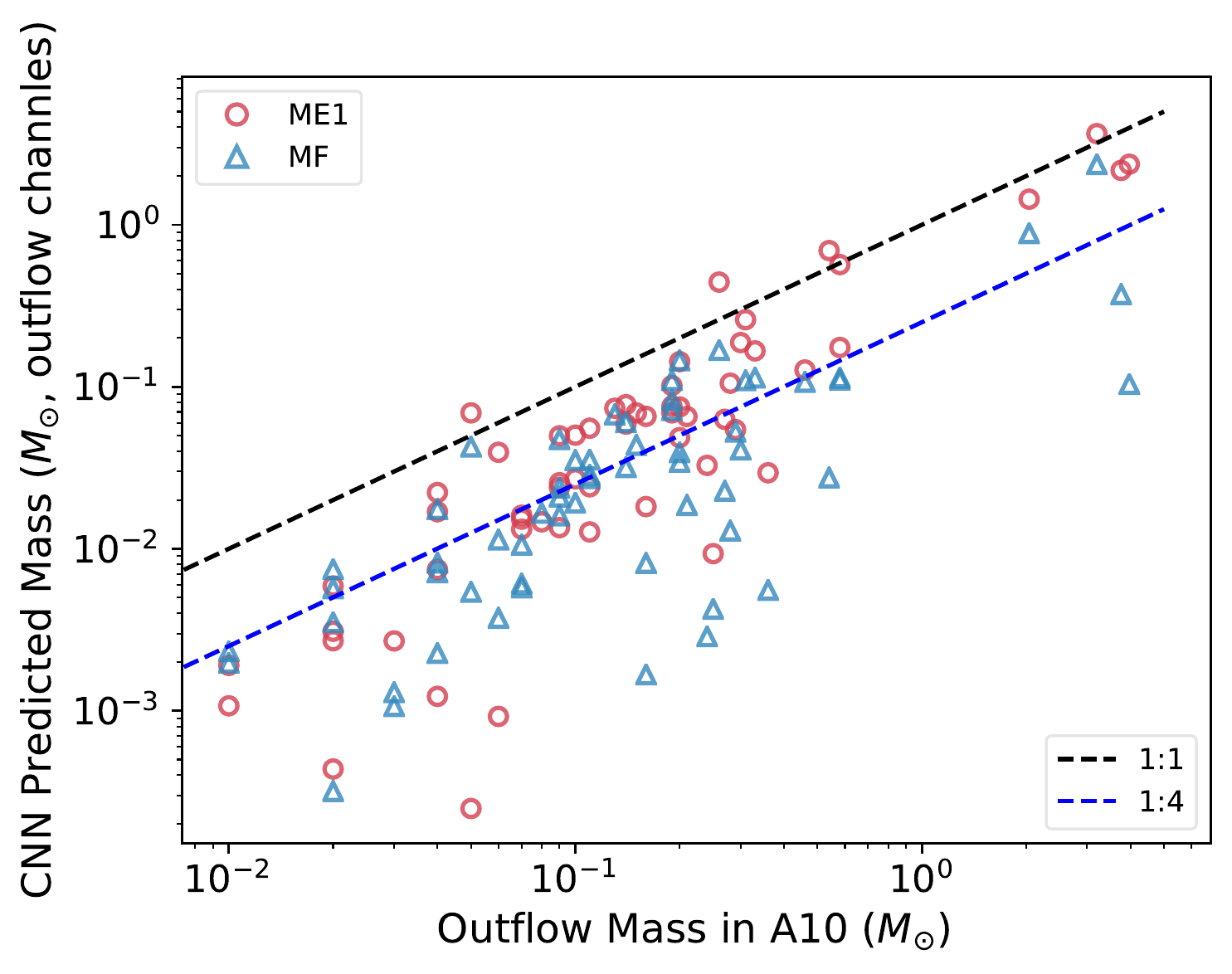}
\caption{Relation between the outflow mass predicted by \CASItD\ and the outflow mass calculated by \citet{2010ApJ...715.1170A}, where both using only the outflow velocity channels. The circle symbols indicate the mass calculated by model ME1. The triangle symbols represent the mass calculated by model MF. The black dashed line and the blue dashed line have a slope of 1 and 1/4, respectively. }
\label{fig.cnn-perseus-mass-oc}
\end{figure}

Figure~\ref{fig.cnn-perseus-mass-oc} compares the outflow mass estimated by our CNN models and by \citet{2010ApJ...715.1170A}, where both include mass only in the outflow velocity channels determined by \citet{2010ApJ...715.1170A}. \citet{2010ApJ...715.1170A} calculate the mass of each outflow by adding all the emission in the outflow velocity channels within the red box as shown in Figures~\ref{fig.pred-perseus-outflow-7} and \ref{fig.pred-perseus-outflow-57}. Figure~\ref{fig.cnn-perseus-mass-oc} suggests that extra emission from diffuse gas not associated with the outflow overestimates the mass compared to our models. 

For some outflows, the masses predicted by the two CNN models are similar. This is likely because the mass estimates are dominated by emission in the high-velocity channels, which are similar for both models. However, the masses predicted by the two CNN models for other outflows is quite different. We conclude that this is because these outflows span a wider velocity range. The fraction of mass associated with the outflow as predicted by model MF drops in the rest-frame velocity channels. In these channels, model ME1 likely overestimates the mass by a large factor, because it includes cloud material.

If we only consider the outflow mass located in the outflow velocity channels, i.e., where the outflow morphology is distinct, the traditional observational approach to calculate the outflow mass probably overestimates the mass for two reasons. First, it includes extra emission around the outflows that is not associated with them but is enclosed in the boxes indicated in \citet{2010ApJ...715.1170A}. As shown in Figure~\ref{fig.pred-perseus-outflow-7}, the red box encloses a large region where the outflow appears to occupy less than one third of the entire box. Some extended emission near the boundary of the box contributes significantly. Second, apart from the extra emission around the outflow lobe, background or foreground cloud gas not associated with outflows that is located in the outflow velocity channels at the outflow location contributes additional emission. The effect of contamination by cloud gas in model ME1 was previously found by \citet{2020ApJ...890...64X} in an analysis of bubble identifications. Model ME1, which more closely models how humans visually identify data by construction, is not able to accurately separate feedback and non-feedback emission. However, on the other hand, mass associated with feedback that is located outside the outflow velocity channels but is difficult to distinguish by visual examination is apparently missed. 

Our \CASItD\ models are able to identify the outflow structure across the entire velocity range. As indicated in Figure~\ref{fig.pred-perseus-outflow-7-pv}, the visually identified outflow velocity channels are between -0.5 km/s and 2.0 km/s. The model ME1 prediction in the position-velocity diagram shows that the outflow structure extends from -0.5 km/s to 4.8 km/s. Model MF illustrates that the outflow structure extends from -0.5 km/s to 2.5 km/s. Both models demonstrate a wider outflow velocity range than that identified by visual inspection. 

\begin{figure}[hbt!]
\centering
\includegraphics[width=0.98\linewidth]{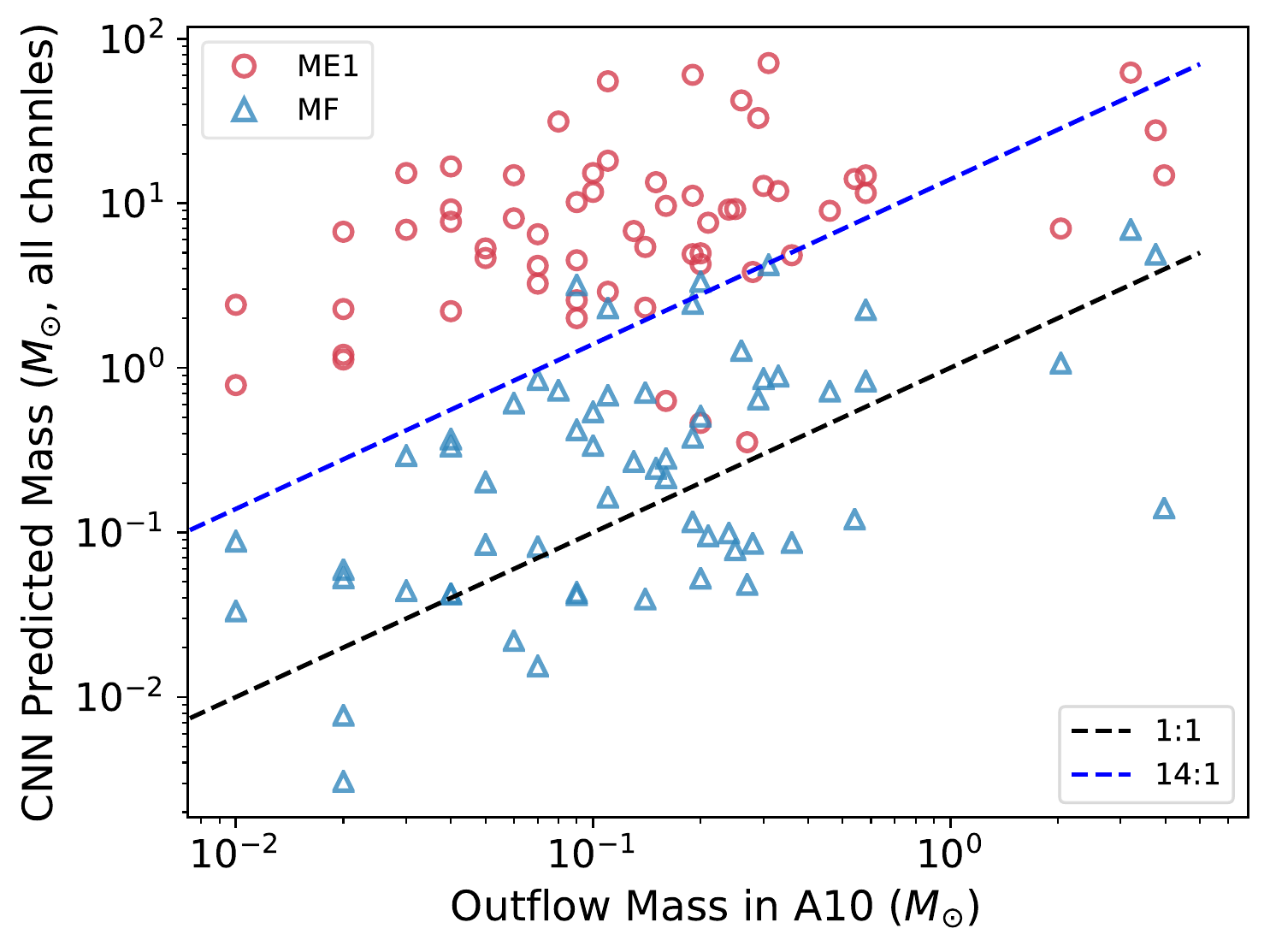}
\caption{Relation between the CNN predicted outflow mass over the entire velocity channels and the outflow mass calculated by \citet{2010ApJ...715.1170A}. The circle symbols indicate the mass calculated by model ME1. The triangle symbols represent the mass calculated by model MF. The black dashed line and the blue dashed line have a slope of 1 and 14, respectively. }
\label{fig.cnn-perseus-mass-ac}
\end{figure}

Figure~\ref{fig.cnn-perseus-mass-ac} shows the relation between the CNN predicted outflow mass over all velocity channels and the outflow mass calculated by \citet{2010ApJ...715.1170A}, which uses only the high velocity channels. Model ME1 overestimates the outflow mass by an order of magnitude compared to that predicted by \citet{2010ApJ...715.1170A}. This is caused by large contamination from foreground or background gas near the cloud rest-frame velocity channels. When correcting for the fraction of mass associated with outflows, the final mass predicted by model MF is comparable to the outflow mass calculated by \citet{2010ApJ...715.1170A}. This similarity is due to a cancelation in errors: \citet{2010ApJ...715.1170A} miss outflow material emitting in the cloud rest-frame channels, however, they compensate for this by over-estimating the amount of mass in the high-velocity channels.

\begin{figure}[hbt!]
\centering
\includegraphics[width=0.98\linewidth]{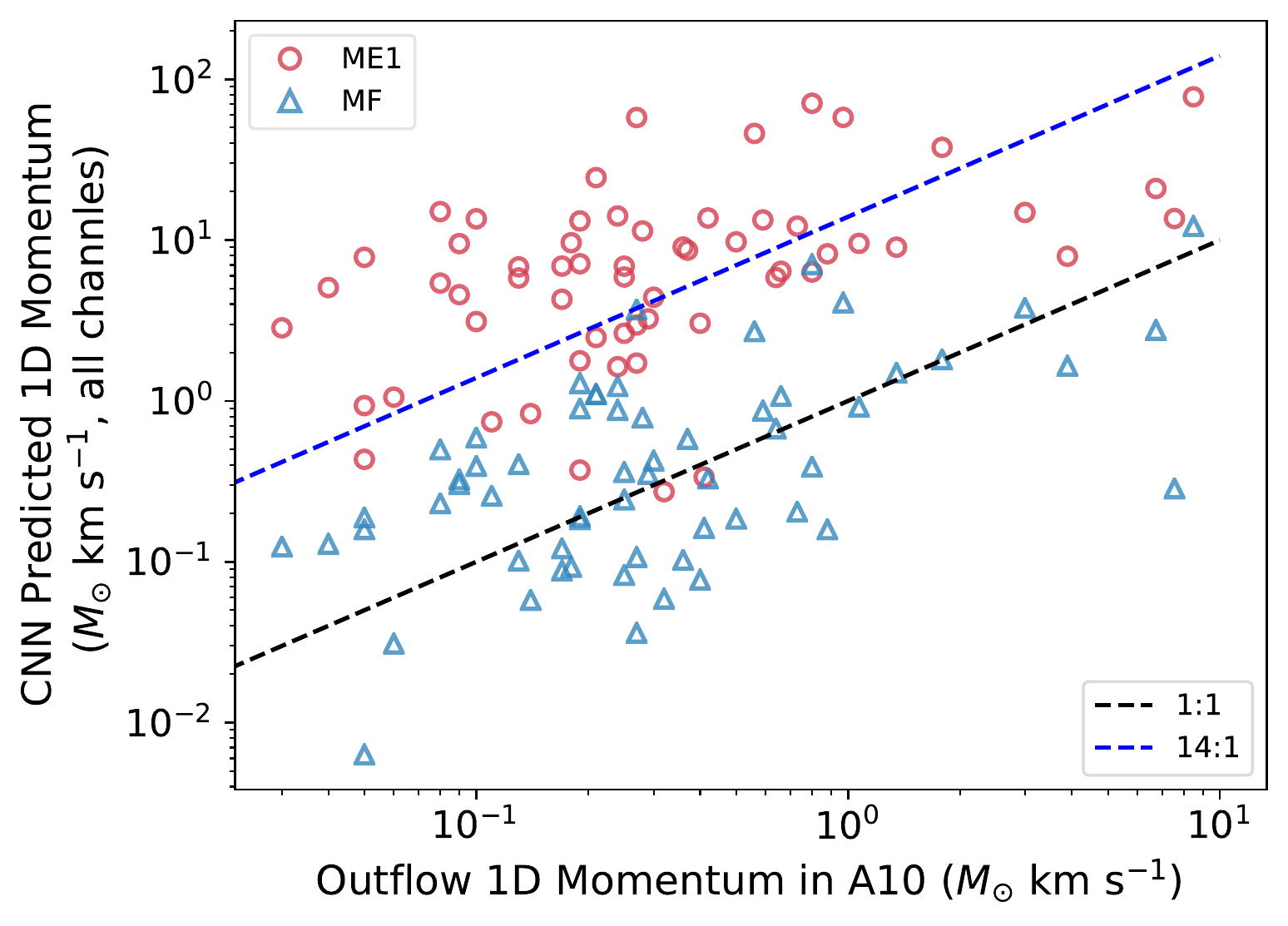}
\caption{Relation between the CNN predicted outflow 1D momentum over the entire velocity channels and the 1D outflow momentum calculated by \citet{2010ApJ...715.1170A}. The circle symbols indicate the momentum calculated by model ME1. The triangle symbols represent the momentum calculated by model MF. The black dashed line and the blue dashed line have a slope of 1 and 14, respectively. }
\label{fig.cnn-perseus-momentum-ac}
\end{figure}

\begin{figure}[hbt!]
\centering
\includegraphics[width=0.98\linewidth]{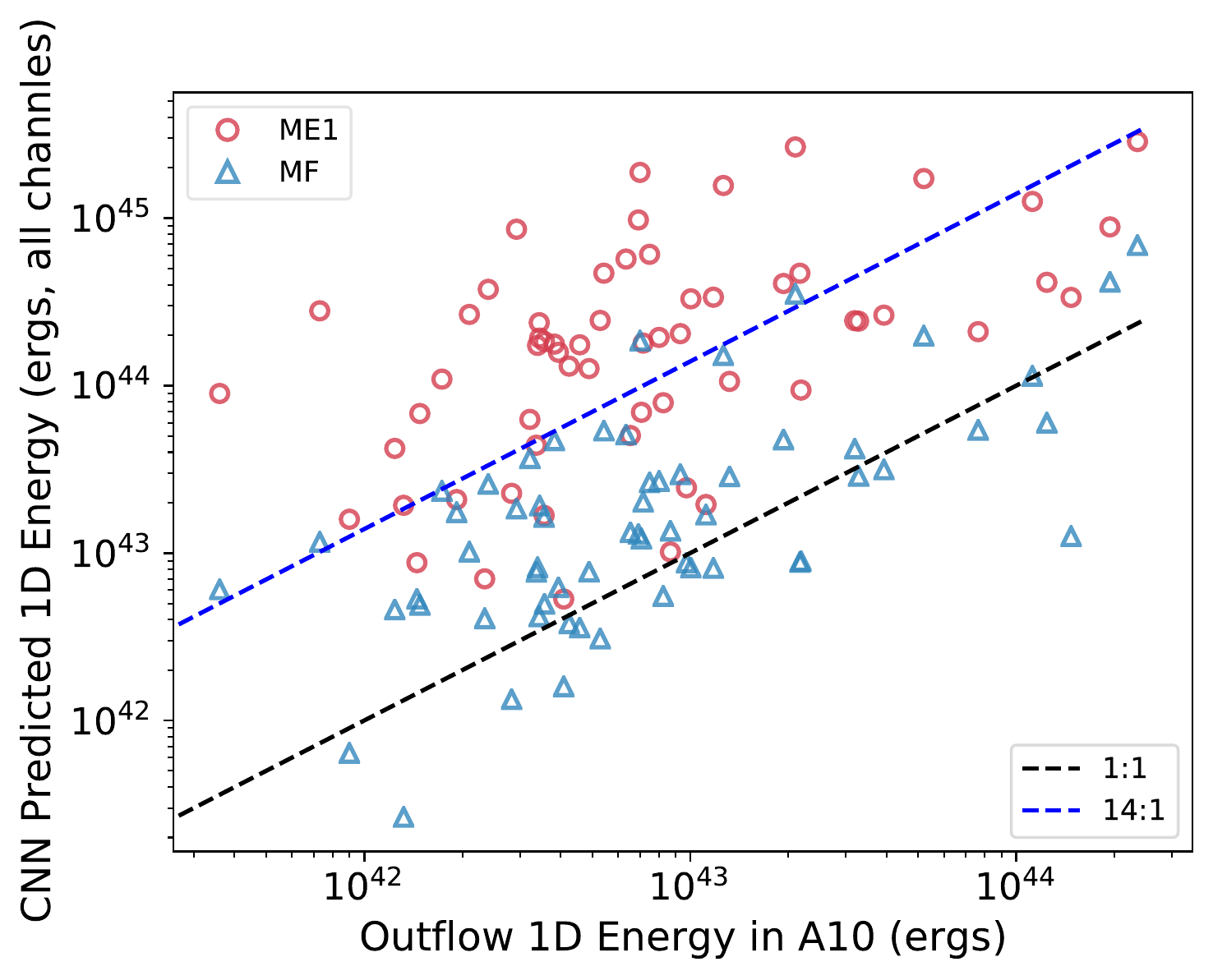}
\caption{Relation between the CNN predicted outflow 1D energy over the entire velocity channels and the 1D outflow energy calculated by \citet{2010ApJ...715.1170A}. The circle symbols indicate the energy calculated by model ME1. The triangle symbols represent the energy calculated by model MF. The black dashed line and the blue dashed line have a slope of 1 and 14, respectively. }
\label{fig.cnn-perseus-energy-ac}
\end{figure}

Figure~\ref{fig.cnn-perseus-momentum-ac} shows the relation between the outflow 1D momentum predicted using all velocity channels and the 1D line-of-sight outflow momentum calculated by \citet{2010ApJ...715.1170A}. Figure~\ref{fig.cnn-perseus-energy-ac} shows the relation between the outflow 1D energy predicted using all velocity channels and the outflow 1D line-of-sight energy calculated by \citet{2010ApJ...715.1170A}. The figures show that the outflow 1D momenta and 1D energies predicted by model MF are comparable to the outflow momenta and energies calculated by \citet{2010ApJ...715.1170A}, respectively. 

We estimate that the total 1D momentum is 56.4 $M_{\odot}$ km/s and the 1D energy is $2.6\times 10^{45}$ erg from outflows. These are the same order of magnitude as the 1D calculations in \citet{2010ApJ...715.1170A}, which are 49.2 $M_{\odot}$ km/s and $1.4\times 10^{45}$ ergs, respectively. The correction factor from 1D momentum/energy to 3D momentum/energy in \citet{2010ApJ...715.1170A} is $\sqrt{2}$. This number is consistent with our argument in Section~\ref{Assessing Model Accuracy Using Synthetic Observations}, where an average outflow inclination of $45^{\circ}$ is adopted. Table~\ref{Table Physical Properties of Outflows in Perseus} lists the physical properties of all 60 previously identified Perseus outflows. It is worth noting that to make a fair comparison, the total outflow mass, 1D momentum and 1D energy calculations above do not include the 20 newly identified outflows. The physical properties of the newly identified outflows are listed in Table~\ref{Table Physical Properties of New Outflow Candidates in Perseus}.

The total mass, 1D momentum and 1D energy including the 20 newly identified are 53.8 $M_{\odot}$, 67.7 $M_{\odot}$ km/s and $3.6\times 10^{45}$ erg, respectively.

\begin{longtable*}{cccccc}
\caption{Physical Properties of Outflows in Perseus$^{a}$  \label{Table Physical Properties of Outflows in Perseus}}\\
\hline
Source  & Velocity  & Velocity &   Mass   &   Momentum   & Energy \\
Name  & (ME1, km/s)  & (MF, km/s)&   ($\msun$)   &   ($\msun$ km/s)   &  ($10^{43}$ ergs)\\ \hline
\endfirsthead

\caption{(\emph{Continued})} \\
\hline
Source  & Velocity  & Velocity &   Mass   &   Momentum   & Energy \\
Name  & (ME1, km/s)  & (MF, km/s)&   ($\msun$)   &   ($\msun$ km/s)   &  ($10^{43}$ ergs)\\ \hline
\endhead
 \hline \multicolumn{6}{@{}r@{}}{\emph{Continued on next page}}
\endfoot
  
\endlastfoot
    CPOC 1 & 5.5-6.8 & 6.3-8.6 & 0.08  & 0.19  & 1.3 \\
    CPOC 2 & 4.3-8.9 & 6.0-8.6 & 0.09  & 0.16  & 0.88 \\
    CPOC 3 & 5.0-7.9 & 5.9-9.0 & 0.06  & 0.23  & 1.84 \\
    CPOC 4 & -0.6-1.5 & -1.8-0.8 & 0.03  & 0.13  & 1.02 \\
    CPOC 5 & 1.3-4.3 & 1.4-2.1 & 0.00  & 0.01  & 0.03 \\
    CPOC 6 & 7.9-10.4 & 8.5-11.0 & 0.05  & 0.12  & 0.61 \\
    CPOC 7 & -0.6-7 & -0.8-3.5 & 0.64  & 1.81  & 11.39 \\
    CPOC 8 & 4.1-9.0 & 6.1-9.4 & 0.68  & 0.79  & 2.02 \\
    CPOC 9 & 8.1-11.1 & 8.6-12.8 & 1.27  & 2.70  & 15.15 \\
    CPOC 10 & 8.4-11.1 & 9.0-11.1 & 0.10  & 0.33  & 2.64 \\
    CPOC 11 & 7.6-11.0 & 8.1-11.4 & 2.31  & 3.70  & 18.51 \\
    CPOC 12 & 1.1-10.8 & -1.4-7.8 & 2.45  & 4.08  & 19.8 \\
    CPOC 13 & 8.5-11.0 & 8.8-11.8 & 4.21  & 7.06  & 35.26 \\
    CPOC 14 & 5.1-10.9 & 9.0-9.9 & 0.04  & 0.16  & 1.16 \\
    CPOC 15 & -0.5-6.9 & 0.6-3.8 & 0.12  & 0.39  & 2.88 \\
    CPOC 16 & 7.9-10.8 & 8.8-10.8 & 0.37  & 0.59  & 2.58 \\
    CPOC 17 & 7.1-13.6 & 8.0-13.6 & 6.91  & 12.19 & 68.79 \\
    CPOC 18 & 7.3-12.4 & 7.9-12.4 & 0.73  & 1.10  & 5.09 \\
    CPOC 19 & 8.3-13.0 & 7.3-14.3 & 0.38  & 0.87  & 4.75 \\
    CPOC 20 & 7.1-10.8 & 8.1-11.3 & 0.33  & 0.50  & 2.34 \\
    CPOC 21 & 7.0-11.1 & 8.4-11.4 & 0.20  & 0.40  & 1.92 \\
    CPOC 22 & 0.9-3.3 & 1.1-2.3 & 0.86  & 0.92  & 3.15 \\
    CPOC 23 & 0.4-2.5 & 0.1-3.0 & 0.09  & 0.19  & 1.64 \\
    CPOC 24 & -1.6-6.1 & -1.6-4.4 & 0.72  & 3.78  & 41.52 \\
    CPOC 25 & -1.6-0.9 & -1.8-1 & 0.01  & 0.06  & 0.87 \\
    CPOC 26 & 6.8-9.6 & 7.5-10.4 & 0.04  & 0.11  & 0.55 \\
    CPOC 27 & -0.8-3.9 & -0.6-2.3 & 0.02  & 0.08  & 0.89 \\
    CPOC 28 & 5.1-10.3 & 5.3-12.4 & 2.25  & 1.49  & 4.19 \\
    CPOC 29 & 6.5-9.4 & 9.1-9.6 & 0.02  & 0.09  & 0.77 \\
    CPOC 30 & 7.1-10.0 & 8.1-10.5 & 0.04  & 0.10  & 0.5 \\
    CPOC 31 & 6.6-10.0 & 7.6-11.3 & 0.24  & 0.58  & 2.95 \\
    CPOC 32 & 6.5-9.3 & 7.0-9.8 & 0.08  & 0.12  & 0.38 \\
    CPOC 33 & 6.4-10.3 & 9.0-9.9 & 0.09  & 0.18  & 0.81 \\
    CPOC 34 & 7.1-9.5 & 8.9-9.9 & 0.04  & 0.08  & 0.36 \\
    CPOC 35 & 7.3-9.6 & 7.8-10.5 & 0.34  & 0.88  & 5.37 \\
    CPOC 36 & 5.9-10.9 & 8.9-10.0 & 0.16  & 0.42  & 2.68 \\
    CPOC 37 & 6.9-10.1 & 7.3-10.1 & 0.28  & 0.18  & 0.41 \\
    CPOC 38 & 6.4-10.0 & 5.9-9.0 & 0.04  & 0.03  & 0.06 \\
    CPOC 39 & 8.3-12.1 & 8.4-13.1 & 0.54  & 0.90  & 4.69 \\
    CPOC 40 & 6.5-10.8 & 5.6-10.9 & 3.18  & 1.23  & 1.69 \\
    CPOC 41 & 8.4-12.0 & 9.3-11.8 & 0.71  & 1.10  & 3.67 \\
    CPOC 42 & 7.1-10.0 & 6.5-10.8 & 3.33  & 1.29  & 1.75 \\
    CPOC 43 & 8.5-11.5 & 9.3-11.9 & 1.06  & 1.65  & 5.43 \\
    CPOC 44 & 9.8-10.9 & 10.8-11.6 & 0.04  & 0.09  & 0.42 \\
    CPOC 45 & 8.8-11.8 & 10.8-11.5 & 0.12  & 0.20  & 0.82 \\
    CPOC 46 & 7.6-11.1 & 9.8-10.8 & 0.89  & 1.07  & 2.86 \\
    CPOC 47 & 5.8-9.0 & 4.6-8.4 & 0.05  & 0.16  & 1.35 \\
    CPOC 48 & 8.8-11.4 & 10.9-11.3 & 0.21  & 0.24  & 0.62 \\
    CPOC 49 & 6.0-6.9 & 5.1-5.9 & 0.29  & 0.32  & 0.82 \\
    CPOC 50 & 9.5-11.1 & 10.9-11.3 & 0.08  & 0.10  & 0.31 \\
    CPOC 51 & 9.4-11.5 & 9.3-12.5 & 0.05  & 0.06  & 0.16 \\
    CPOC 52 & 7.8-10.4 & 6.8-8.6 & 0.27  & 0.35  & 1.33 \\
    CPOC 53 & 9.4-11.6 & 8.6-12.3 & 0.84  & 0.39  & 0.46 \\
    CPOC 54 & 9.3-11.9 & 10.4-12.0 & 0.61  & 0.30  & 0.49 \\
    CPOC 55 & 9.1-13.1 & 11.6-12.6 & 0.14  & 0.29  & 1.26 \\
    CPOC 56 & 6.5-10.9 & 5.9-13.5 & 4.88  & 2.76  & 5.99 \\
    CPOC 57 & 9.4-12.0 & 9.6-12.5 & 0.42  & 0.26  & 0.53 \\
    CPOC 58 & 7.5-11.8 & 9.9-11.0 & 0.09  & 0.04  & 0.13 \\
    CPOC 59 & 7.6-10.9 & 6.1-12.1 & 0.51  & 0.36  & 0.77 \\
    CPOC 60 & 7.9-11.9 & 8.0-9.9 & 0.83  & 0.67  & 1.22 \\   
    \hline
    Total & - & - & 44.2  & 56.4  & 260 \\        
\hline
\multicolumn{6}{p{1\linewidth}}{Notes:}\\
\multicolumn{6}{p{1\linewidth}}{
$^{a}$ Outflow name, velocity range indicated by model ME1, velocity range indicated by model MF, mass predicted by model MF, 1D momentum predicted by model MF, and 1D energy predicted by model MF. }\\
\end{longtable*}

\subsection{Relation Between Outflow Properties and Candidate Driving Sources}
\label{Relation Between Outflow Properties and Candidate Driving Sources}

In this section, we discuss the relation between outflow properties and the type of the source that likely drives each outflow. We first divided the previously identified outflows into several categories based on the YSOs nearby. Some outflows are located near multiple sources, so the true driving source is ambiguous. The four categories are:
\begin{itemize}
\item Younger YSOs: The outflow is likely driven by an early evolutionary stage YSO, e.g. protostars with cold, dusty, quasi-spherical envelopes \citep{Gutermuth.in.prep,2020arXiv200505466P}. 

\item Older YSOs with disks: The outflow is likely driven by a later evolutionary stage YSO, e.g. pre-main sequence stars with protoplanetary disks \citep{Gutermuth.in.prep,2020arXiv200505466P}. 

\item Multiple YSOs: Multiple YSOs are nearby. Either an early or a late evolutionary stage YSO may drive the outflow. 

\item No YSOs: The outflow has distinct coherent high-velocity features but has no known YSOs in close proximity. 

\item No Spitzer: The outflow is located outside the YSOs catalog coverage \citep{Gutermuth.in.prep,2020arXiv200505466P}. 

\end{itemize}

We note that there are many more YSOs than identified outflows. This may mean the outflows of these sources are small and/or relatively low-velocity, perpendicular to the line of sight such that no high velocity gas is apparent or not distinct from the identified outflow gas. The difference in outflow and source number may also be because the outflows of multiple sources are blended together and difficult to cleanly separate (see Section~\ref{Outflows in Star Clusters} for additional discussion).

Among the 60 outflows identified by \citet{2010ApJ...715.1170A}, 12 outflows are outside the Spitzer YSO catalog coverage as indicated in Figure~\ref{fig.pred-perseus-lm-region-mask}. Of the remaining 48 outflows, 8 (17\%) are associated with younger YSOs,12 (25\%) are associated with older YSOs with disks, 19 (39\%) are associated with multiple YSOs and 9 (19\%) have no nearby source. There are two possibilities to explain the latter category of outflows: either the YSO catalog is not complete or the outflows without YSOs are false detections. The SESNA YSO catalog is derived from Spitzer mid-IR data, and thus it is quite sensitive to most YSOs.  However, the youngest type, Class 0, can be missed. For outflow identifications with no YSO catalog member, we visually examined Herschel 70 and 160~$\mu$m images where those particularly young protostars should be very bright. We found no evidence for hidden and highly embedded protostars in those data. However, there is still a chance of overlooking faint/young sources due to the sensitivity. We also note that it is implausible for YSOs to move out of the field of view on the dynamical time of the outflow, which is usually of order $10^{3}-10^{4}$ years. On the other hand, outflows are likely to extend beyond the box circumscribing them, making it difficult to identify which part of the outflow is closest to the driving source. Thus it is possible the driving source is outside the box.

We stress that the outflows with no nearby source are identified by both models and by eye as well. We confirm these outflow candidates do visually look like other outflows with nearby YSOs. Moreover, our test set has an extremely low false positive rate; no false outflows are identified in the simulation data. Considering that morphologies of these high velocity blobs are indistinguishable from confirmed outflows, it is impractical to rule out these candidates by examining only the 3D data without complete YSO catalogs and the ability to match an outflow to a particular YSO source. Correspondingly, we still consider these outflows to be high-confidence candidates. 

\begin{figure}[hbt!]
\centering
\includegraphics[width=0.98\linewidth]{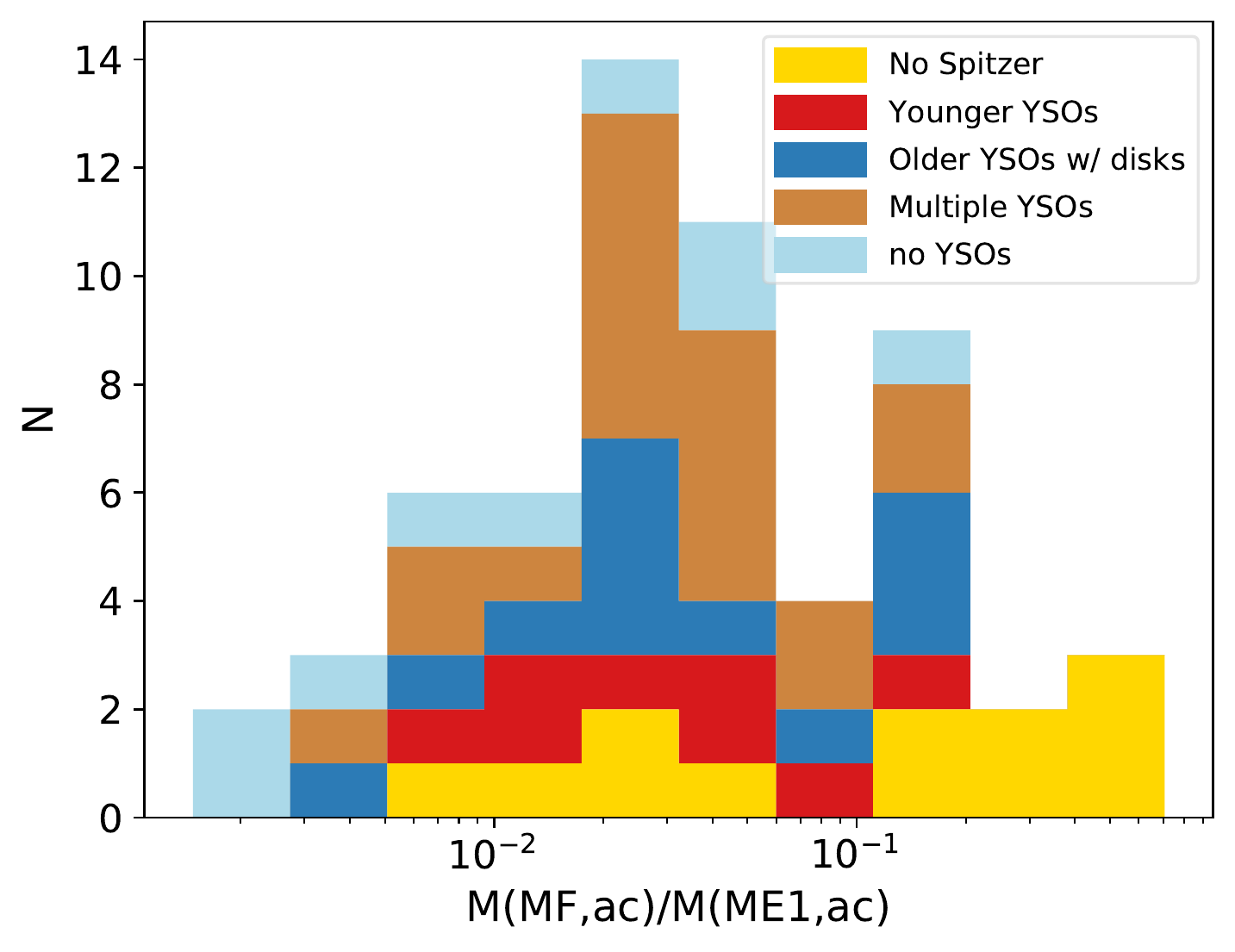}
\caption{Distribution of the ratio between the mass predicted by model MF and that by model ME1 over all velocity channels for different categories of outflows. }
\label{fig.stat-perseus-yso-ME1-MF-ac}
\end{figure}

Figure~\ref{fig.stat-perseus-yso-ME1-MF-ac} shows the distribution of the ratio between the mass predicted by model MF and that by model ME1 over all velocity channels for the different categories of outflows. Since model ME1 includes more mass associated with the cloud, while model MF is a more direct measure of the actual outflow mass, this ratio roughly indicates the fraction of mass associated with feedback. A smaller ratio suggests high contamination from ambient gas in model ME1 that is not associated with the outflow. Generally, at rest-frame velocity channels, this ratio drops quickly compared to that at high velocity channels. In other words, the ratio between the mass predicted from model MF and that from model ME1 across all velocity channels reveals at what velocity more outflow mass is located. If the ratio is high, more mass is likely associated with high velocity channels. If the ratio is low, more mass is likely associated with the rest-frame velocity channels. Figure~\ref{fig.stat-perseus-yso-ME1-MF-ac} suggests that outflows likely driven by late evolutionary stage YSOs (older YSOs with disks) have slightly higher ratios compared to those driven by early evolutionary stage YSOs. This has an intuitive explanation: at early evolutionary stages, YSOs have not ejected a significant amount of high-velocity gas due to the presence of the envelope. The high velocity gas is likely slowed by interaction with dense gas. At late evolutionary stages, YSOs can efficiently expel gas at higher velocities. However, due to the limited sample size, the difference between the two distributions is not significant.

\begin{figure}[hbt!]
\centering
\includegraphics[width=0.98\linewidth]{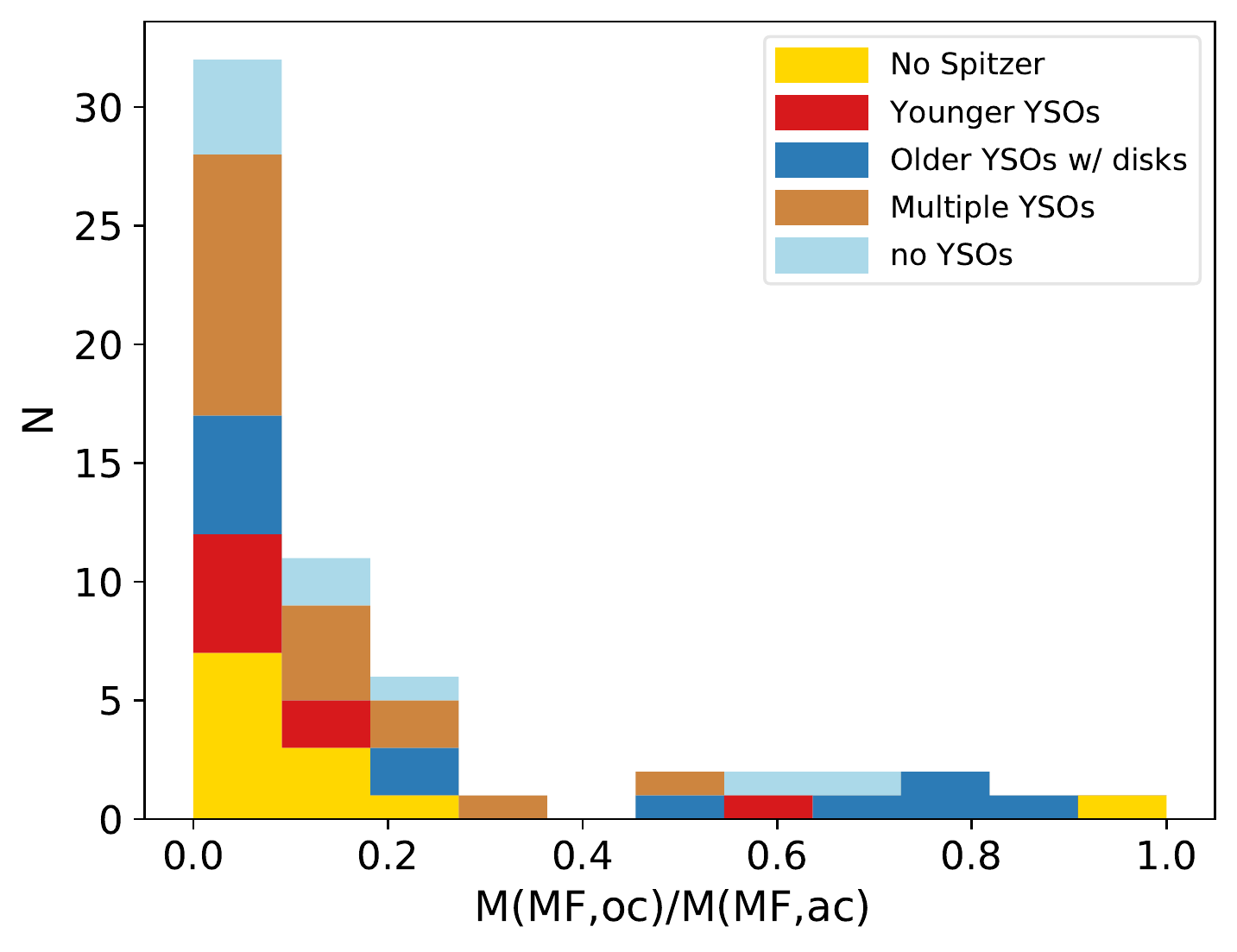}
\caption{Distribution of the ratio between the mass predicted by model MF over only the outflow velocity channels from \citet{2010ApJ...715.1170A} and that over all velocity channels for different categories of outflows. }
\label{fig.stat-perseus-yso-MF-oc-ac}
\end{figure}

Figure~\ref{fig.stat-perseus-yso-MF-oc-ac} shows the distribution of the ratio between the mass predicted by model MF over  only the subset of outflow velocity channels and that predicted over all velocity channels for different categories of outflows. This ratio indicates at which velocities the mass associated with feedback is located. If more mass emits in high velocity channels, where the outflow morphology is the most distinct, this ratio increases. If more mass emits near the rest-frame velocity channels, where the outflow morphology vanishes in the diffuse gas emission, this ratio drops. Similar to the ratio between the mass predicted from model MF and that from model ME1 for all velocity channels, the ratio shown in Figure~\ref{fig.stat-perseus-yso-MF-oc-ac} is also sensitive to the evolutionary stage of the driving YSOs. Early evolutionary stage YSOs are likely to have less gas at high velocities compared to that of late evolutionary stage YSOs. Figure~\ref{fig.stat-perseus-yso-MF-oc-ac}  shows that for older YSOs, more gas is likely observable in \co\ emission at high velocities.

The distribution of the outflows outside the YSO catalog coverage in both Figure~\ref{fig.stat-perseus-yso-ME1-MF-ac} and \ref{fig.stat-perseus-yso-MF-oc-ac} spans a large range of ratios. This may indicate that the outflows outside the YSO catalog coverage include both early evolutionary stage and late evolutionary stage driving YSOs. 

The distribution of the outflows that have no nearby source in both Figure~\ref{fig.stat-perseus-yso-ME1-MF-ac} and \ref{fig.stat-perseus-yso-MF-oc-ac} is similar to those associated with younger YSOs. This in turn demonstrates that there is a chance the SESNA YSO catalog overlooks some faint/young sources. 

In both Figure~\ref{fig.stat-perseus-yso-ME1-MF-ac} and \ref{fig.stat-perseus-yso-MF-oc-ac}, most of the outflows have small ratios. This is because most of our sources are young and located in relatively dense gas regions. It takes $\sim$~50 kyr for the outflow to break out of the core and appear at high velocities in \co\ \citep{2014ApJ...784...61O}.

\section{Discussion}
\label{Discussion}

\subsection{Outflows in Star Clusters}
\label{Outflows in Star Clusters}

In this section, we present the performance of models ME1 and MF on the star cluster NGC 1333. NGC 1333 is one of the most active star forming regions in Perseus with multiple outflows driven by the protostellar cluster \citep{1996AJ....111.1964L,2000A&A...361..671K,2008ApJ...674..336G}. Astronomers have identified tens of outflows in this region \citep{1996ApJ...473L..49B,2000A&A...361..671K,2010ApJ...715.1170A}. \citet{2010ApJ...715.1170A} identified 16 individual outflows in \co\ (1-0), while \citet{2010MNRAS.408.1516C} identified 27 distinct outflows in \co\ (3-2) in NGC 1333. In total, there are over one hundred YSOs in this region \citep{2008ApJ...674..336G,2009ApJS..184...18G,Gutermuth.in.prep}. Due to the intense star formation going on, most outflows in \co\ (1-0) blend with each other, making them difficult to separate. 
\CASItD\ provides a means to separate the outflow emission from the rest of the cloud.

\begin{figure*}[hbt!]
\centering
\includegraphics[width=0.98\linewidth]{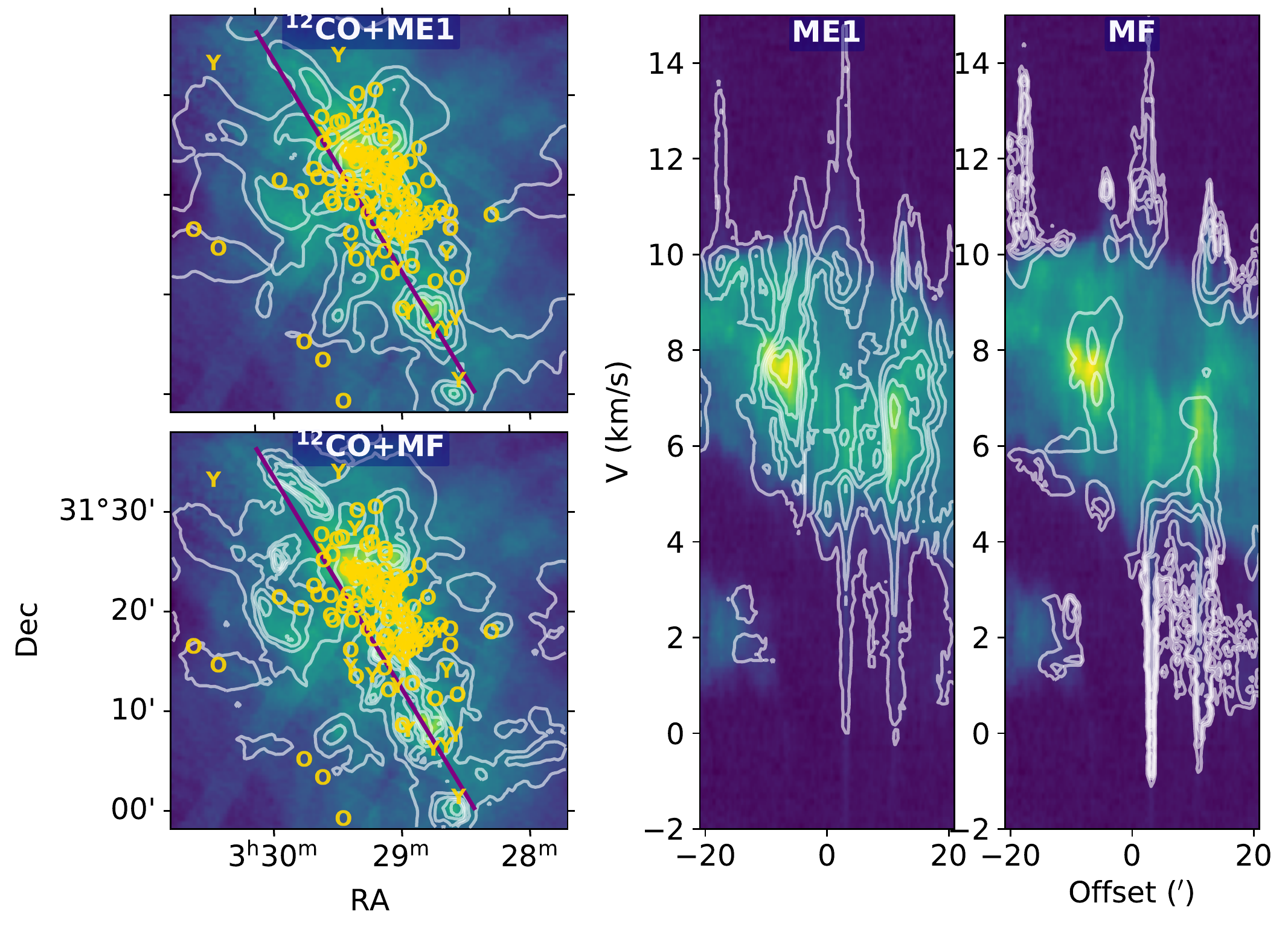}
\caption{Position-velocity diagram of \co\ emission toward NGC 1333. Left panel: integrated intensity of \co\ over the the full velocity range (from -2 km/s to 15 km/s) overlaid with the model ME1 and MF predictions in white contours. Letters ``Y'' and ``O'' mark YSO positions, as described in Figure~\ref{fig.pred-perseus-lm-region-mask}. The purple line illustrates the cut direction of the position-velocity diagram. Middle and right panel: position-velocity diagram of \co\ emission overlaid with the model ME1 and MF predictions in white contoursx.}
\label{fig.perseus-lm-pv-new-5-ngc1333}
\end{figure*}

Figure~\ref{fig.perseus-lm-pv-new-5-ngc1333} demonstrates the performance of model ME1 and MF on NGC 1333. Both models identify coherent high-velocity structures that are similar to outflows in the position-velocity diagram in Figure~\ref{fig.perseus-lm-pv-new-5-ngc1333}. In Figure~\ref{fig.perseus-lm-pv-new-5-ngc1333}, we randomly choose a position-velocity cut direction that spans as many YSOs as possible. The right panels show that the predictions by models ME1 and MF are similar in the high velocity channels, but the prediction by model ME1 is more extended towards the cloud rest frame velocity. The fraction of gas associated with outflows drops significantly at cloud velocities, which explains the different predictions by the models. As expected, in projection the identified outflow emission covers most of NGC 1333, but model MF successfully excludes most of the cloud emission. Other position-velocity cuts show similar results in NGC 1333.

The position-velocity diagram in Figure~\ref{fig.perseus-lm-pv-new-5-ngc1333} indicates that both models predict that the left bottom blob is part of an outflow, although this blob is isolated in velocity. This feature is likely a high velocity component of the cloud. However, the morphology exhibits velocity variation, where a flipped ``V'' structure appears. This shape is a signature of feedback as discussed in \citet{2011ApJ...742..105A}. We cannot distinguish the difference between outflow structures and high velocity cloud components visually or using the models with the \co\ data alone.

\subsection{Outflows without Driving Sources and False Detections}
\label{Outflows without Driving Sources and False Detections}

\begin{figure*}[hbt!]
\centering
\includegraphics[width=0.98\linewidth]{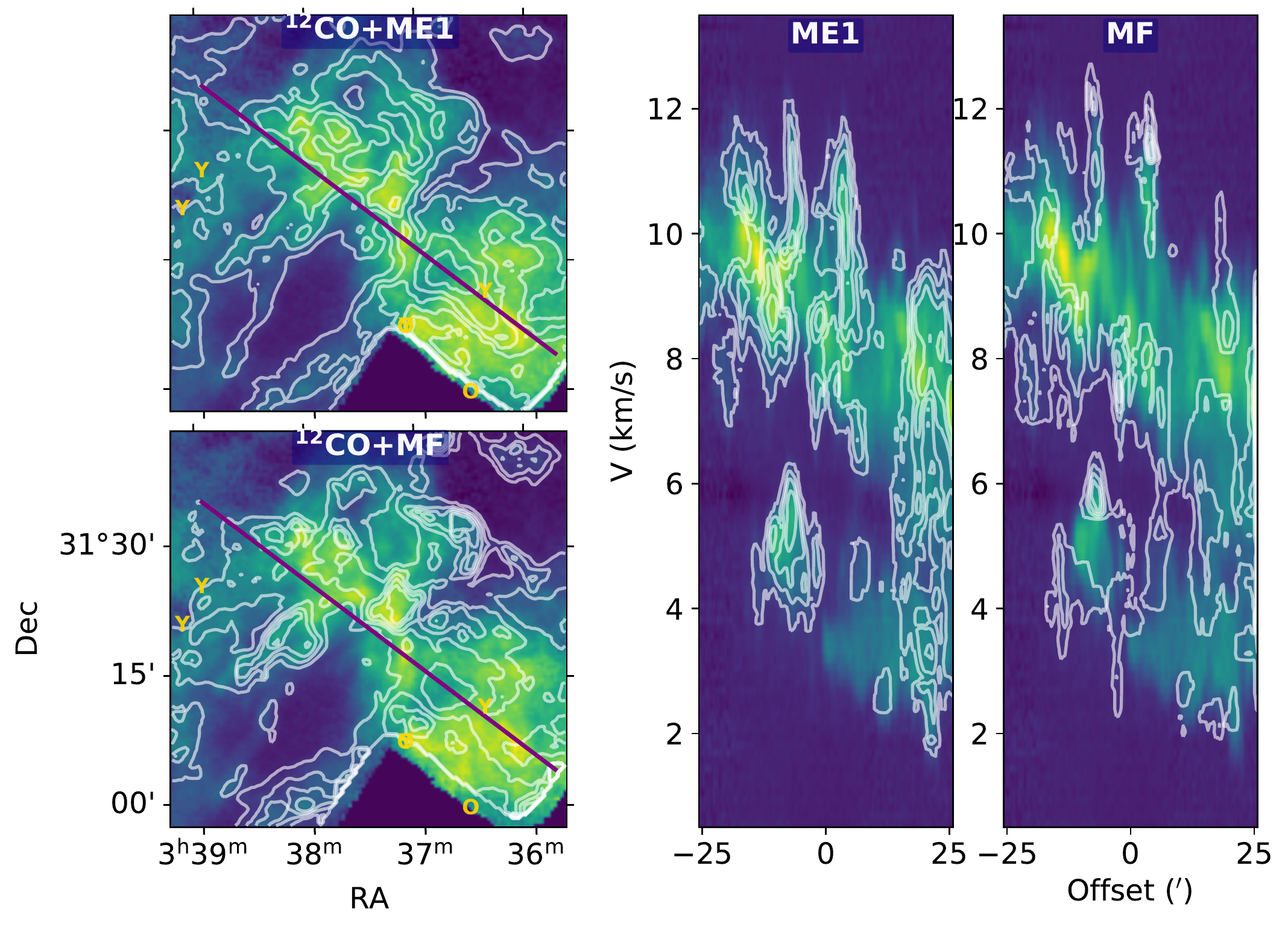}
\caption{Position-velocity diagram of \co\ emission toward a region with few YSOs. Left panel: integrated intensity of \co\ over the the full velocity range (from -2 km/s to 15 km/s) overlaid with the model ME1 and MF predictions in white contours. Letters ``Y'' and ``O'' mark YSO positions, as described in Figure~\ref{fig.pred-perseus-lm-region-mask}. The purple line illustrates the cut direction of the position-velocity diagram. Middle and right panel: position-velocity diagram of \co\ emission overlaid with the model ME1 and MF predictions in white contours.}
\label{fig.perseus-lm-pv-new-1-few-yso}
\end{figure*}

In this section, we present the performance of models ME1 and MF on a region with few YSOs. We select the region to the east of B1, where the YSO density is relatively low. Figure~\ref{fig.perseus-lm-pv-new-1-few-yso} presents the performance of model ME1 and MF on this low YSO density region. Although there is only one YSO along the cut direction, both models identify coherent high-velocity structures whose morphologies are consistent with outflows far away from the YSO. The predictions by models ME1 and MF are indistinguishable from the morphology of high-confidence outflows that have obvious driving sources. We cannot identify if these are true outflows without ancillary data. 

As indicated in Figure~\ref{fig.perseus-lm-pv-new-1-few-yso}, this region has a sharp velocity gradient from 10~km/s to 6~km/s. It bridges two subregions, B1 and B3, which have two different central velocities. The position-velocity diagram in Figure~\ref{fig.perseus-lm-pv-new-1-few-yso} illustrates a significant number of high velocity features. Many mechanisms may cause this, including but not limited to cloud formation, cloud-cloud collision or gas phase transition due to radiation \citep{2014A&A...571A..32M,2014ApJ...791L..23N}. Most of these coherent high velocity structures are very similar to confirmed outflow structures with obvious driving sources. It is possible that both models may have false detections that are not produced by feedback but by other mechanisms. This illustrates that clouds have high-velocity features that are indistinguishable -- either visually or using our method -- from high-confident outflow signatures. Consequently, we caution that machine learning models are not ``magic bullets'' and must be applied with care.

\subsection{Case Studies}
\label{Case Studies}

\subsubsection{Discrepancies Between the Two Model Predictions}
\label{Discrepancies Between the Two Model Predictions}

\begin{figure*}[hbt!]
\centering
\includegraphics[width=0.98\linewidth]{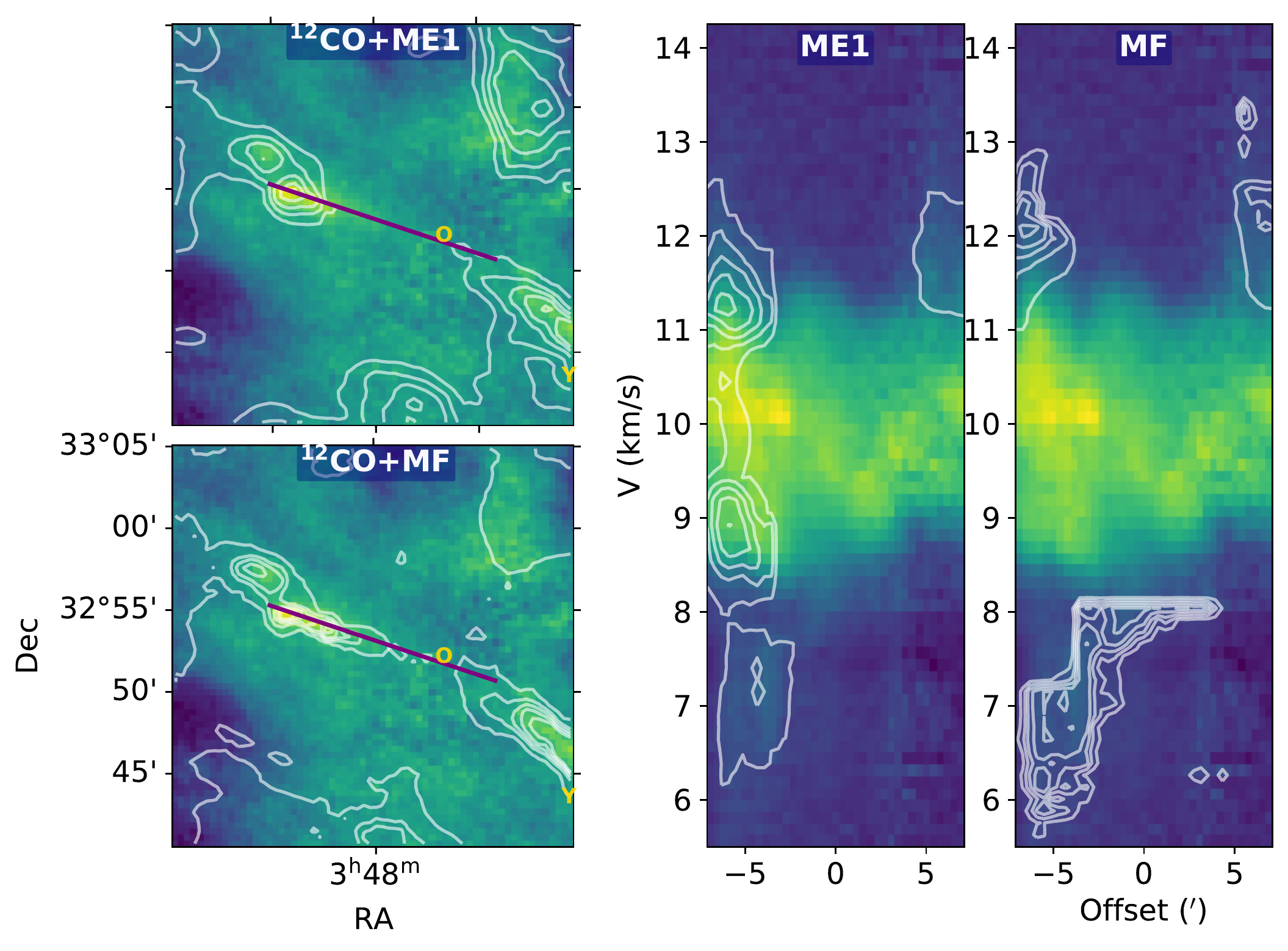}
\caption{Position-velocity diagram of \co\ emission toward a region where the predictions by model ME1 and MF are discrepant. Left panel: integrated intensity of \co\ over the the full velocity range (from -2 km/s to 15 km/s) overlaid with the model ME1 and MF predictions in white contours. Letters ``Y'' and ``O'' mark YSO positions, as described in Figure~\ref{fig.pred-perseus-lm-region-mask}. The purple line illustrates the cut direction of the position-velocity diagram. Middle and right panel: position-velocity diagram of \co\ emission overlaid with the model ME1 and MF predictions in white contours.}
\label{fig.perseus-lm-pv-new-0-diff-pred}
\end{figure*}

In this section, we discuss a case where the two models make significantly different predictions. In most regions, the predictions by the two models are similar at high velocities. The model ME1 prediction is often more extended towards the cloud rest frame velocity compared to that by model MF, but the overall coherent high velocity structures identified by the two models are similar. However, there are some regions where the predictions by the models are different. Figure~\ref{fig.perseus-lm-pv-new-0-diff-pred} shows an example of the performance of models ME1 and MF toward a region where the predictions are discrepant. Model MF predicts a more extended outflow structure compared to model ME1. In the position-velocity diagram, the two predictions are similar except in the middle. A late evolutionary stage YSO located on the cut direction likely drives the two outflow features on each side. It is ambiguous whether the feature identified by model MF but not by model ME1 is a true outflow. The high-velocity feature is not as distinct as other outflows. We can recognize some faint diffuse emission highlighted by model MF only in the center around 8 km/s. This diffuse emission seems associated with the left side blob that is identified by both models. Model MF is likely to identify a more extended outflow structure than model ME1, but not necessarily a new individual outflow. The presence of the YSO lends confidence to the model MF prediction, which appears as part of the outflow. However, due to the discrepancy, we only consider the left feature identified by both models as a high-confidence outflow candidate.

\subsubsection{A Previously Identified Outflow: IRAS~03282+3035}
\label{A Previously Identified Outflow: IRAS 03282+3035}

In this section, we discuss the performance of the two models on previously identified outflow IRAS~03282+3035. \citet{2014ApJ...783...29D} conducted a \co\ (2-1) and \co\ (3-2) survey towards 28 molecular outflows driven by low-mass protostars that are isolated spatially and kinematically. Among these outflows, IRAS 03282+3035 is located in Perseus and identified by our two models but missed by \citet{2010ApJ...715.1170A}. 

\begin{figure*}[hbt!]
\centering
\includegraphics[width=0.98\linewidth]{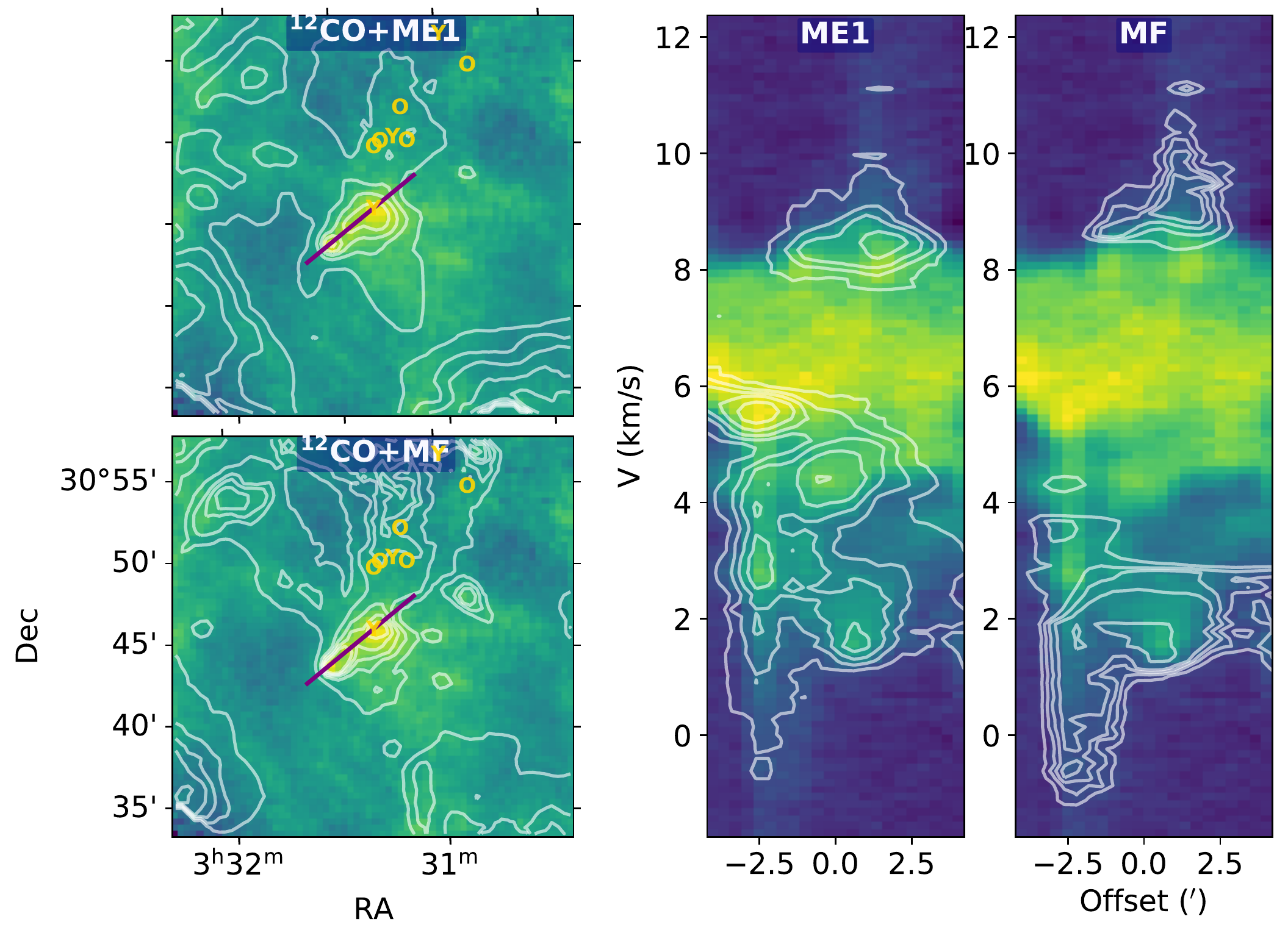}
\caption{Position-velocity diagram of \co\ emission toward outflow IRAS~03282+3035. Left panel: integrated intensity of \co\ over the the full velocity range (from -2 km/s to 15 km/s) overlaid with the model ME1 and MF predictions in white contours. Letters ``Y'' and ``O'' mark YSO positions, as described in Figure~\ref{fig.pred-perseus-lm-region-mask}. The purple line illustrates the cut direction of the position-velocity diagram. Middle and right panel: position-velocity diagram of \co\ emission overlaid with the model ME1 and MF predictions in white contours.}
\label{fig.perseus-lm-pv-new-1-iras}
\end{figure*}

Figure~\ref{fig.perseus-lm-pv-new-1-iras} shows the performance of models ME1 and MF toward IRAS~03282+3035. The morphology of the \co\ (1-0) is similar but more extended compared to that in the \co\ (2-1) and the \co\ (3-2) emission in \citet{2014ApJ...783...29D}. Both models identify coherent high-velocity features. Since this outflow is close to a cluster of YSOs and \co\ (1-0) has better sensitivity toward diffuse gas, the prediction by both models extends to a wider area than the \co\ (2-1) and \co\ (3-2) emission. 

Next, we compare the physical properties calculated by our models and those reported by \citet{2014ApJ...783...29D}. \citet{2014ApJ...783...29D} conscientiously calculate the outflow mass, energy and momentum by considering several correction factors, including opacity and outflow emission at low velocities confused with ambient cloud emission. We adopt the same box size as in \citet{2014ApJ...783...29D} to constrain the outflow area. Model MF corrects contamination by the cloud in low-velocity channels, which is similar to the method in \citet{2014ApJ...783...29D}. The outflow mass predicted by Model MF is 0.2~\msun, while \citet{2014ApJ...783...29D} calculate it to be 0.43~\msun. This is only a factor of two difference. \citet{2014ApJ...783...29D} adopt an excitation temperature of 50 K to calculate the outflow mass, however, we adopt 25 K for the calculation. A higher excitation temperature indicates an approximately linearly increased mass. If we adopt 50 K as the excitation temperature in the calculation, we get an outflow mass 0.4~\msun, which is consistent with the result, 0.43~\msun, in \citet{2014ApJ...783...29D}. 
The result that using different transition lines return a similar result is promising, which in turn gives confidence in our approach and assumptions. 

The outflow 1D momentum and 1D energy predicted by Model MF are 0.4 $M_{\odot}$ km/s and $1.9\times 10^{43}$ ergs, while \citet{2014ApJ...783...29D} finds 2.1 $M_{\odot}$ km/s and $1.4\times 10^{44}$ ergs. This is a factor of 5 difference in momentum and a factor of 7 difference in energy. The main reason for the difference is due to the velocity range. As pointed out in \citet{2014ApJ...783...29D}, the minimum velocity of IRAS~03282+3035 is 6.0~km/s and the maximum velocity is 25.9~km/s. While in our analysis, the \co\ (1-0) has a velocity coverage between -2~km/s and 15~km/s. A small amount of gas located in extremely high-velocity channels contributes a significant amount of momentum ($\propto v$) and energy ($\propto v^{2}$). On the other hand, \citet{2014ApJ...783...29D} observed \co\ (2-1) with a RMS of 0.04~K, but our resampled \co\ (1-0) has a RMS of 0.17~K. The outflow emission vanishes into the noise in the high-velocity channels of \co\ (1-0). This may also explain the differences in the outflow properties.

\section{Conclusions}
\label{Conclusions}

We adopt the deep learning method \CASItD\ to identify protostellar outflows in \co\ spectral cubes. By creating different training sets, we develop two deep machine learning models. Model ME1 predicts the position of the outflows. Model MF predicts the fraction of the mass associated with the outflows.  Our main findings are the following:

\begin{enumerate}

\item We apply \CASItD\ to Perseus and successfully identify 60 previously visually identified outflows.

\item We identify 20 new high-confidence outflows in Perseus using \CASItD. All of these have coherent high-velocity structure and nearby YSOs.

\item The outflow mass in Perseus predicted by model MF is comparable to the outflow mass calculated by \citet{2010ApJ...715.1170A}. This similarity is due to a cancelation in errors: \citet{2010ApJ...715.1170A} miss outflow material emitting in the cloud rest-frame channels, however, they compensate for this by over-estimating the amount of mass in high-velocity channels that have foreground and background contamination.

\item The total 1D momentum, 56.4 $M_{\odot}$ km/s, and 1D energy, $2.6\times 10^{45}$ ergs, from outflows in Perseus are on the same order of magnitude as the 1D calculations in \citet{2010ApJ...715.1170A}, which are 49.2 $M_{\odot}$ km/s and $1.4\times 10^{45}$ ergs, respectively. 

\item We find outflows likely driven by older YSOs have more gas ejected at high velocities compared to those driven by younger YSOs.

\item We use \CASItD\ to identify an extended amount of outflow gas around the NGC 1333 region, which is difficult to visually identify individual outflows due to the intense star formation. 

\end{enumerate}

In future work, we plan to apply \CASItD\ to more active star forming regions where it is not possible to cleanly separate outflow signatures visually.

We thank Hector Arce for the helpful discussion. {We also thank the anonymous referee for comments that improved this manuscript.} D.X., S.S.R.O., R.A.G. and C.V.O. were supported by NSF grant AST-1812747. S.S.R.O. also acknowledges support from NSF Career grant AST-1748571. R.A.G. also acknowledges support from NASA ADAP grant NNX17AF24G. The authors acknowledge the Texas Advanced Computing Center (TACC) at the University of Texas at Austin for providing HPC resources that have contributed to the research results reported within this paper.

\appendix
\section{Exploring Different Outflow Definitions}
\label{Exploring Different Outflow Definitions}

\begin{figure*}[hbt!]
\centering
\includegraphics[width=0.98\linewidth]{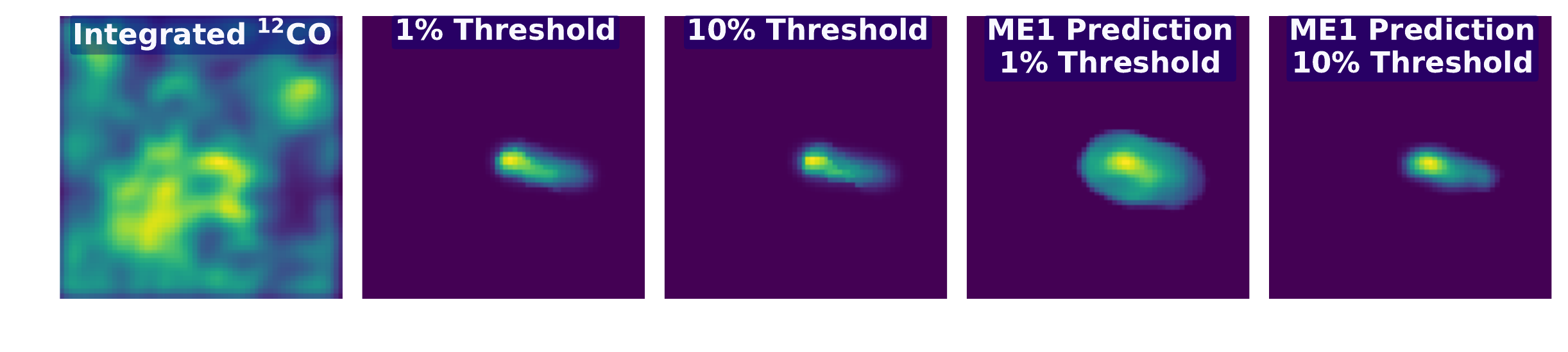}
\caption{Results of two ME1 models with different tracer fraction thresholds applied to a synthetic outflow. First panel: integrated intensity of \co. Second panel: integrated tracer field with a 1\% minimum tracer fraction. Third panel: integrated tracer field with a 10\% minimum tracer fraction. Fourth panel: predicted integrated intensity from model ME1 trained using the 1\% threshold to identify outflows. Fifth panel: predicted integrated intensity by model ME1 trained using a 10\% threshold. }
\label{fig.pred-synth-outflow-4}
\end{figure*}

\begin{figure*}[hbt!]
\centering
\includegraphics[width=0.98\linewidth]{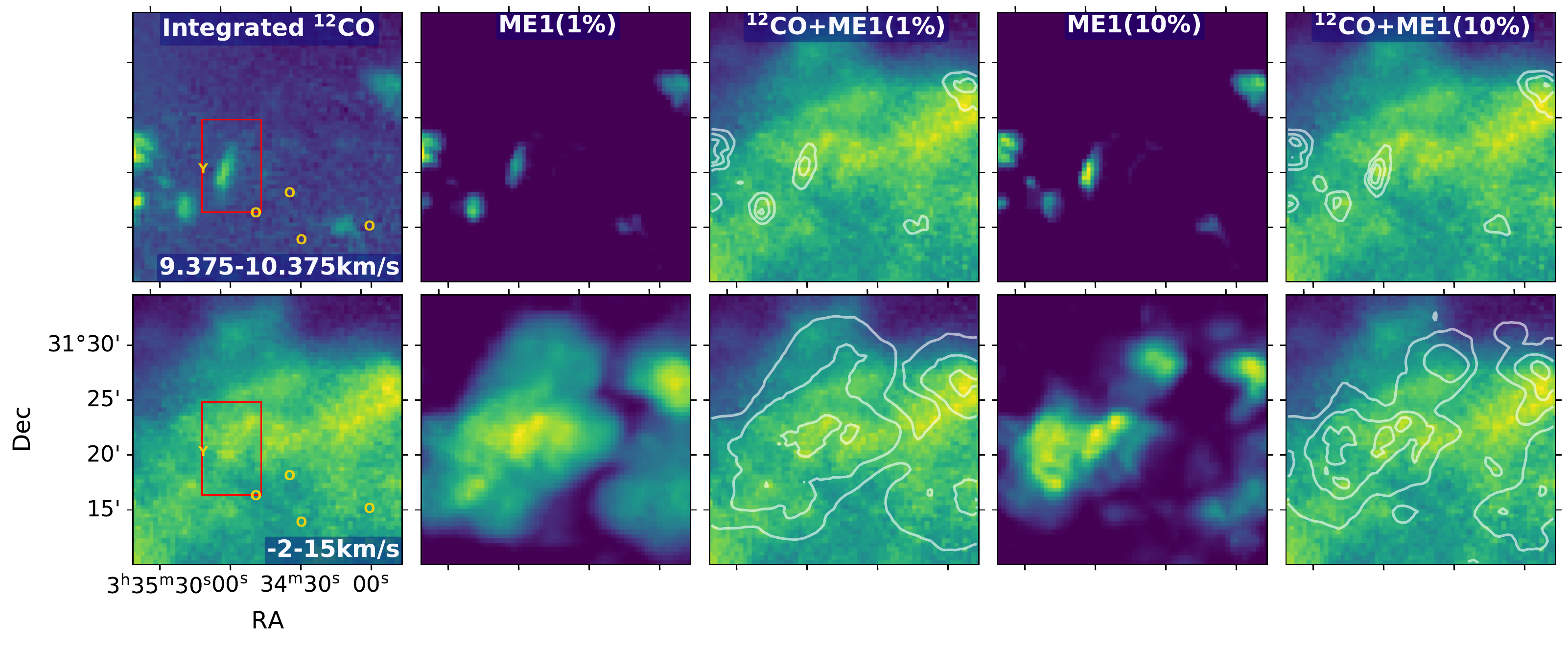}
\caption{Results of two ME1 models with different thresholds applied to the perviously identified Perseus outflow CPOC 35. In the upper row, first panel: integrated intensity of \co\ over the outflow velocity channels. Letters ``Y'' and ``O'' mark YSO positions, as described in Figure~\ref{fig.pred-perseus-lm-region-mask}. Second panel: predicted intensity integrated along the velocity axis from model ME1 with a 1\% threshold. The third panel: integrated intensity of \co\ over only the outflow velocity channels overlaid with the model ME1 (1\%) prediction (white contour). Fourth and fifth panels: same as the second and third panels but for model ME1 trained using a 10\% threshold. The lower row is the same as the upper row but is integrated over the full velocity range.}
\label{fig.pred-perseus-outflow-comp-35}
\end{figure*}

In this section, we assess the impact of different thresholds on the derived outflow mass. We examine thresholds of values of 1\% and 10\% for the minimum tracer fraction for which material is defined as part of an outflow. Figure~\ref{fig.pred-synth-outflow-4} shows an example for two ME1 models with different thresholds applied to a synthetic outflow. The prediction by model ME1 using a 1\% minimum tracer fraction is more extended compared to that trained using 10\%. The morphology of the tracer field with a 1\% threshold and that with a 10\% threshold is almost identical. However, the prediction by model ME1 trained using a 10\% minimum tracer fraction better reproduces the morphology of the outflow without contamination around the outflow boundary. On average, the model ME1 trained using a 1\% minimum tracer fraction overestimates the mass by a factor of two compared to that trained using a 10\% minimum tracer fraction. Figure~\ref{fig.pred-perseus-outflow-comp-35} demonstrates the performance of the two ME1 models with different thresholds on a perviously identified outflow in Perseus. In the outflow channels, the predictions by the two ME1 models are almost identical. In contrast, the integrated prediction over the entire velocity range from model ME1 with a 1\% threshold is more extended than that with a 10\% threshold. We conclude that using a 10\% threshold reduces diffuse contamination from ambient gas. Consequently, we adopt model ME1 with a 10\% threshold as the fiducial model.

 \section{Training Results}
\label{Training Results}

\begin{figure}[hbt!]
\centering
\includegraphics[width=0.48\linewidth]{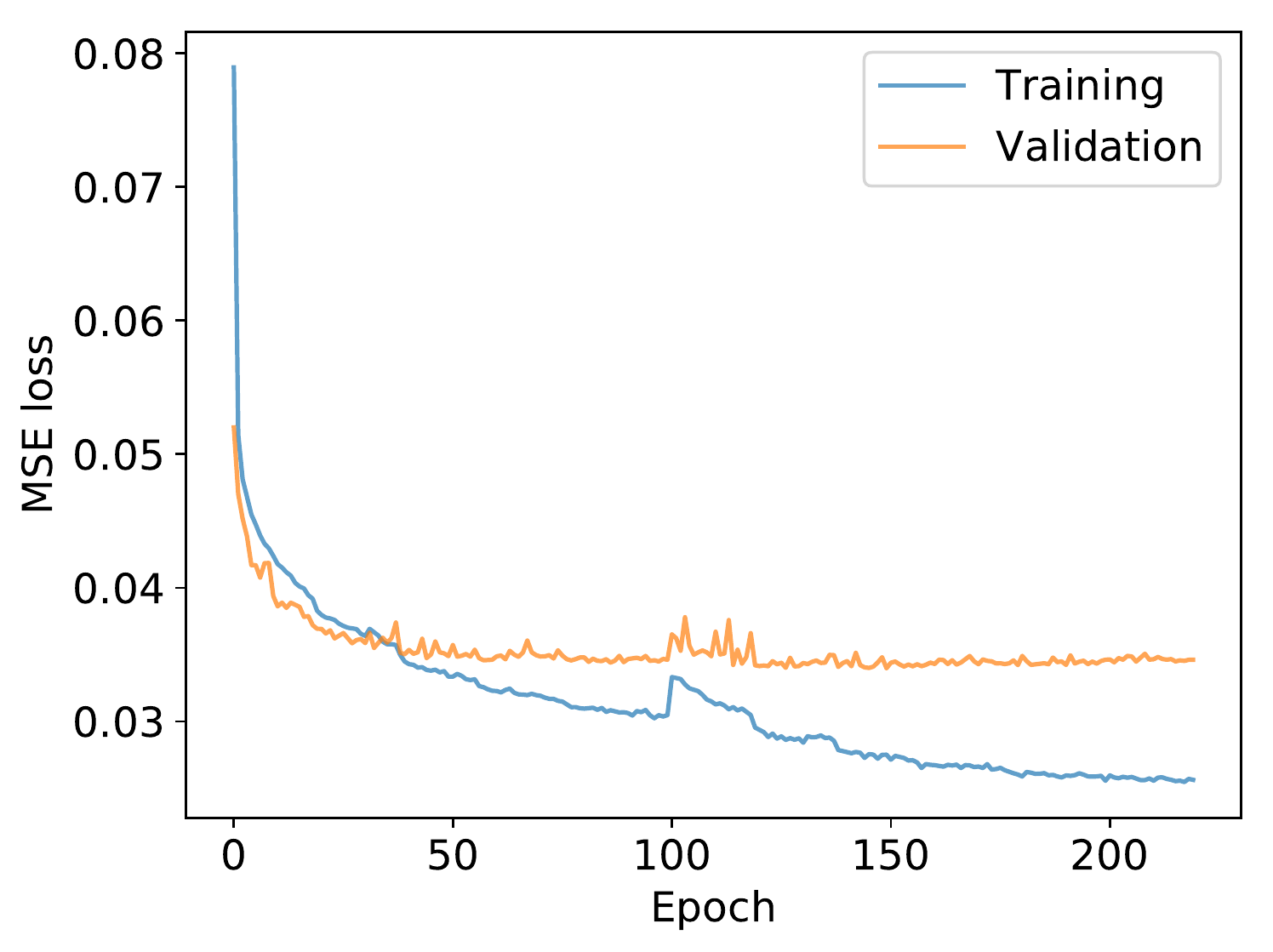}
\caption{Training and validation errors of model ME1 during training.}
\label{fig.outflow-lossfunction-ME1}
\end{figure}

\begin{figure}[hbt!]
\centering
\includegraphics[width=0.48\linewidth]{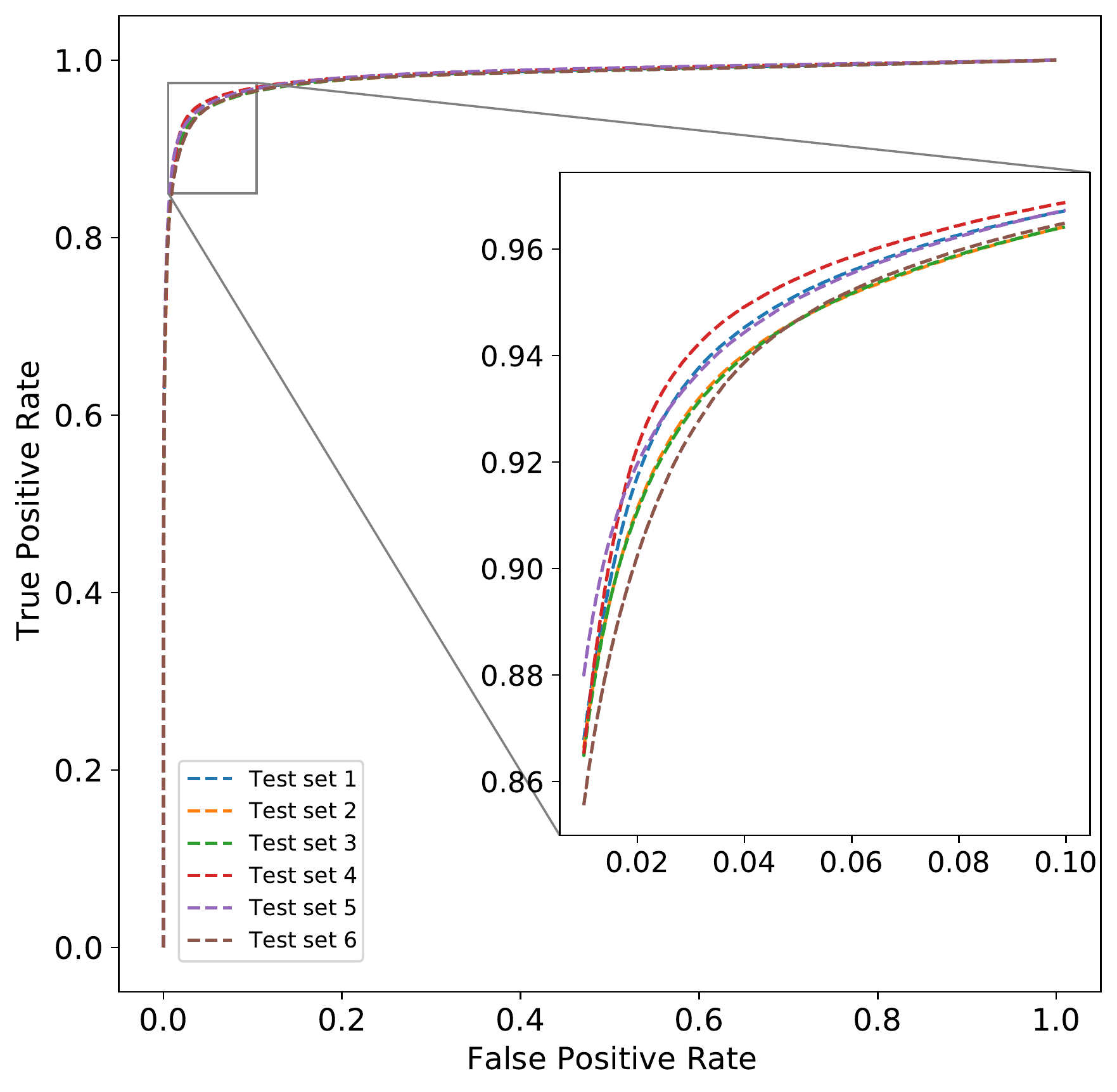}
\caption{Receiver operating characteristic curve of model MF tested on different test sets.}
\label{fig.outflow-ROC-MF}
\end{figure}

After training, we find model ME1 converges to a mean-square-error (MSE) below 0.05. Figure~\ref{fig.outflow-lossfunction-ME1} shows the training and validation errors of model ME1. After 220 epochs, this model converges to a MSE of 0.035. Since the validation error flattens, we stop training after 220 epochs. Figure~\ref{fig.outflow-ROC-MF} indicates the receiver operating characteristic (ROC) curve of model MF. We assess the model MF performance on six different test sets. We achieve 95\% accuracy within a 5\% false positive rate on all six test sets.

{ 
 
\section{Quantitatively Evaluating Model Performance}
\label{Quantitatively Evaluating Model Performance}

\subsection{Performance on Test Set Data}
\label{Performance on Test Set Data}

In this section, we quantitatively evaluate the performance of both models on the outflow mass estimates under different physical and chemical conditions. We calculate the mean mass and its uncertainty for each simulated outflow with different kinetic temperatures and different \co\ abundances in the test set. Figure~\ref{fig.errorbar-comp-cnn2-mass} shows the relation between the \CASItD\ predicted outflow mass and the true mass for different outflows, where error bars indicate the standard deviation of the mass prediction on a simulated outflow with two different kinetic temperatures (10 and 14 K) and three different \co\ abundances ($10^{-4}$, $5\times 10^{-5}$, $10^{-5}$). On average, model ME1 overestimates the outflow mass by a factor of 5, while model MF is able to correctly predict the outflow mass within a reasonable uncertainty under different physical and chemical conditions. 

\begin{figure}[hbt!]
\centering
\includegraphics[width=0.48\linewidth]{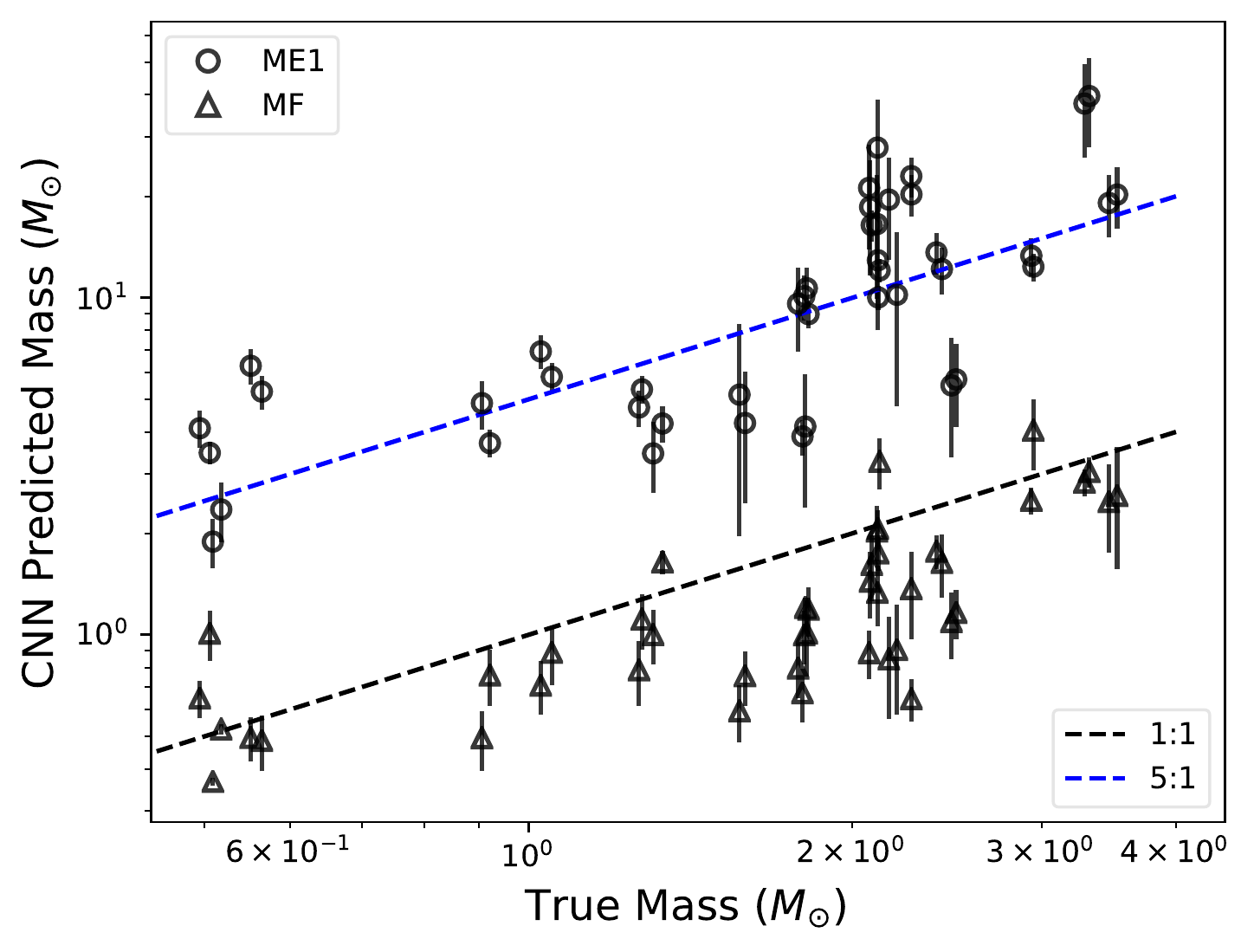}
\caption{Reproduced from Figure~\ref{fig.tacer-comp-cnn2}, relation between the \CASItD\ predicted outflow mass and the true mass for different outflows. Circle symbols indicate the mass calculated by model ME1. Triangle symbols represent the mass calculated by model MF. Error bars indicate the standard deviation of the mass prediction on a simulating outflow under 6 different physical and chemical conditions. The black dashed line indicates where \CASItD\ correctly predicts the true mass. The blue dashed line has a slope of 5.}
\label{fig.errorbar-comp-cnn2-mass}
\end{figure}

\subsection{Performance on New Conditions}
\label{Performance on New Conditions}

In this section, we explore the performance of both models on synthetic outflows whose conditions are not included in our training set. This exercise examines the uncertainty when applying both models to different observations under different conditions.

We test the two models performance on a simulated outflow with different beam sizes than we used in our training set: 100$^{\prime\prime}$, 25$^{\prime\prime}$, and 10$^{\prime\prime}$. Both models are trained on a training set that only includes synthetic observations with a beam size of 50$^{\prime\prime}$. Figure~\ref{fig.outflow-comp-beamsize} shows the two models performance on a simulated outflow with different beam sizes. Both models are able to capture the morphology of the outflow even though images with these beam sizes are not included in the training set.

We then quantitatively assess the performance of both models on outflow mass estimates under different conditions. The training set only includes synthetic observations with these parameters: a beam size of 50$^{\prime\prime}$, a noise level of 0.17 K, kinetic temperatures of 10 K and 14 K, and \co/\h2 abundances of $10^{-4}$, $5\times 10^{-5}$, and $10^{-5}$. We explore the two models performance on synthetic outflows under more conditions: beam sizes of 100$^{\prime\prime}$, 50$^{\prime\prime}$, 25$^{\prime\prime}$, and 10$^{\prime\prime}$,  noise levels of 0.06, 0.17, and 0.51 K, kinetic temperatures of 10, 14, and 20 K, and \co/\h2 abundances of $10^{-4}$, $5\times 10^{-5}$, and $10^{-5}$. In total, we test the two models on 108 synthetic outflows with different combinations of parameters, among which only 6 combinations of parameters are included in the training set. Figure~\ref{fig.mass-stat-test-one} shows the distribution of the mass predicted by the two models on 108 synthetic outflows with different combinations of parameters. Both models are able to predict the outflow mass consistently under different conditions. The mean mass predicted by model ME1 is 5.94 \msun, with a standard deviation of 2 \msun. The mean mass predicted by model MF is 0.76 \msun, with a standard deviation of 0.32 \msun. The true mass of these synthetic outflow is 1.04 \msun. Model MF is able to predict the outflow mass under different conditions within a factor of two. Table~\ref{Prediction on Different Conditions} gives the masses predicted by the two models for synthetic outflows under several different conditions. We adopt the mean masses predicted by both models on the 6 synthetic outflows whose parameters are included in the training set as the fiducial values for comparison. We also calculate the mean and the standard deviation of the masses of the synthetic outflows with different parameters. 
 
As indicated in Table~\ref{Prediction on Different Conditions}, different parameters have different effects on the outflow mass prediction. We discuss them separately as follows.

Model ME1 overestimates the outflow mass when the beam size is large. When the beam size is large, more gas located near the rest-frame velocity that is not associated with the outflow is included in the model ME1 prediction, which yields a larger mass estimate. In contrast, Model MF underestimates the mass when the beam size is large. Model MF is more sensitive to outflow emission in the high-velocity channels where the the fraction of mass associated with feedback is high. The relatively faint outflow emission in the high-velocity channels vanishes into the background noise due to beam smearing. Model MF fails to capture the morphology of the outflow in the high-velocity channels, which yields a smaller mass estimate. We calculate the mean and the standard deviation of the masses of synthetic outflows with different beam sizes, which are 0.95 \msun and 0.30 \msun. There is a factor of two between the maximum and the minimum predicted mass of the synthetic outflows with different beam sizes.

The estimated mass increases for both lower and higher noise levels. As indicated in Table~\ref{Prediction on Different Conditions}, when the noise level is increased by a factor of 3, the masses predicted by both models increase. This is due to the contamination by the noise. When the noise level is decreased by a factor of 3, the masses predicted by both models also increase. This is because that the outflow emission is more distinct, and both models are able to identify a more complete morphology. We calculate the mean and the standard deviation of the masses of synthetic outflows with different noise levels, which are 1.02 \msun and 0.24 \msun. There is a factor of 1.6 between the maximum and the minimum predicted mass of synthetic outflows with different noise levels.

When the \co/\h2 abundance drops, the mass predicted by model MF decreases. Since we use a constant \co/\13co ratio, when \co/\h2 decreases by a factor of 10, the abundance of \13co\ also drops by a factor of 10. Under this circumstance, \13co\ emission of the outflow is fainter and difficult to detect. In most voxels of the low \co\ and \13co\ abundance synthetic outflow, we can only rely on \co\ emission to calculate the outflow mass rather than combining both \co\ and \13co. \co\ is usually optical thick, so we underestimate the outflow mass based on \co\ only. We calculate the mean and the standard deviation of the masses of synthetic outflows with different \co/\h2 abundances, which are 0.64 \msun and 0.15 \msun. There is a factor of 1.6 between the maximum and the minimum predicted mass of the synthetic outflows with different \co/\h2 abundances.

There is only a weak dependence between the cloud kinetic temperature and the mass estimates by the two models. The averaged kinetic temperature of launched gas is higher than the mean cloud kinetic temperature. We adopt a constant excitation temperature of 25 K when we calculate outflow mass as discussed in Section~\ref{Assessing Model Accuracy Using Synthetic Observations}. Consequently, the cloud kinetic temperature plays a limited role in setting the emission of the gas associated with feedback. We calculate the mean and the standard deviation of the masses of synthetic outflows with different kinetic temperatures, which is 0.83 \msun and 0.03 \msun.

We find that the beam sizes, noise levels and \co/\h2 abundances dominate the uncertainty of outflow mass estimates. Kinetic temperatures do not significantly affect the outflow mass estimation. These variations at most introduce a factor of 2 change in the estimated mass. To conclude, we demonstrate that \CASItD\ performs well on other observations whose conditions are not included in the training set.

\begin{figure}[hbt!]
\centering
\includegraphics[width=0.98\linewidth]{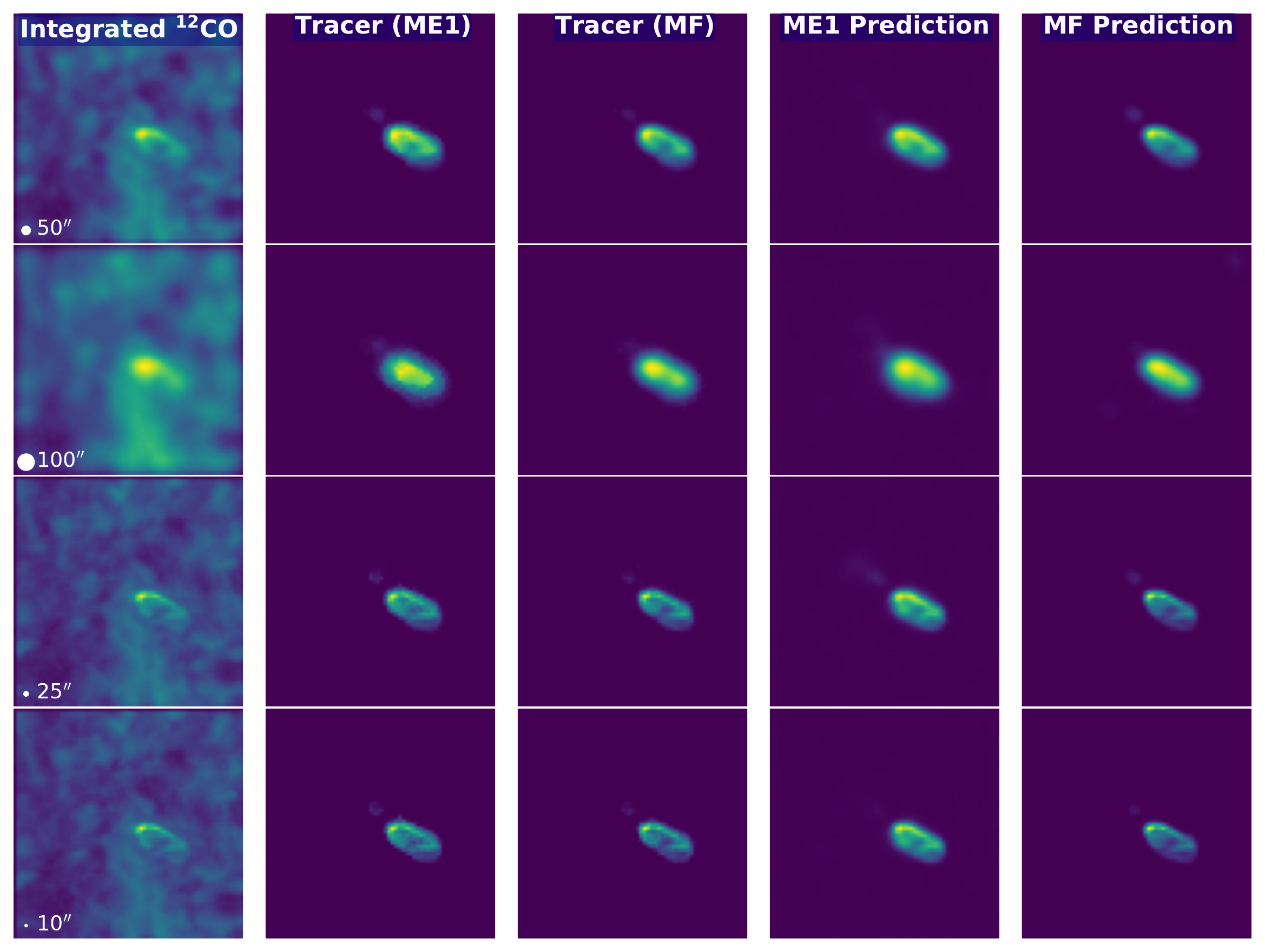}
\caption{Performance of the two models on a simulated outflow with different beam sizes.}
\label{fig.outflow-comp-beamsize}
\end{figure}

\begin{figure}[hbt!]
\centering
\includegraphics[width=0.58\linewidth]{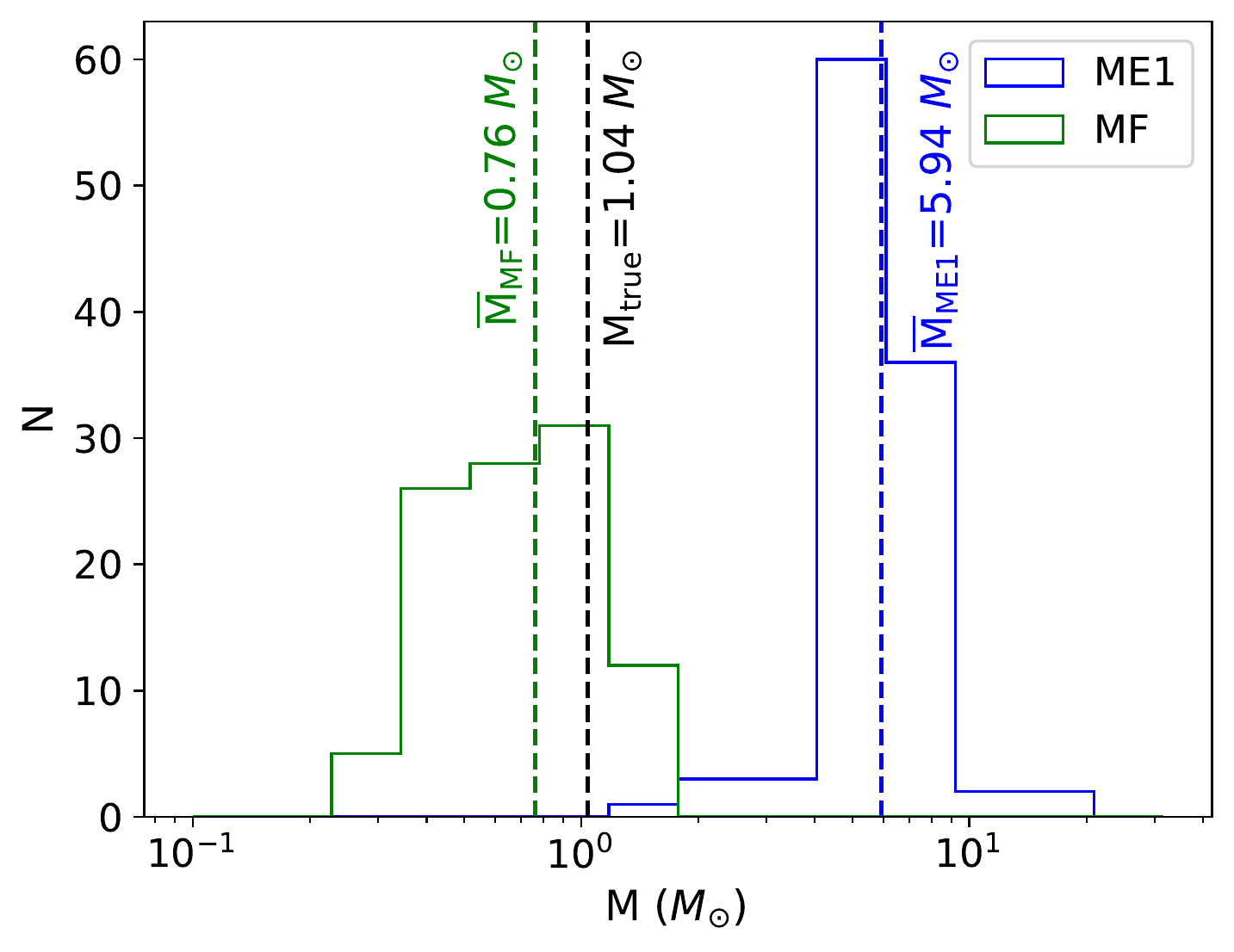}
\caption{Distribution of the mass predicted by the two models on 108 synthetic outflows with different combinations of parameters.}
\label{fig.mass-stat-test-one}
\end{figure}

\begin{table*}[]
\begin{center}
\caption{\CASItD\ Predictions on Synthetic Outflows with Different Conditions$^{a}$ \label{Prediction on Different Conditions}}
\begin{tabular}{ccccccccc}
\hline
No. &  $T_{\rm k} (K)$ & \co/\h2 & Beam ($^{\prime\prime}$) & Noise (K)   & M$_{\rm ME1}$(\msun) & M$_{\rm MF}$(\msun) & $\frac{\rm M_{ME1}-M_{ME1,fid}}{\rm M_{ME1,fid}}$ & $\frac{\rm M_{MF}-M_{MF,fid}}{\rm M_{MF,fid}}$ \\
 \hline

    fiducial &(10, 14) & (10, 5, 1)$\times 10^{-5}$       & 50    & 0.17  & 5.19  & 0.65  &  -     & - \\
    1     & 14    & $10^{-4}$ & 50    & 0.17  & 5.55  & 0.82  & 6.91\% & 26.27\% \\
    2     & 14    & $10^{-4}$ & 100   & 0.17  & 7.10  & 0.53  & 36.72\% & -17.81\% \\
    3     & 14    & $10^{-4}$ & 25    & 0.17  & 5.55  & 1.13  & 6.96\% & 73.65\% \\
    4     & 14    & $10^{-4}$ & 10    & 0.17  & 5.00  & 1.31  & -3.71\% & 101.25\% \\
    5     & 14    & $10^{-4}$ & 50    & 0.51  & 6.29  & 1.36  & 21.14\% & 109.36\% \\
    6     & 14    & $10^{-4}$ & 50    & 0.06  & 5.78  & 0.87  & 11.34\% & 33.37\% \\
    7     & 14    & $5 \times 10^{-5}$ & 50    & 0.17  & 5.74  & 0.65  & 10.55\% & -0.27\% \\
    8     & 14    & $10^{-5}$ & 50    & 0.17  & 5.64  & 0.45  & 8.66\% & -30.44\% \\
    9     & 10    & $10^{-4}$ & 50    & 0.17  & 4.55  & 0.86  & -12.41\% & 32.71\% \\
    10    & 20    & $10^{-4}$ & 50    & 0.17  & 5.97  & 0.80  & 15.06\% & 22.69\% \\
 \hline \hline
    (1,2,3,4) & 14  &$10^{-4}$   &-& 0.17  & 5.80$\pm$0.78  & 0.95$\pm$0.30 & - & -\\
    (1,5,6) & 14    &$10^{-4}$  & 50    &-& 5.87$\pm$0.31  & 1.02$\pm$0.24 & - & -\\
    (1,7,8) & 14    &-       & 50    & 0.17  & 5.64$\pm$0.08  & 0.64$\pm$0.15 & - & -\\
    (1,9,10) &-& $10^{-4}$& 50    & 0.17  & 5.36$\pm$0.60  & 0.83$\pm$0.03 & - & -\\
    (1-10)  & - & -  & -   & -  & 5.72$\pm$0.65  & 0.88$\pm$0.29 & - & - \\
    
 \hline
 
\multicolumn{9}{p{0.99\linewidth}}{Notes:}\\
\multicolumn{9}{p{0.99\linewidth}}{
$^{a}$ Synthetic observation list, kinetic temperature, \co\ to \h2\ abundance, beam size, noise level, mass predicted by model ME1, mass predicted by model MF, the relative error of the mass predicted by model ME1 compared with the fiducial ME1 mass, the relative error of the mass predicted by model MF compared with the fiducial MF mass. The true mass for the simulated outflow is 1.04 \msun.}\\
\end{tabular}
\end{center}
\end{table*}

}

\bibliographystyle{aasjournal}
\bibliography{references}

\begin{thebibliography}{}
\expandafter\ifx\csname natexlab\endcsname\relax\def\natexlab#1{#1}\fi
\providecommand{\url}[1]{\href{#1}{#1}}
\providecommand{\dodoi}[1]{doi:~\href{http://doi.org/#1}{\nolinkurl{#1}}}
\providecommand{\doeprint}[1]{\href{http://ascl.net/#1}{\nolinkurl{http://ascl.net/#1}}}
\providecommand{\doarXiv}[1]{\href{https://arxiv.org/abs/#1}{\nolinkurl{https://arxiv.org/abs/#1}}}

\bibitem[{{Arce} {et~al.}(2011){Arce}, {Borkin}, {Goodman}, {Pineda}, \&
  {Beaumont}}]{2011ApJ...742..105A}
{Arce}, H.~G., {Borkin}, M.~A., {Goodman}, A.~A., {Pineda}, J.~E., \&
  {Beaumont}, C.~N. 2011, \apj, 742, 105, \dodoi{10.1088/0004-637X/742/2/105}

\bibitem[{{Arce} {et~al.}(2010){Arce}, {Borkin}, {Goodman}, {Pineda}, \&
  {Halle}}]{2010ApJ...715.1170A}
{Arce}, H.~G., {Borkin}, M.~A., {Goodman}, A.~A., {Pineda}, J.~E., \& {Halle},
  M.~W. 2010, \apj, 715, 1170, \dodoi{10.1088/0004-637X/715/2/1170}

\bibitem[{{Bachiller}(1996)}]{1996ARA&A..34..111B}
{Bachiller}, R. 1996, \araa, 34, 111, \dodoi{10.1146/annurev.astro.34.1.111}

\bibitem[{{Bachiller} {et~al.}(1991){Bachiller}, {Martin-Pintado}, \&
  {Fuente}}]{1991A&A...243L..21B}
{Bachiller}, R., {Martin-Pintado}, J., \& {Fuente}, A. 1991, \aap, 243, L21

\bibitem[{{Bally}(2016)}]{2016ARA&A..54..491B}
{Bally}, J. 2016, \araa, 54, 491, \dodoi{10.1146/annurev-astro-081915-023341}

\bibitem[{{Bally} {et~al.}(1996){Bally}, {Devine}, \&
  {Reipurth}}]{1996ApJ...473L..49B}
{Bally}, J., {Devine}, D., \& {Reipurth}, B. 1996, \apjl, 473, L49,
  \dodoi{10.1086/310381}

\bibitem[{{Beaumont} {et~al.}(2014){Beaumont}, {Goodman}, {Kendrew},
  {Williams}, \& {Simpson}}]{2014ApJS..214....3B}
{Beaumont}, C.~N., {Goodman}, A.~A., {Kendrew}, S., {Williams}, J.~P., \&
  {Simpson}, R. 2014, \apjs, 214, 3, \dodoi{10.1088/0067-0049/214/1/3}

\bibitem[{{Beaumont} {et~al.}(2011){Beaumont}, {Williams}, \&
  {Goodman}}]{2011ApJ...741...14B}
{Beaumont}, C.~N., {Williams}, J.~P., \& {Goodman}, A.~A. 2011, \apj, 741, 14,
  \dodoi{10.1088/0004-637X/741/1/14}

\bibitem[{{Cunningham} {et~al.}(2011){Cunningham}, {Klein}, {Krumholz}, \&
  {McKee}}]{2011ApJ...740..107C}
{Cunningham}, A.~J., {Klein}, R.~I., {Krumholz}, M.~R., \& {McKee}, C.~F. 2011,
  \apj, 740, 107, \dodoi{10.1088/0004-637X/740/2/107}

\bibitem[{{Curtis} {et~al.}(2010){Curtis}, {Richer}, {Swift}, \&
  {Williams}}]{2010MNRAS.408.1516C}
{Curtis}, E.~I., {Richer}, J.~S., {Swift}, J.~J., \& {Williams}, J.~P. 2010,
  \mnras, 408, 1516, \dodoi{10.1111/j.1365-2966.2010.17214.x}

\bibitem[{{Cyganowski} {et~al.}(2008){Cyganowski}, {Whitney}, {Holden},
  {Braden}, {Brogan}, {Churchwell}, {Indebetouw}, {Watson}, {Babler},
  {Benjamin}, {Gomez}, {Meade}, {Povich}, {Robitaille}, \&
  {Watson}}]{2008AJ....136.2391C}
{Cyganowski}, C.~J., {Whitney}, B.~A., {Holden}, E., {et~al.} 2008, \aj, 136,
  2391, \dodoi{10.1088/0004-6256/136/6/2391}

\bibitem[{{Draine}(2011)}]{2011piim.book.....D}
{Draine}, B.~T. 2011, {Physics of the Interstellar and Intergalactic Medium}

\bibitem[{{Dullemond} {et~al.}(2012){Dullemond}, {Juhasz}, {Pohl}, {Sereshti},
  {Shetty}, {Peters}, {Commercon}, \& {Flock}}]{2012ascl.soft02015D}
{Dullemond}, C.~P., {Juhasz}, A., {Pohl}, A., {et~al.} 2012, {RADMC-3D: A
  multi-purpose radiative transfer tool}, Astrophysics Source Code Library.
\newblock \doeprint{1202.015}

\bibitem[{{Dunham} {et~al.}(2014){Dunham}, {Arce}, {Mardones}, {Lee},
  {Matthews}, {Stutz}, \& {Williams}}]{2014ApJ...783...29D}
{Dunham}, M.~M., {Arce}, H.~G., {Mardones}, D., {et~al.} 2014, \apj, 783, 29,
  \dodoi{10.1088/0004-637X/783/1/29}

\bibitem[{{Federrath} {et~al.}(2014){Federrath}, {Schr{\"o}n}, {Banerjee}, \&
  {Klessen}}]{2014ApJ...790..128F}
{Federrath}, C., {Schr{\"o}n}, M., {Banerjee}, R., \& {Klessen}, R.~S. 2014,
  \apj, 790, 128, \dodoi{10.1088/0004-637X/790/2/128}

\bibitem[{{Flaherty} {et~al.}(2007){Flaherty}, {Pipher}, {Megeath}, {Winston},
  {Gutermuth}, {Muzerolle}, {Allen}, \& {Fazio}}]{2007ApJ...663.1069F}
{Flaherty}, K.~M., {Pipher}, J.~L., {Megeath}, S.~T., {et~al.} 2007, \apj, 663,
  1069, \dodoi{10.1086/518411}

\bibitem[{{Foster} {et~al.}(2009){Foster}, {Rosolowsky}, {Kauffmann}, {Pineda},
  {Borkin}, {Caselli}, {Myers}, \& {Goodman}}]{2009ApJ...696..298F}
{Foster}, J.~B., {Rosolowsky}, E.~W., {Kauffmann}, J., {et~al.} 2009, \apj,
  696, 298, \dodoi{10.1088/0004-637X/696/1/298}

\bibitem[{{Frank} {et~al.}(2014){Frank}, {Ray}, {Cabrit}, {Hartigan}, {Arce},
  {Bacciotti}, {Bally}, {Benisty}, {Eisl{\"o}ffel}, {G{\"u}del}, {Lebedev},
  {Nisini}, \& {Raga}}]{2014prpl.conf..451F}
{Frank}, A., {Ray}, T.~P., {Cabrit}, S., {et~al.} 2014, in Protostars and
  Planets VI, ed. H.~{Beuther}, R.~S. {Klessen}, C.~P. {Dullemond}, \&
  T.~{Henning}, 451, \dodoi{10.2458/azu_uapress_9780816531240-ch020}

\bibitem[{{Goldsmith} \& {Langer}(1978)}]{1978ApJ...222..881G}
{Goldsmith}, P.~F., \& {Langer}, W.~D. 1978, \apj, 222, 881,
  \dodoi{10.1086/156206}

\bibitem[{{Gutermuth} {et~al.}(2009){Gutermuth}, {Megeath}, {Myers}, {Allen},
  {Pipher}, \& {Fazio}}]{2009ApJS..184...18G}
{Gutermuth}, R.~A., {Megeath}, S.~T., {Myers}, P.~C., {et~al.} 2009, \apjs,
  184, 18, \dodoi{10.1088/0067-0049/184/1/18}

\bibitem[{{Gutermuth} {et~al.}(2008){Gutermuth}, {Myers}, {Megeath}, {Allen},
  {Pipher}, {Muzerolle}, {Porras}, {Winston}, \& {Fazio}}]{2008ApJ...674..336G}
{Gutermuth}, R.~A., {Myers}, P.~C., {Megeath}, S.~T., {et~al.} 2008, \apj, 674,
  336, \dodoi{10.1086/524722}

\bibitem[{{Gutermuth et al.}(in prep)}]{Gutermuth.in.prep}
{Gutermuth et al.} in prep

\bibitem[{{Hansen} {et~al.}(2012){Hansen}, {Klein}, {McKee}, \&
  {Fisher}}]{2012ApJ...747...22H}
{Hansen}, C.~E., {Klein}, R.~I., {McKee}, C.~F., \& {Fisher}, R.~T. 2012, \apj,
  747, 22, \dodoi{10.1088/0004-637X/747/1/22}

\bibitem[{He {et~al.}(2016)He, Zhang, Ren, \& Sun}]{He2016}
He, K., Zhang, X., Ren, S., \& Sun, J. 2016, in Proceedings of the IEEE
  conference on computer vision and pattern recognition, 770--778

\bibitem[{{Jayasinghe} {et~al.}(2019){Jayasinghe}, {Dixon}, {Povich}, {Binder},
  {Velasco}, {Lepore}, {Xu}, {Offner}, {Kobulnicky}, {Anderson}, {Kendrew}, \&
  {Simpson}}]{2019MNRAS.488.1141J}
{Jayasinghe}, T., {Dixon}, D., {Povich}, M.~S., {et~al.} 2019, \mnras, 488,
  1141, \dodoi{10.1093/mnras/stz1738}

\bibitem[{{Knee} \& {Sandell}(2000)}]{2000A&A...361..671K}
{Knee}, L.~B.~G., \& {Sandell}, G. 2000, \aap, 361, 671

\bibitem[{{Krumholz} {et~al.}(2004){Krumholz}, {McKee}, \&
  {Klein}}]{2004ApJ...611..399K}
{Krumholz}, M.~R., {McKee}, C.~F., \& {Klein}, R.~I. 2004, \apj, 611, 399,
  \dodoi{10.1086/421935}

\bibitem[{{Lada} {et~al.}(1996){Lada}, {Alves}, \&
  {Lada}}]{1996AJ....111.1964L}
{Lada}, C.~J., {Alves}, J., \& {Lada}, E.~A. 1996, \aj, 111, 1964,
  \dodoi{10.1086/117933}

\bibitem[{{Li} {et~al.}(2015){Li}, {Li}, {Qian}, {Xu}, {Goldsmith},
  {Noriega-Crespo}, {Wu}, {Song}, \& {Nan}}]{2015ApJS..219...20L}
{Li}, H., {Li}, D., {Qian}, L., {et~al.} 2015, \apjs, 219, 20,
  \dodoi{10.1088/0067-0049/219/2/20}

\bibitem[{{Li} {et~al.}(2012){Li}, {Martin}, {Klein}, \&
  {McKee}}]{2012ApJ...745..139L}
{Li}, P.~S., {Martin}, D.~F., {Klein}, R.~I., \& {McKee}, C.~F. 2012, \apj,
  745, 139, \dodoi{10.1088/0004-637X/745/2/139}

\bibitem[{{Mac Low}(1999)}]{1999ApJ...524..169M}
{Mac Low}, M.-M. 1999, \apj, 524, 169, \dodoi{10.1086/307784}

\bibitem[{{Machida} \& {Hosokawa}(2013)}]{2013MNRAS.431.1719M}
{Machida}, M.~N., \& {Hosokawa}, T. 2013, \mnras, 431, 1719,
  \dodoi{10.1093/mnras/stt291}

\bibitem[{{Martin-Pintado} {et~al.}(1992){Martin-Pintado}, {Bachiller}, \&
  {Fuente}}]{1992A&A...254..315M}
{Martin-Pintado}, J., {Bachiller}, R., \& {Fuente}, A. 1992, \aap, 254, 315

\bibitem[{{Matzner}(2007)}]{2007ApJ...659.1394M}
{Matzner}, C.~D. 2007, \apj, 659, 1394, \dodoi{10.1086/512361}

\bibitem[{{Matzner} \& {Jumper}(2015)}]{2015ApJ...815...68M}
{Matzner}, C.~D., \& {Jumper}, P.~H. 2015, \apj, 815, 68,
  \dodoi{10.1088/0004-637X/815/1/68}

\bibitem[{{McKee} \& {Ostriker}(2007)}]{2007ARA&A..45..565M}
{McKee}, C.~F., \& {Ostriker}, E.~C. 2007, \araa, 45, 565,
  \dodoi{10.1146/annurev.astro.45.051806.110602}

\bibitem[{{Motte} {et~al.}(2014){Motte}, {Nguy{\^e}n Luong}, {Schneider},
  {Heitsch}, {Glover}, {Carlhoff}, {Hill}, {Bontemps}, {Schilke}, {Louvet},
  {Hennemann}, {Didelon}, \& {Beuther}}]{2014A&A...571A..32M}
{Motte}, F., {Nguy{\^e}n Luong}, Q., {Schneider}, N., {et~al.} 2014, \aap, 571,
  A32, \dodoi{10.1051/0004-6361/201323001}

\bibitem[{{Myers} {et~al.}(2014){Myers}, {Klein}, {Krumholz}, \&
  {McKee}}]{2014MNRAS.439.3420M}
{Myers}, A.~T., {Klein}, R.~I., {Krumholz}, M.~R., \& {McKee}, C.~F. 2014,
  \mnras, 439, 3420, \dodoi{10.1093/mnras/stu190}

\bibitem[{{Myers}(2008)}]{2008ApJ...687..340M}
{Myers}, P.~C. 2008, \apj, 687, 340, \dodoi{10.1086/591664}

\bibitem[{{Nakamura} {et~al.}(2014){Nakamura}, {Sugitani}, {Tanaka},
  {Nishitani}, {Dobashi}, {Shimoikura}, {Shimajiri}, {Kawabe}, {Yonekura},
  {Mizuno}, {Kimura}, {Tokuda}, {Kozu}, {Okada}, {Hasegawa}, {Ogawa}, {Kameno},
  {Shinnaga}, {Momose}, {Nakajima}, {Onishi}, {Maezawa}, {Hirota}, {Takano},
  {Iono}, {Kuno}, \& {Yamamoto}}]{2014ApJ...791L..23N}
{Nakamura}, F., {Sugitani}, K., {Tanaka}, T., {et~al.} 2014, \apjl, 791, L23,
  \dodoi{10.1088/2041-8205/791/2/L23}

\bibitem[{{Narayanan} {et~al.}(2012){Narayanan}, {Snell}, \&
  {Bemis}}]{2012MNRAS.425.2641N}
{Narayanan}, G., {Snell}, R., \& {Bemis}, A. 2012, \mnras, 425, 2641,
  \dodoi{10.1111/j.1365-2966.2012.21579.x}

\bibitem[{{Offner} \& {Arce}(2014)}]{2014ApJ...784...61O}
{Offner}, S. S.~R., \& {Arce}, H.~G. 2014, \apj, 784, 61,
  \dodoi{10.1088/0004-637X/784/1/61}

\bibitem[{{Offner} \& {Chaban}(2017)}]{2017ApJ...847..104O}
{Offner}, S. S.~R., \& {Chaban}, J. 2017, \apj, 847, 104,
  \dodoi{10.3847/1538-4357/aa8996}

\bibitem[{{Offner} {et~al.}(2009){Offner}, {Klein}, {McKee}, \&
  {Krumholz}}]{2009ApJ...703..131O}
{Offner}, S. S.~R., {Klein}, R.~I., {McKee}, C.~F., \& {Krumholz}, M.~R. 2009,
  \apj, 703, 131, \dodoi{10.1088/0004-637X/703/1/131}

\bibitem[{{Pokhrel} {et~al.}(2020){Pokhrel}, {Gutermuth}, {Betti}, {Offner},
  {Myers}, {Megeath}, {Sokol}, {Ali}, {Allen}, {Allen}, {Dunham}, {Fischer},
  {Henning}, {Heyer}, {Hora}, {Pipher}, {Tobin}, \&
  {Wolk}}]{2020arXiv200505466P}
{Pokhrel}, R., {Gutermuth}, R.~A., {Betti}, S.~K., {et~al.} 2020, arXiv
  e-prints, arXiv:2005.05466.
\newblock \doarXiv{2005.05466}

\bibitem[{{Ridge} {et~al.}(2006{\natexlab{a}}){Ridge}, {Schnee}, {Goodman}, \&
  {Foster}}]{2006ApJ...643..932R}
{Ridge}, N.~A., {Schnee}, S.~L., {Goodman}, A.~A., \& {Foster}, J.~B.
  2006{\natexlab{a}}, \apj, 643, 932, \dodoi{10.1086/502957}

\bibitem[{{Ridge} {et~al.}(2006{\natexlab{b}}){Ridge}, {Di Francesco}, {Kirk},
  {Li}, {Goodman}, {Alves}, {Arce}, {Borkin}, {Caselli}, {Foster}, {Heyer},
  {Johnstone}, {Kosslyn}, {Lombardi}, {Pineda}, {Schnee}, \&
  {Tafalla}}]{2006AJ....131.2921R}
{Ridge}, N.~A., {Di Francesco}, J., {Kirk}, H., {et~al.} 2006{\natexlab{b}},
  \aj, 131, 2921, \dodoi{10.1086/503704}

\bibitem[{Ronneberger {et~al.}(2015)Ronneberger, Fischer, \&
  Brox}]{Ronneberger2015}
Ronneberger, O., Fischer, P., \& Brox, T. 2015, in International Conference on
  Medical image computing and computer-assisted intervention, Springer,
  234--241

\bibitem[{{Shallue} \& {Vanderburg}(2018)}]{2018AJ....155...94S}
{Shallue}, C.~J., \& {Vanderburg}, A. 2018, \aj, 155, 94,
  \dodoi{10.3847/1538-3881/aa9e09}

\bibitem[{{Van Oort} {et~al.}(2019){Van Oort}, {Xu}, {Offner}, \&
  {Gutermuth}}]{2019ApJ...880...83V}
{Van Oort}, C.~M., {Xu}, D., {Offner}, S. S.~R., \& {Gutermuth}, R.~A. 2019,
  \apj, 880, 83, \dodoi{10.3847/1538-4357/ab275e}

\bibitem[{{Xu} \& {Offner}(2017)}]{2017ApJ...851..149X}
{Xu}, D., \& {Offner}, S. S.~R. 2017, \apj, 851, 149,
  \dodoi{10.3847/1538-4357/aa9a42}

\bibitem[{{Xu} {et~al.}(2020){Xu}, {Offner}, {Gutermuth}, \&
  {Oort}}]{2020ApJ...890...64X}
{Xu}, D., {Offner}, S. S.~R., {Gutermuth}, R., \& {Oort}, C.~V. 2020, \apj,
  890, 64, \dodoi{10.3847/1538-4357/ab6607}

\bibitem[{{Zhang} {et~al.}(2005){Zhang}, {Hunter}, {Brand}, {Sridharan},
  {Cesaroni}, {Molinari}, {Wang}, \& {Kramer}}]{2005ApJ...625..864Z}
{Zhang}, Q., {Hunter}, T.~R., {Brand}, J., {et~al.} 2005, \apj, 625, 864,
  \dodoi{10.1086/429660}

\bibitem[{{Zhang} {et~al.}(2020){Zhang}, {Yang}, {Xu}, {Chen}, {Su}, {Sun},
  {Zhou}, {Li}, \& {Lu}}]{2020ApJS..248...15Z}
{Zhang}, S., {Yang}, J., {Xu}, Y., {et~al.} 2020, \apjs, 248, 15,
  \dodoi{10.3847/1538-4365/ab879a}

\end{thebibliography}

\end{CJK*}

\end{document}